\documentclass[10pt]{emulateapj}

\usepackage{epsfig}
\usepackage{amsmath}
\usepackage{natbib}

\begin{document}

\author{Rowin Meijerink$^1$}
\author{Marco Spaans$^1$}
\email{meijerink@astro.rug.nl, spaans@astro.rug.nl}
\author{Inga Kamp$^1$}
\author{Giambattista Aresu$^1$}
\author{Wing-Fai Thi$^2$}
\author{Peter Woitke$^3$}

\affiliation{$^1$ University of Groningen, Kapteyn Astronomical Institute, Postbus 800, 9700 AV, Groningen, The Netherlands}
\affiliation{$^2$ UJF-Grenoble, CNRS-INSU, Institute de Plan\`etologie et d'Astrophysique (IPAG) UMR 5274}
\affiliation{$^3$ SUPA, School of Physics and Astronomy, University of St. Andrews, KY16 9SS, UK}

\title{Tracing the Physical Conditions in
  Active Galactic Nuclei with Time-Dependent Chemistry}

\begin{abstract}

  We present an extension of the code ProDiMo that allows for a
  modeling of processes pertinent to active galactic nuclei and to an
  ambient chemistry that is time dependent. We present a
  proof-of-concept and focus on a few astrophysically relevant
  species, e.g., H$^+$, H$_2^+$ and H$_3^+$; C$^+$ and N$^+$; C and O;
  CO and H$_2$O; OH$^+$, H$_2$O$^+$ and H$_3$O$^+$; HCN and
  HCO$^+$. We find that the freeze-out of water is strongly suppressed
  and that this affects the bulk of the oxygen and carbon chemistry
  occurring in AGN. The commonly used AGN tracer HCN/HCO$^+$ is
  strongly time-dependent, with ratios that vary over orders of
  magnitude for times longer than $10^4$ years. Through ALMA
  observations this ratio can be used to probe how the narrow-line
  region evolves under large fluctuations in the SMBH accretion rate.
  Strong evolutionary trends, on time scales of $10^4-10^8$ years, are
  also found in species such as H$_3$O$^+$, CO, and H$_2$O.  These
  reflect, respectively, time dependent effects in the ionization
  balance, the transient nature of the production of molecular gas,
  and the freeze-out/sublimation of water.

\end{abstract}


\section{Introduction}

Actively star-forming galaxies, like ultra-luminous infrared galaxies
(ULIRGs), are generally highly obscured and may host an accreting
super-massive black hole (SMBH).  Observations at optical, UV, and
X-ray wavelengths do not allow one to probe the inner regions of these
systems. This is possible at far-infrared and millimeter wavelengths,
though. A major step forward is being made with the current generation
of ground and space based telescopes.  The Herschel Space Observatory
probes the window at wavelengths ranging from $\lambda= 50$ and
600~$\mu$m that is blocked by the Earth's atmosphere. It contains CO
and H$_2$O transitions that are suitable to diagnose the gas chemical
and thermal properties. These are affected by ultra-violet and X-ray
photons, which are emitted by newly formed stars and the central
SMBH. High spatial resolution (better than a few pc) observations of
the molecular gas around an active galactic nucleus (AGN) have not
been possible yet, but progress is foreseen when the Atacama Large
Millimeter Array (ALMA) becomes fully functional in 2013. This array
will consist of 66 telescope dishes, and will observe on scales
smaller than one parsec in the nearest active galaxies, resolving at
least the narrow line region around the SMBH.

One of the lessons learned from the Science Demonstration Phase of
Herschel was that the presence of both SMBH accretion and star
formation in Mrk 231 can be deduced from the CO spectral line energy
distribution (SLED). A SPIRE FTS spectrum with CO lines up to J=13-12
between wavelengths $\lambda=303 - 670$~$\mu$m \citep{VdWerf2010} was
obtained within the Open Time Key Program HerCULES (PI: Van der
Werf). The spectrum shows CO, [CI], and [NII] lines, and surprisingly
also strong lines of H$_2$O and the ionic species OH$^+$ and
H$_2$O$^+$. For the interpretation they used UV and X-ray dominated
semi-infinite slabs models \citep{Meijerink2005,Meijerink2007}. These
models calculate the thermal and chemical balance of gas irradiated by
UV or X-rays, and predict the molecular emission of, e.g., CO and
H$_2$O. They showed that the excitation of the CO ladder in the
circumnuclear molecular disk (CMD) is dominated by the nuclear X-ray
radiation field, out to a radius of $\sim$160 pc. In a companion
paper \citep{Gonzalez2010} the modeling was done for the H$_2$O lines
and the conclusions was that the gas excitation and chemistry in Mrk
231 is by affected shocks/cosmic rays, an XDR chemistry, and/or an
``undepleted chemistry'' where grain mantles are evaporated.

Although some lessons were drawn from the model results for Mrk 231,
it is not satisfactory to need separate modeling for, e.g., the CO and
H$_2$O lines. Also, we would like to extract more information than
ambient densities and radiation field strengths, such as the spatial
origin of the different observed lines. Therefore, it is required to
make more sophisticated models, which are able to deal with (1) the
new inventory of diagnostic lines obtained with Herschel and (2) the
high spatial resolution data from ALMA. Such a model should consider
the geometry of the system that consists of gas, stars and an
accreting SMBH in a single simulation. Indeed, the narrow-line region
has a (puffed-up) disk structure with the radiation from the
broad-line region very close to the SMBH impinging under a range of
angles. Our aim in this work is to provide such a model, but also one
that is fast enough to allow for a large parameter study, such that it
can be used by the scientific community to understand the nature of
the ISM located in the centers of nearby and high redshift, active
galaxies.

The chemical relaxation timescales (see the section on chemical
relaxtion timescales below for the precise definition we adopted) in
the UV and X-ray exposed warm, partly molecular layer are typically of
order $10^4$ yrs and can be as short as $\sim 1-100$ yrs for highly
ionized gas, $x_e \sim 1$, and moderate densities $n >
10^3-10^4$~cm$^{-3}$). The chemical relaxation timescale in the
regions where the ISM is shielded from both UV and X-ray radiation
($A_{\rm v} >> 1$ and $\tau(1 {\rm keV}) >> 1$, see
Fig. \ref{chem_time_scale}) can be as large as $10^8$ yrs. These long
timescales will occur at low dust and gas temperatures, and requires
that both FUV and X-ray ionization and heating are greatly reduced
from its initial value. Here, the gas is slowly freezing out on dust
grains, yielding a slow grain-surface chemistry. These timescales are
larger than the typical dynamical (orbital) timescales of $\tau_{\rm
  dyn}<10^6-10^7$ years at $<100$ pc from a $>10^7$ M$_\odot$ SMBH
with a surrounding torus of $\sim 10^9$ M$_{\odot}$, while supernova
are expected to stir up the medium on timescales of $\tau_{\rm SN}\sim
10^4$ yrs \citep{Wada2009}; a duration over which one would expect
significant changes in the (column) density, gas kinematics and
irradiation of orbiting molecular clouds. Work on AGN  \citep{Wada2009,
  Perez2011} has shown that an inhomogeneous swelled-up disk (on a 10
pc scale) is created by supernovae, with interstellar material
transiting in between cold/dense and warm/diffuse phases on a
dynamical time, while enjoying a range in UV and X-ray irradiations.
Furthermore, supernova explosions, aided by a possible outflow, can
drive shocks and cloud-cloud collisions that affect the gas phase
chemistry and cause sputtering/evaporation of dust grains, releasing
their icy mantles.

In light of the above, it is therefore likely that H$_2$O, crucial in
interstellar chemistry, is not entirely frozen out in dense clouds.
Also, important species like CO and H$_2$ may not be in chemical
equilibrium, while ion-molecule chemistry (e.g., C$^+$, OH$^+$,
HCO$^+$) can be boosted by large fluctuations in ionization rate.  All
these effects strongly alter the chemistry in the narrow line region
and thus the interpretation of observations that use molecules to
probe, a.o., the accretion flow towards the SMBH, the nature of star
formation in AGN, and the importance of mechanical and radiative
feedback.  To make progress in this field, it is thus necessary to
calculate the disk thermal and chemical balance as a function of time.

In this paper, as a first proof-of-concept step, we present a
two-dimensional cylindrically symmetric thermal and chemical model
with a parametrized distribution of gas and stars around a SMBH. We
first calculate the equilibrium thermal and chemical structure for a
generic model, and show that the chemical relaxation timescale in the
regions where the gas is cold ($T < 50$~K) and molecular ($x_{\rm H_2}
\sim 0.5$) is of order $10^8$~yrs, and completely depleted from CO and
H$_2$O. Given the dynamical and supernova timescale described above,
it is expected that the chemistry is not in equilibrium and that
grains are (partially) stripped from their ice-layers. Therefore, we
follow up with a calculation in which we follow the chemical
relaxation of the gas phase and surface chemistry. In that calculation
the thermal and hydrostatic structure is fixed. We compare the
chemical structures of the equilibrium and time-dependent solutions,
and discuss the implications on the observed line emission. This model
is calculated using a modified version of the ProDiMo \citep{Woitke2009,
  Woitke2011} (Protoplanetary Disk Model), that has been developed
originally to interpret observations of protoplanetary disks around
newly formed T Tauri and Herbig Ae/Be stars.

\section{Modifying the ProDiMo code to black hole environments}
 
The original ProDiMo code is a code that is used to study the
physical, chemical and thermal structure of gas and dust around T
Tauri and Herbig Ae/Be stars. The original code includes (1) frequency
dependent continuum radiative transfer (2) kinetic gas-phase and FUV
photochemistry, (3) ice-formation, and (4) detailed non-local
thermodynamic equilibrium (non-LTE) heating and cooling with (5) a
consistent calculation of the hydrostatic disk structure. The models
are characterized by a high degree of consistency among the various
physical, chemical, and radiative processes, since the mutual
feedbacks are solved iteratively. X-ray heating and chemical processes
are included, based on work of several authors \citep{Maloney1996,
  Glassgold2004, Meijerink2005}. It includes the most recent X-ray
physics \citep{Adamkovics2011} with an extension of the chemical network
with species such as Ne, Ar, and their doubly and singly ionized
species, as well as other heavy elements.

The model makes predictions of observables such as [CII] and [OI]
fine-structure lines, rotational lines of CO, and ro-vibrational lines
of H$_2$O, using an escape probability method \citep{Woitke2011}. The
code is set-up in a modular way. It allows to use either a prescribed
density structure, or to calculate the hydrostatic structure
self-consistently with the thermal-chemical structure. A full
parametrization of the structure is important to calculate large
model grids, while the self-consistent structure is more appropriate
in the case of detailed model for a specific object. Examples are the
modeling of HD 163296  \citep{Tilling2012} and a large grid with 300000
models  \citep{Woitke2011}. In two recent papers, discussing a grid of
240 models, the combined effects of X-ray and FUV radiation on the
disk physical structure were calculated for a range of radiation
fields, dust properties and surface density profiles
 \citep{Meijerink2012} and the statistical properties of diagnostic gas
lines were studied  \citep{Aresu2012}. The code allows to start from a
converged physical structure solution and then evolve the chemistry in
time after, e.g., changing the input spectrum.

\section{The generic torus model}
 
\begin{figure*}
  \centering
  \includegraphics[width=5.4cm]{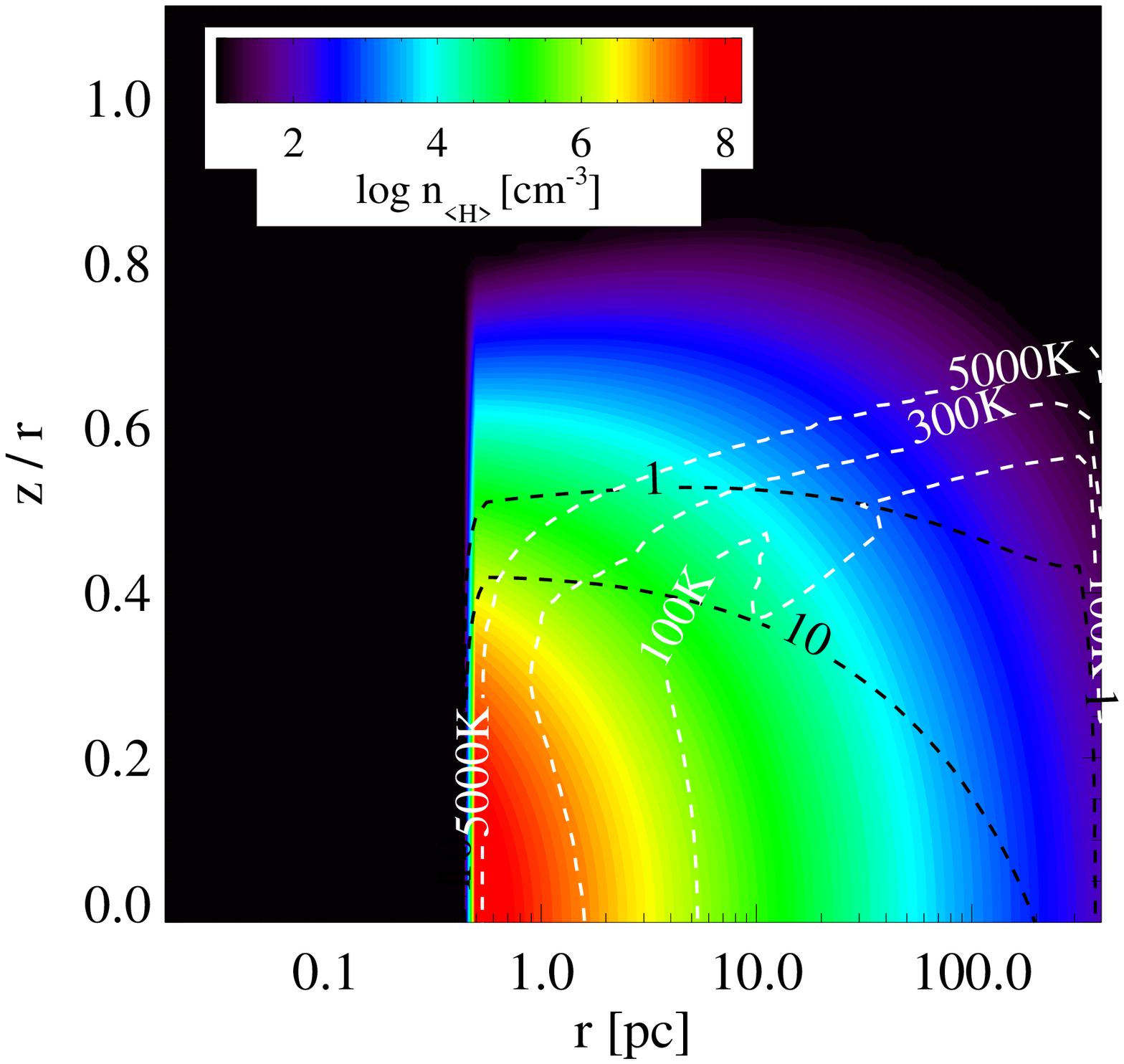}
  \includegraphics[width=5.4cm]{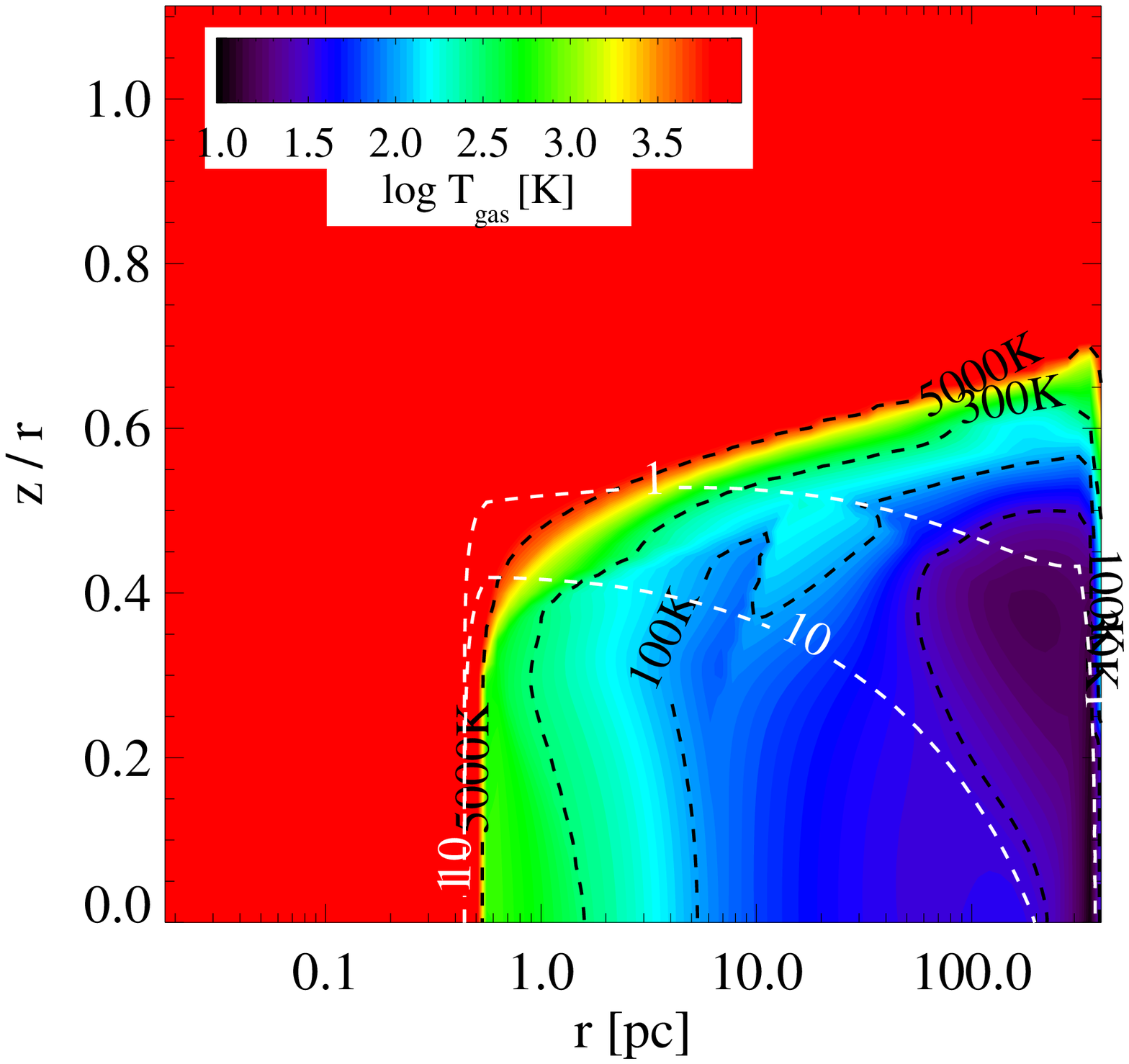}
  \includegraphics[width=5.4cm]{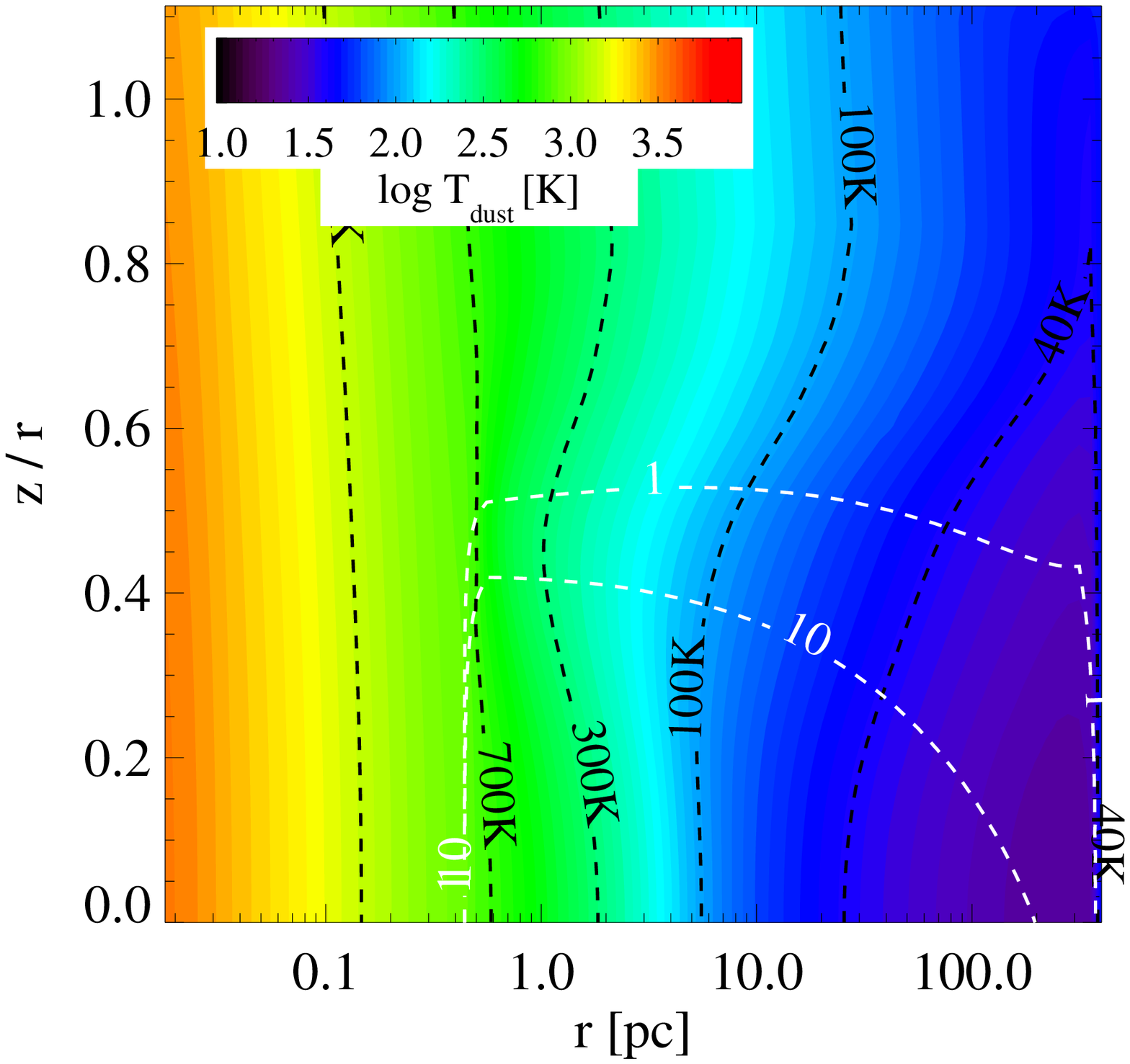}
  \caption{Gas number density (left), and gas temperature (middle) and
    dust temperature (right) for a model with disk mass $M_{\rm disk}
    = 10^9$~M$_{\odot}$ and $L_{\rm X} = 10^{43}$~erg~s$^{-1}$. The
    visual extinction $A_{\rm v}=1$ and 10 and temperatures $T_{\rm
      gas} = 100$, 300 and 5000 K are indicated with contours.Note:
    The vertical axis depicts the relative height, allowing a better
    visualization of the inner regions.}
  \label{density_temperature}
\end{figure*}

We scaled up the ProDiMo models in such a way that they represent the
circumnuclear molecular disk (CMD) around an accreting black
hole. Currently, supernova feedback and radiation pressure are not
included. For this reason, we do not use the iteration of global disk
structure to reach hydrostatic equilibrium, but we define a
pre-prescribed density structure, with a power law surface density
distribution, $\Sigma\propto R^{-1}$. The density distribution is
decoupled from the thermal structure of the torus, implicitly assuming
that the vertical scale-height is maintained by non-thermal processes,
such as turbulent motions. These turbulent motions would result in a
clumpy density structure, which is beyond the scope of this paper and
therefore not taken into account. The penetration depth of especially
UV (and to a lesser extent X-rays that have a lower absorption cross
section) is on average larger in a clumpy structure, and would result
in a less stratified structure \citep{Spaans1996}. Our generic torus
model has a gas mass, $M_{\rm disk}=10^9$~M$_{\odot}$, which is in the
range of gas masses found for ULIRGs in the nearby
universe \citep{Papadopoulos2012}. We set the inner radius $R_{\rm
  in}=0.5$~pc, and outer radius $R_{\rm out} = 400$~pc, the latter
being comparable to the inner disk size found with interferometric
imaging of CO $J=1-0$ and $J=2-1$ for Mrk 231 \citep{Downes1998}. We
note here that ``torus'' is used in a broad sense here, given that the
outer radius is three orders of magnitude larger than the inner radius
of the model. This dynamical range implies that this torus is not
meant to be seen as a cohesive structure. The scale height is $H_0 =
0.5$~pc at a reference radius of 3.0~pc. This gives a puffed up disk,
as suggested by numerical simulations \citep{Wada2009}. The total
luminosity of the central source is $L=10^{10}$~L$_\odot$. The X-ray
luminosity between $0.1$ and $10$~keV of the accreting black hole is
$L_{\rm X}=10^{43}$ erg~s$^{-1}$, which is a typical luminosity for a
sample of local Syfert 1 and 2 galaxies \citep{Lutz2004}, and has a
power law distribution with index $\beta=-0.9$. The disk has a flaring
index of $\alpha=1.1$ (where flaring implies the increase of $H(r)/r$
with radius). The global interstellar radiation field is set to
$\chi=10^3$ (the Draine field \citep{Draine1978}). This is representing
the additional radiation in the UV produced by the stars in the AGN
environment. It needs to be pointed out that the impact of the UV, and
especially that originating from the nucleus, on the molecular part of
the torus is small, since most of the UV is already absorbed given the
large $A_V$. The cosmic ray ionization rate to $\zeta = 5\times
10^{-17}$~s$^{-1}$. Below, we will discuss the results for an
equilibrium solution and a time-dependent solution. We would like to
point out that although these parameters are chosen in such a way that
they are representative for a typical AGN, but that it is still a
proof of concept paper. A study varying parameters such as the torus
mass, and UV and X-ray luminosities from the central source and the
stars, is postponed to another paper.

\subsection{The equilibrium solution}

{\it Density and temperatures:} density and gas and dust temperature
structures are shown for a generic disk
(Fig. \ref{density_temperature}). To guide the eye, we overplot $A_V=1$ and
10, which is defined as the smallest optical depth in either the
vertical or radial direction toward the 'surface' of the torus. The
FUV radiation is dominated by the central source, when the visual
extinction is smallest in the radial direction, dominated by the
stellar radiation field when it is smallest in the vertical
direction. Densities range from $n_{\rm H} < 10$~cm$^{-3}$ at large
altitudes in the disk, to $n_{\rm H}=10^{8}$~cm$^{-3}$ at the
mid-plane ($z/r\sim 0$) and close to the inner rim ($r\sim 0.5$~pc) of
the disk. The gas temperature is decoupled from the dust temperature,
except for the regions of the disk where densities are high $n=10^5 -
10^8$~cm$^{-3}$, and both the UV and X-ray radiation field is largely
attenuated. The unshielded parts of the disk show gas temperatures $T
> 5000$~K, much larger than the dust temperatures, which range from $T
\sim 40 - 1400$~K in the same region. Figure 1 shows a contour where
the dust temperature is $T_{\rm dust}=100$~K. Below this temperature,
formation of H$_{\rm 2}$ is most efficient.

\begin{figure*}
  \centering
  \includegraphics[width=5.4cm]{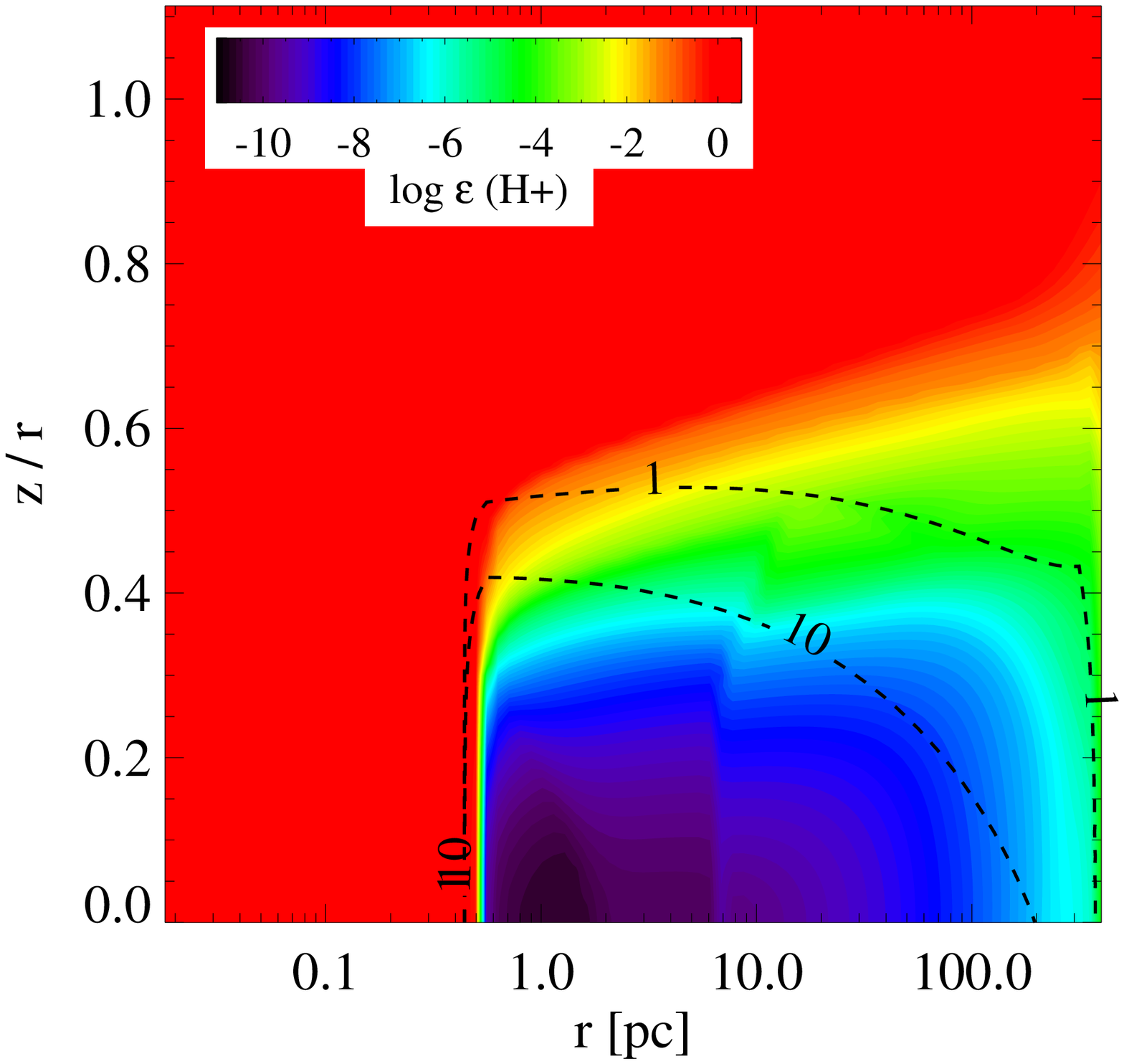}
  \includegraphics[width=5.4cm]{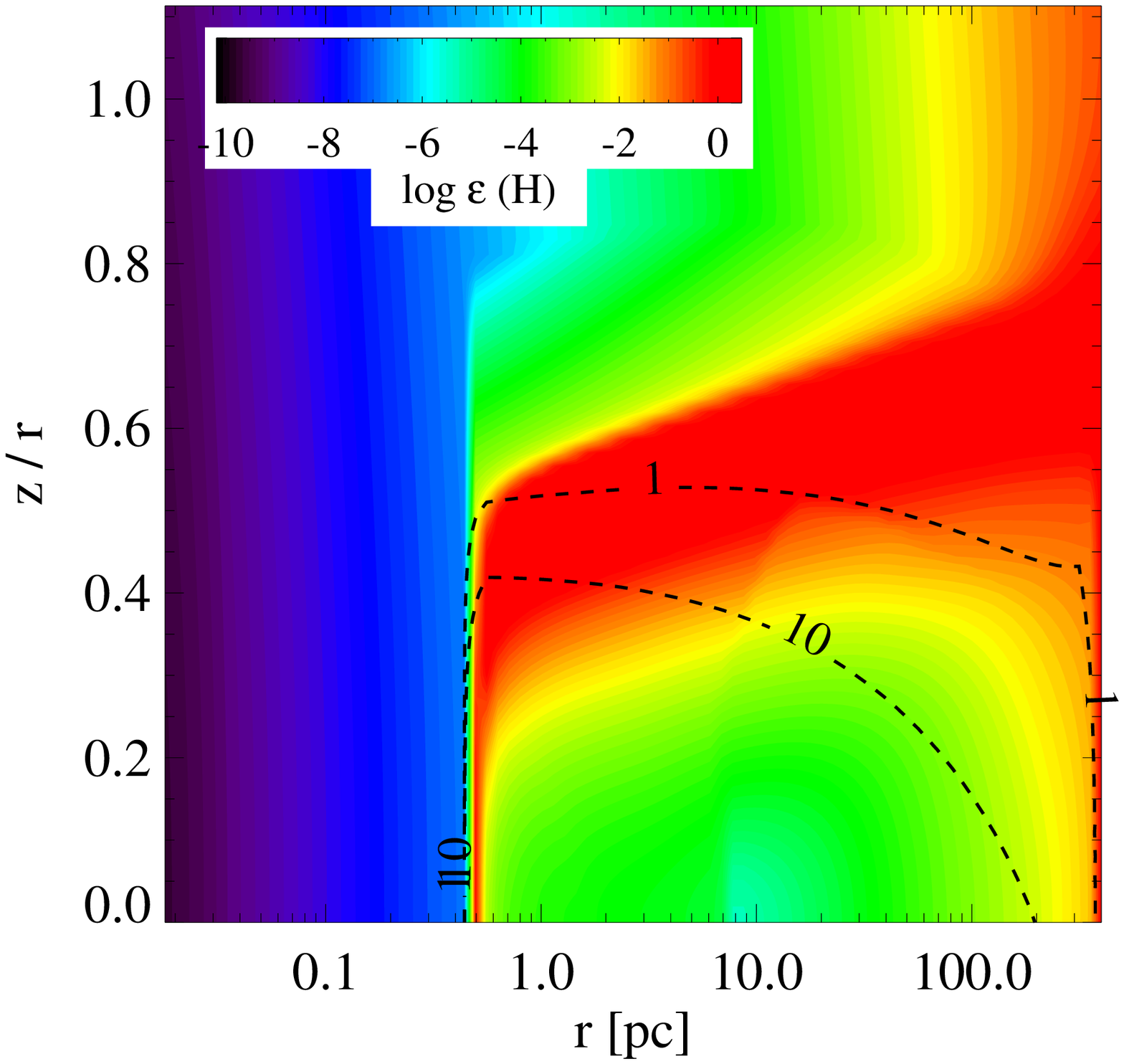}
  \includegraphics[width=5.4cm]{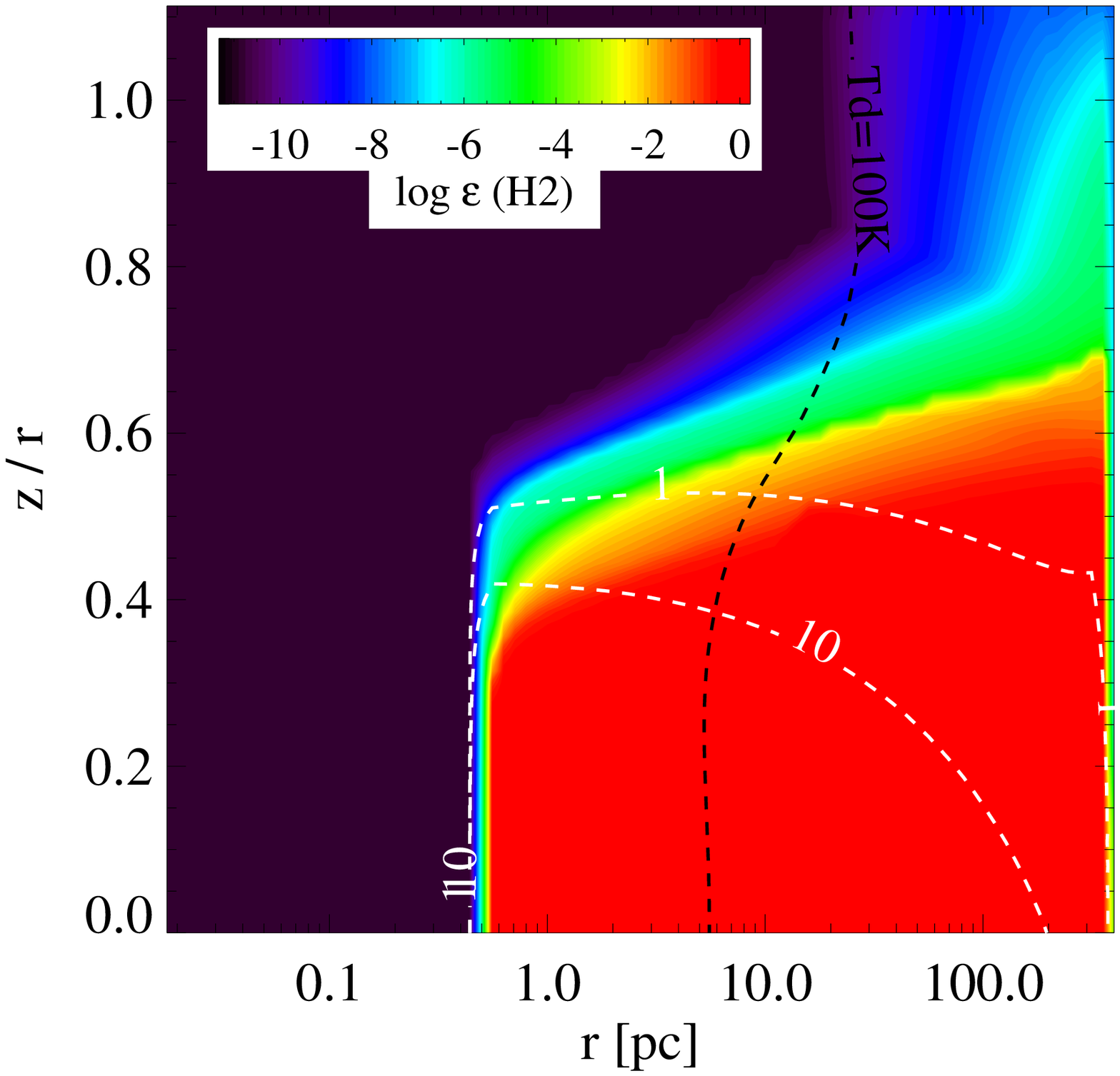}
  \includegraphics[width=5.4cm]{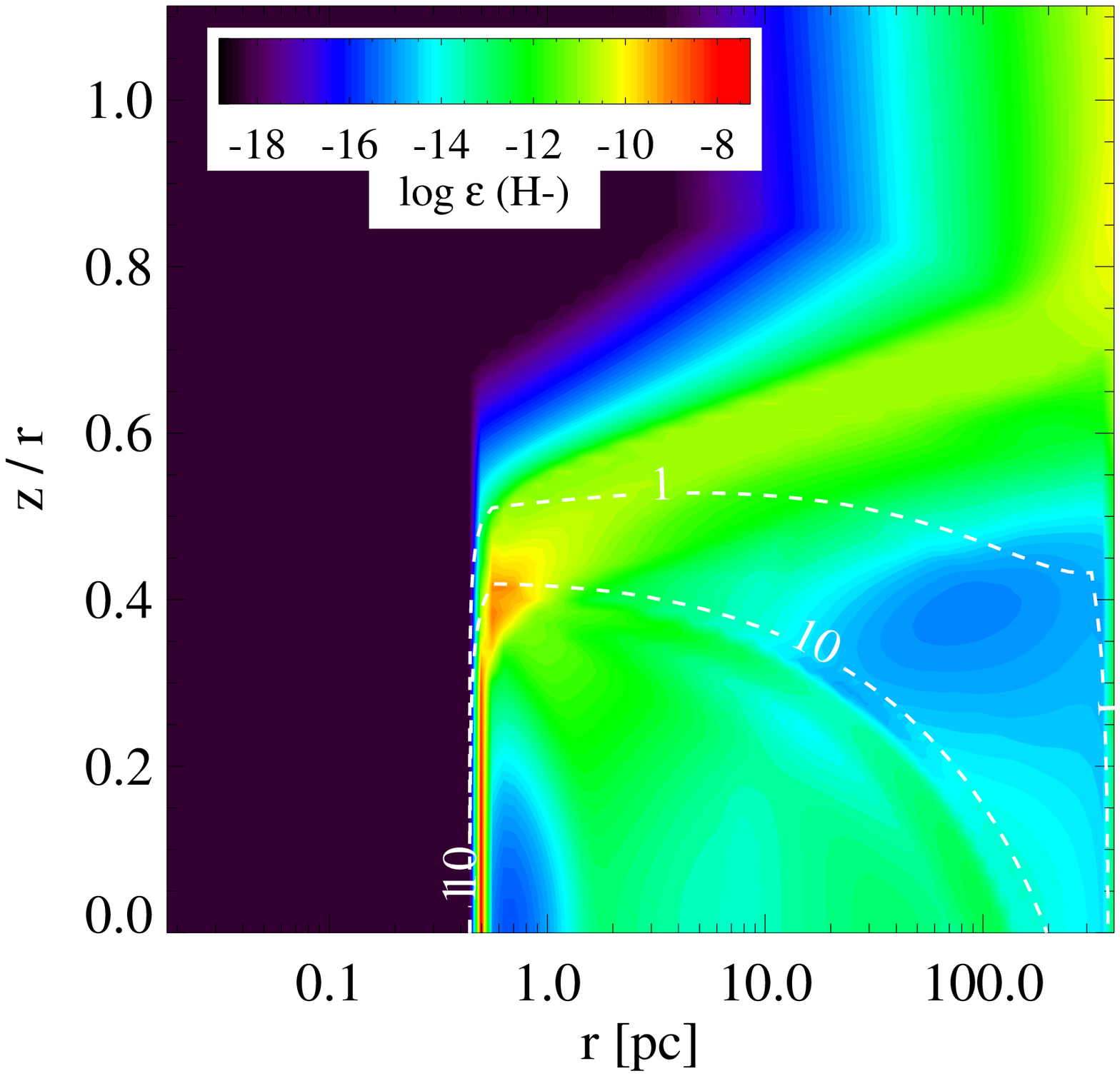}
  \includegraphics[width=5.4cm]{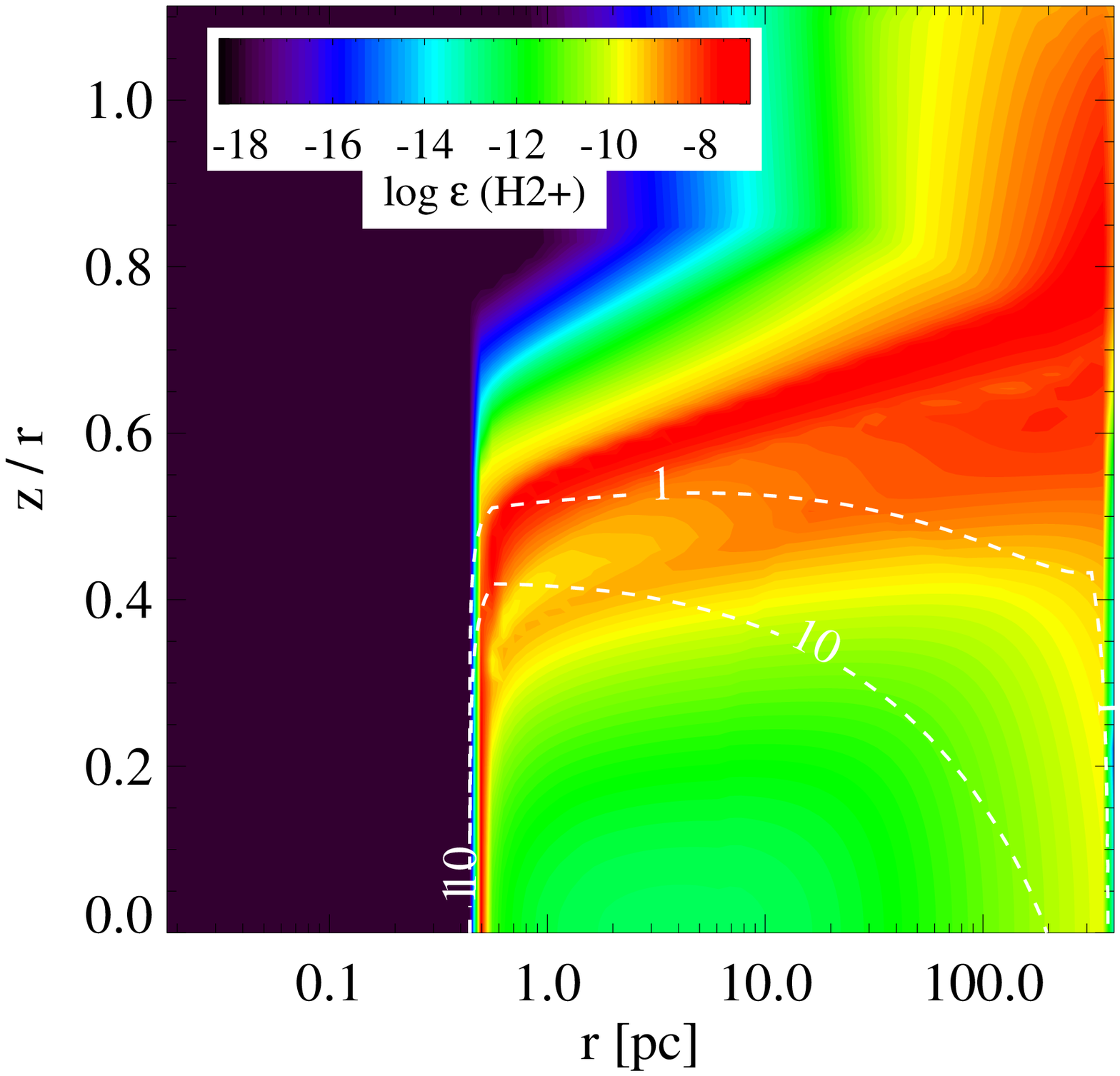}
  \includegraphics[width=5.4cm]{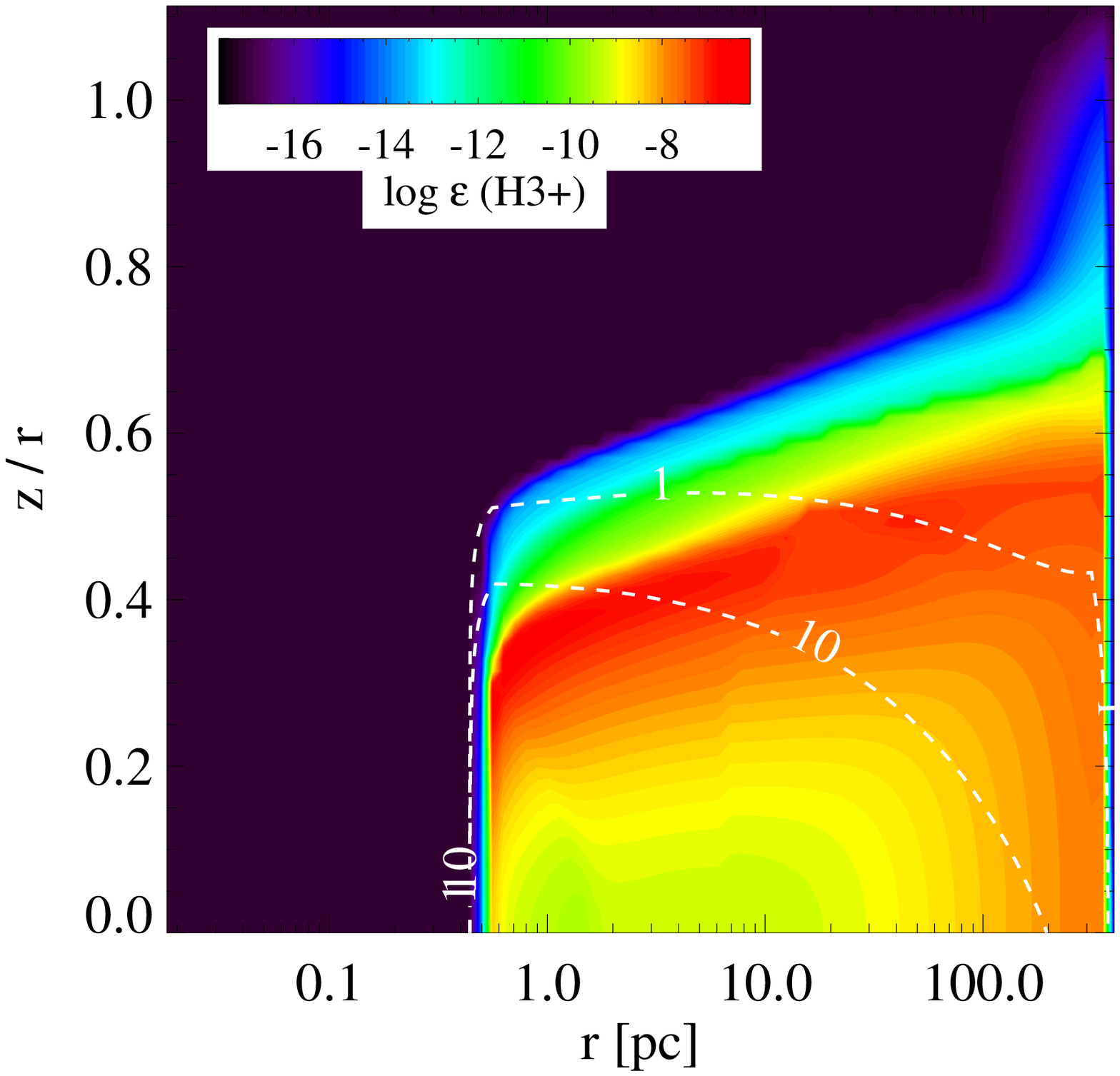}
  \caption{Abundances of H$^+$, H, H$_2$, H$^-$, H$_2^+$ and
    H$_3^+$. Contours for $A_{\rm v}=1$ and 10 and $T_{\rm dust} =
    100$~K are overplotted.}
  \label{hydrogen_species}
\end{figure*}

{\it H$^+$/H/H$_2$:} We show the H$^+$/H/H$_2$ transition in
Fig. \ref{hydrogen_species}. While in an UV dominated chemistry the H$^+$
is produced by cosmic ray ionization or by charge exchange with O$^+$,
X-rays are able to efficiently ionize hydrogen through ionization by
photoelectrons and produce ionization fractions as large as $x_{\rm
  e}\sim 1$. In the low density region of the disk, hydrogen is
completely ionized, because the gas is directly exposed to X-rays, and
the recombination rates are relatively slow due to the low densities
($n_{\rm H}<10^3$~cm$^{-2}$) and high temperatures $T > 5000$~K. This
is followed by a transition layer dominated by atomic hydrogen, but
where the abundance of $x_{\rm H^+}\sim 10^{-4}$ and $x_{\rm H_2}\sim
10^{-2}$. This is an active layer, where transient species, such as
OH$^+$ and H$_2$O$^+$, are highly abundant (see
Fig. \ref{oxygen_species}). H$_2$ is dominant in the shielded part of
the disk, when dust temperatures are below $T\sim 100$~K. Note that
the H$_2$ formation efficiency drops rapidly at $T_{\rm dust} > 100$~K
and is tiny when $T_{\rm dust} > 1000$~K. At these temperatures H$_2$
is formed through ${\rm H}^- + {\rm H} \rightarrow {\rm H}_2 + {\rm
  e}^-$, a route which is especially efficient when X-rays are
present. The electron abundances drop below $x_{\rm e} < 10^{-8}$
where H$_2$ is dominant. Interestingly, we see a sudden increase of
the electron abundance with radius at $R \sim 6$~pc. Beyond this
radius water freezes out, and depletes oxygen from the gas-phase. The
charge exchange reactions $\rm H^+ + O \leftrightarrow H + O^+$ sets
the ratio between O$^+$ and H$^+$. When oxygen is depleted, H$^+$ will
not be neutralized by this reaction resulting in a higher abundance
$x_{\rm H^+}$.

{\it H$^{-}$, H$_2^+$, and H$_3^+$:} H$^-$ is abundant ($x_{\rm H^-} >
10^{-10}$) at the transition from H$^+$ to H. At this location the
formation of H$^-$ is optimal, since there are many electrons
available as well as neutral atomic hydrogen to allow $\rm H + e^-
\rightarrow H^- + photon$. Another region of the disk where H$^-$
abundances are high ($x_{\rm H^-} \sim 10^{-13}$) is for optical
extinction $A_{\rm V} > 10$ mag, in which case the interstellar
radiation field is not able to destroy H$^-$ efficiently. Slightly
beyond the region where the H$^-$ abundance is highest, there is a
layer with a high H$_2^+$ abundance ($x_{\rm H_2^+} \sim 10^{-8}$). At
this location, the temperature is high enough to allow for the
radiative association reaction $\rm H^+ + H \rightarrow H_2^+ +
photon$. Another maximum occurs where the transition from H to H$_2$
occurs, and then H$_2^+$ is formed by X-ray and cosmic ray
ionization. When the full transition to H$_2$ has occurred, the
H$_2^+$ abundance drops, since it is transformed into H$_3^+$ through
the $\rm H_2^+ + H_2 \rightarrow H_3^+ + H$.

\begin{figure*}
  \centering
  \includegraphics[width=5.4cm]{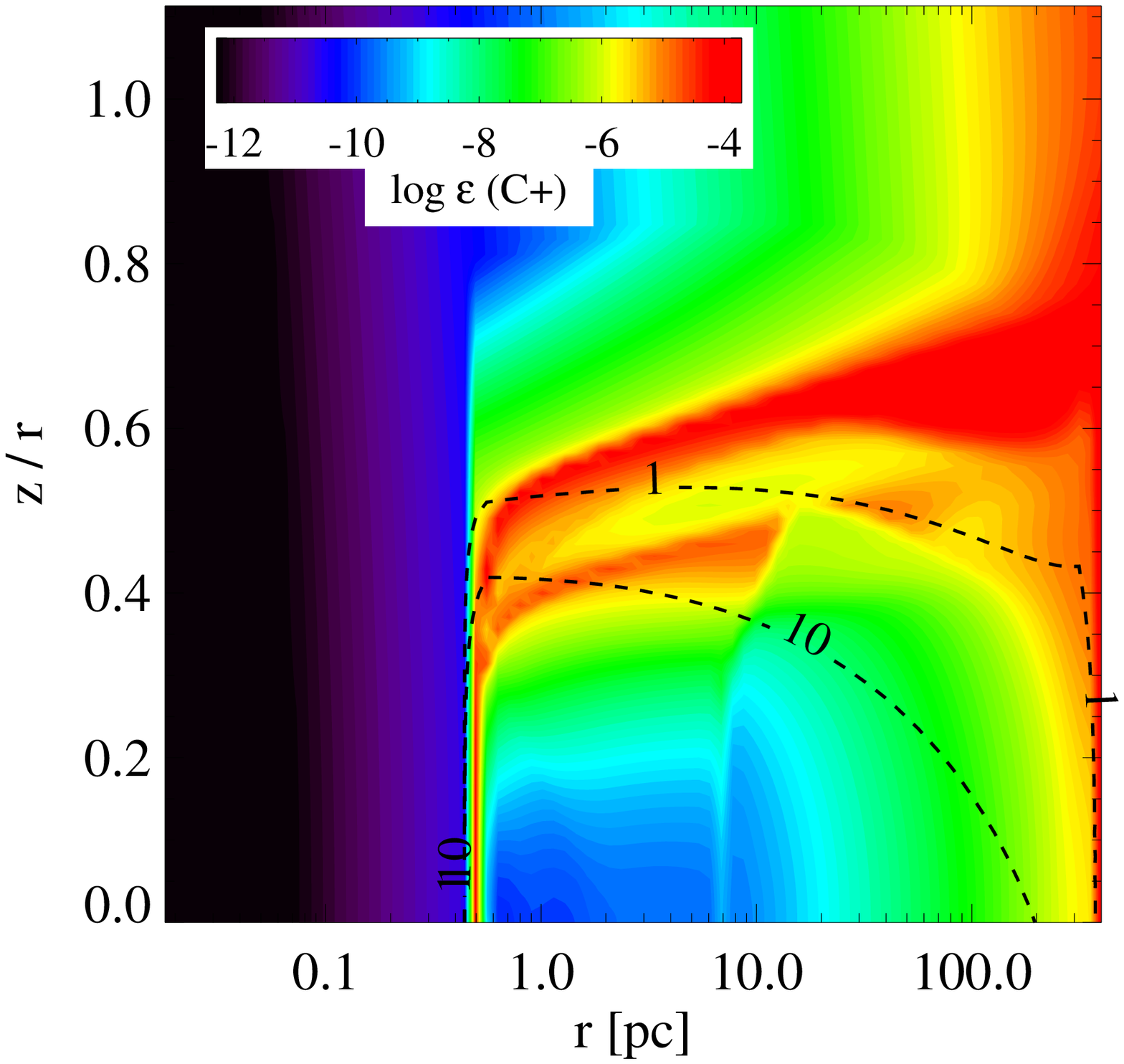}
  \includegraphics[width=5.4cm]{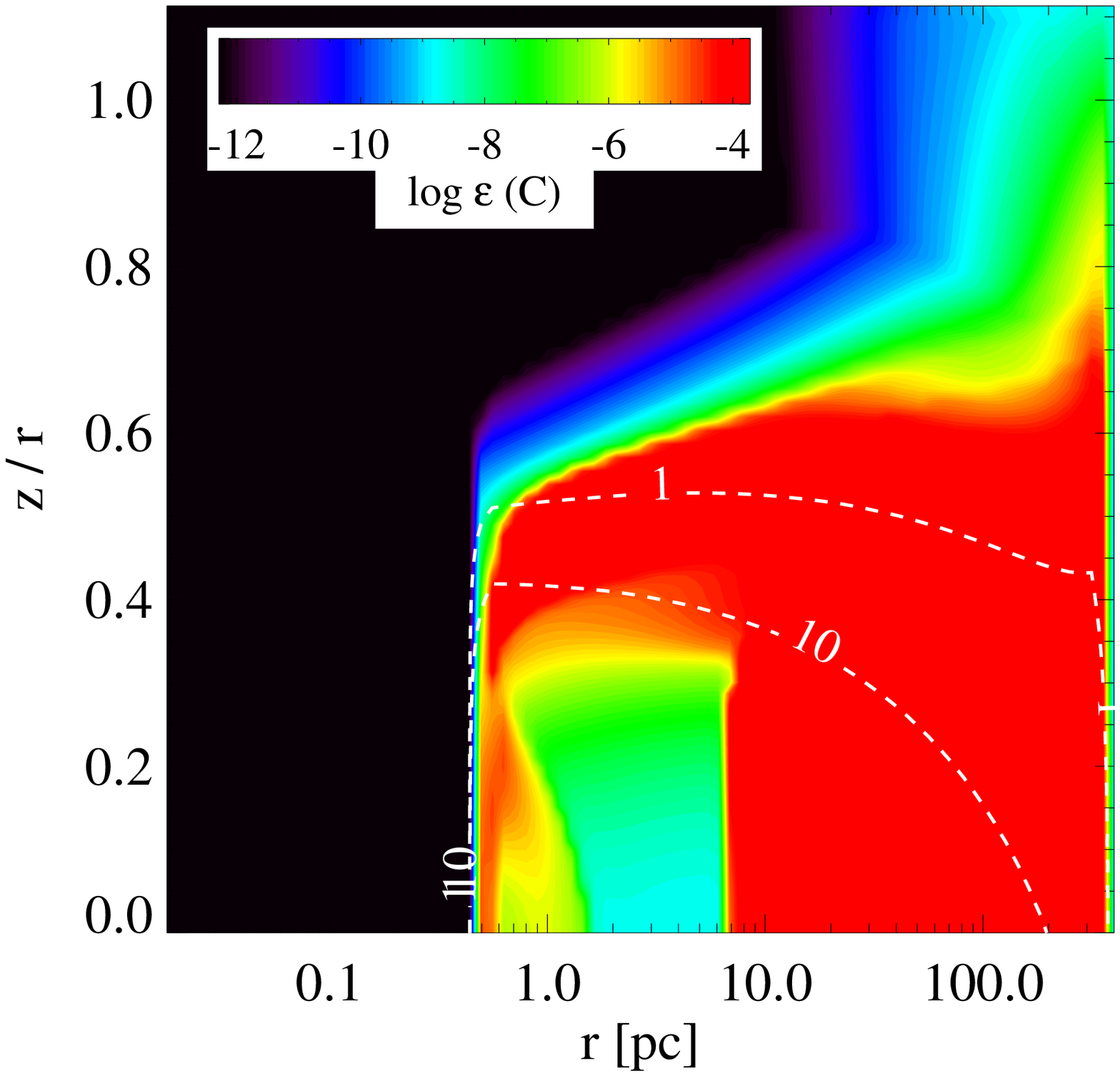}
  \includegraphics[width=5.4cm]{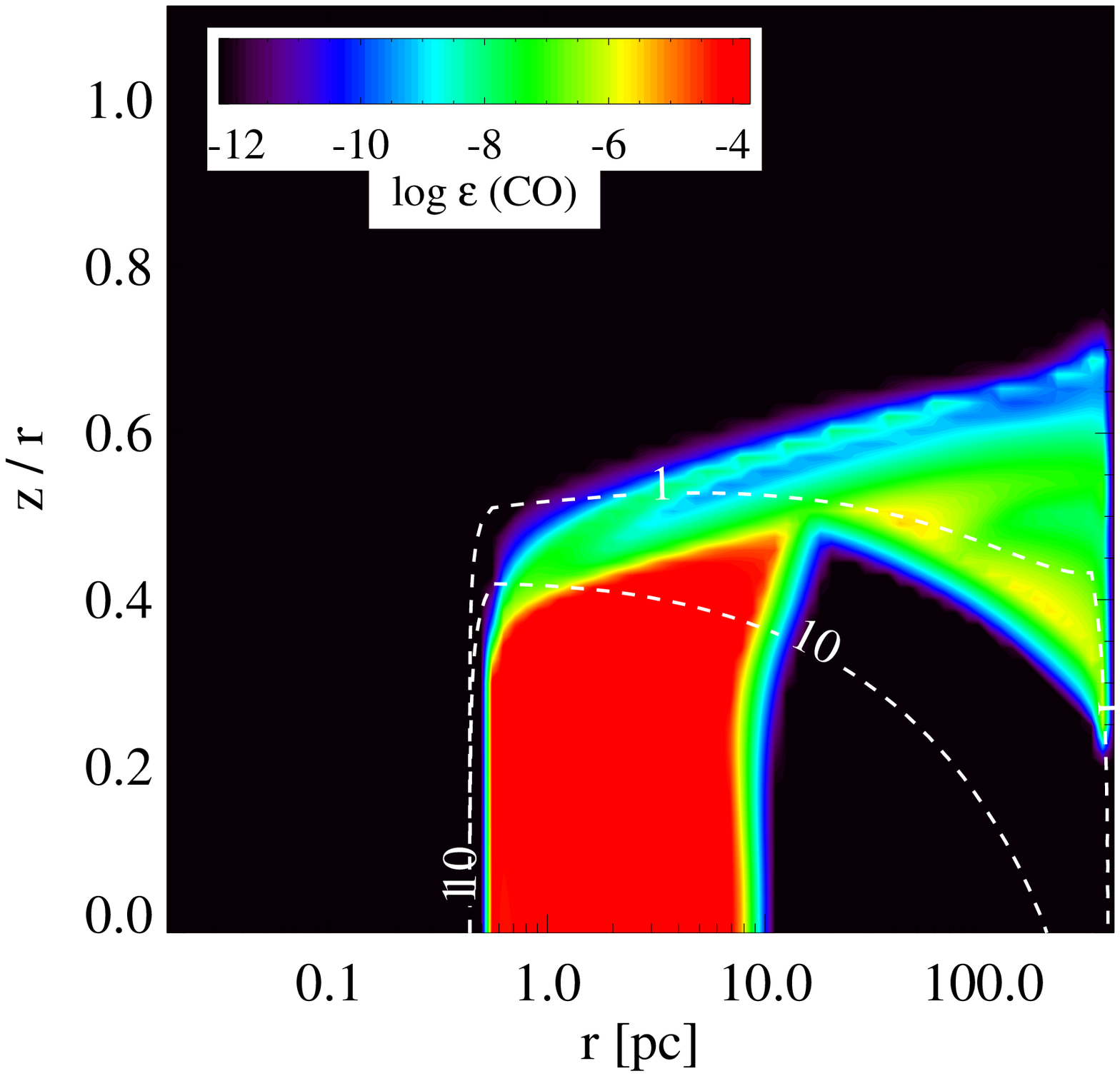}
  \caption{Abundances of C$^+$, C, and CO. Contours for $A_{\rm v}=1$
    and 10 are overplotted.}
  \label{carbon_species}
\end{figure*}

{\it C$^+$/C/CO:} Carbon is ionized to high ionization states high up
in the disk and at small radii ($r < 0.5$~pc). The code allows the
species to be doubly ionized, and the double ionized species serve as
a sink species, and therefore most carbon is in C$^{2+}$. Charge
transfer reactions for three or more times ionized species are
fast \citep{Butler1980}, and generally orders of magnitude larger than
those for singly and doubly ionized species. Therefore, the fact that
we only include doubly ionized species is a reasonable approximation,
since highly ionized species are quickly reduced back to a doubly
ionized state. Fig. \ref{carbon_species} shows the abundances of C$^+$, C,
and CO, which are commonly observed in active galaxies. Although there
is a transition from C$^+$, to C and CO, the structure is much more
complex than in the case of hydrogen. C$^+$ is most abundant in the
upper layers of the disk, and the highest abundances of C$^+$ are
coinciding with the atomic hydrogen layer. This C$^+$ shows a minimum
in the middle of this layer, which is caused by a number of competing
reactions. The abundance of H$_2$ increases over the layer, and makes
the endothermic reaction $\rm C^+ + H_2 \rightarrow CH^+ + H$
possible. As the temperature drops toward the mid-plane, this reaction
becomes less efficient, and the C$^+$ abundance increases again. CO is
most abundant between radii $r\sim 0.5-9$~pc. Beyond these radii,
water is almost fully frozen out, which depletes the gas phase oxygen
abundance entirely, and prevents the formation of CO in the gas
phase. The timescales for freezing out water are very long, and given
the occurrence of supernova and cloud-cloud collisions that remove
molecules from the icy mantles, it is not expected that the CO
entirely freezes out. Also, the size of the molecular disk in the low
($J=1-0$ and $J=2-1$) CO transitions show that the disk is extended
beyond $\sim 10$~pc, as discussed earlier. This implies that a
time-dependent treatment of the chemistry is required.

\begin{figure*}
  \centering
  \includegraphics[width=5.4cm]{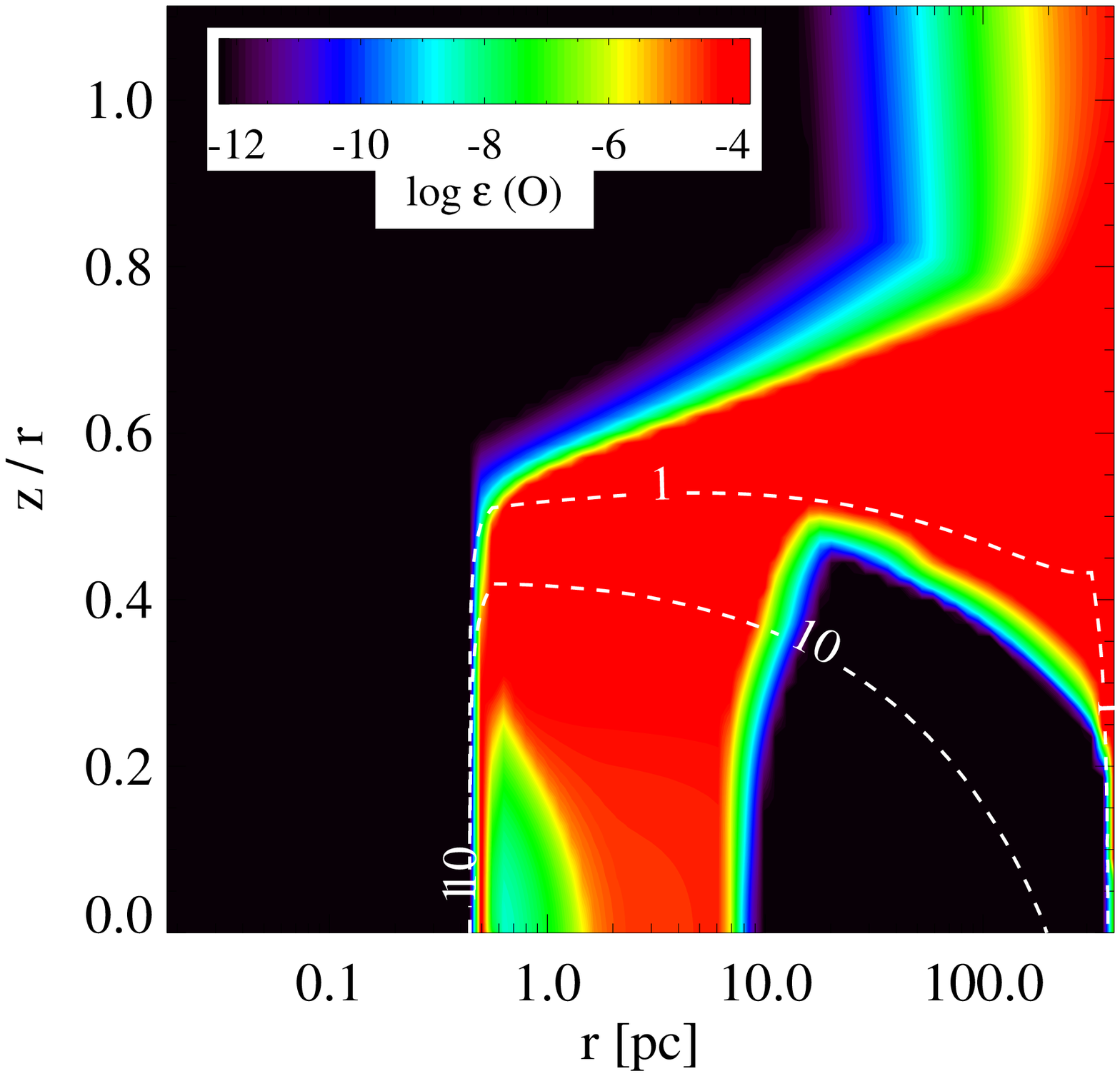}
  \includegraphics[width=5.4cm]{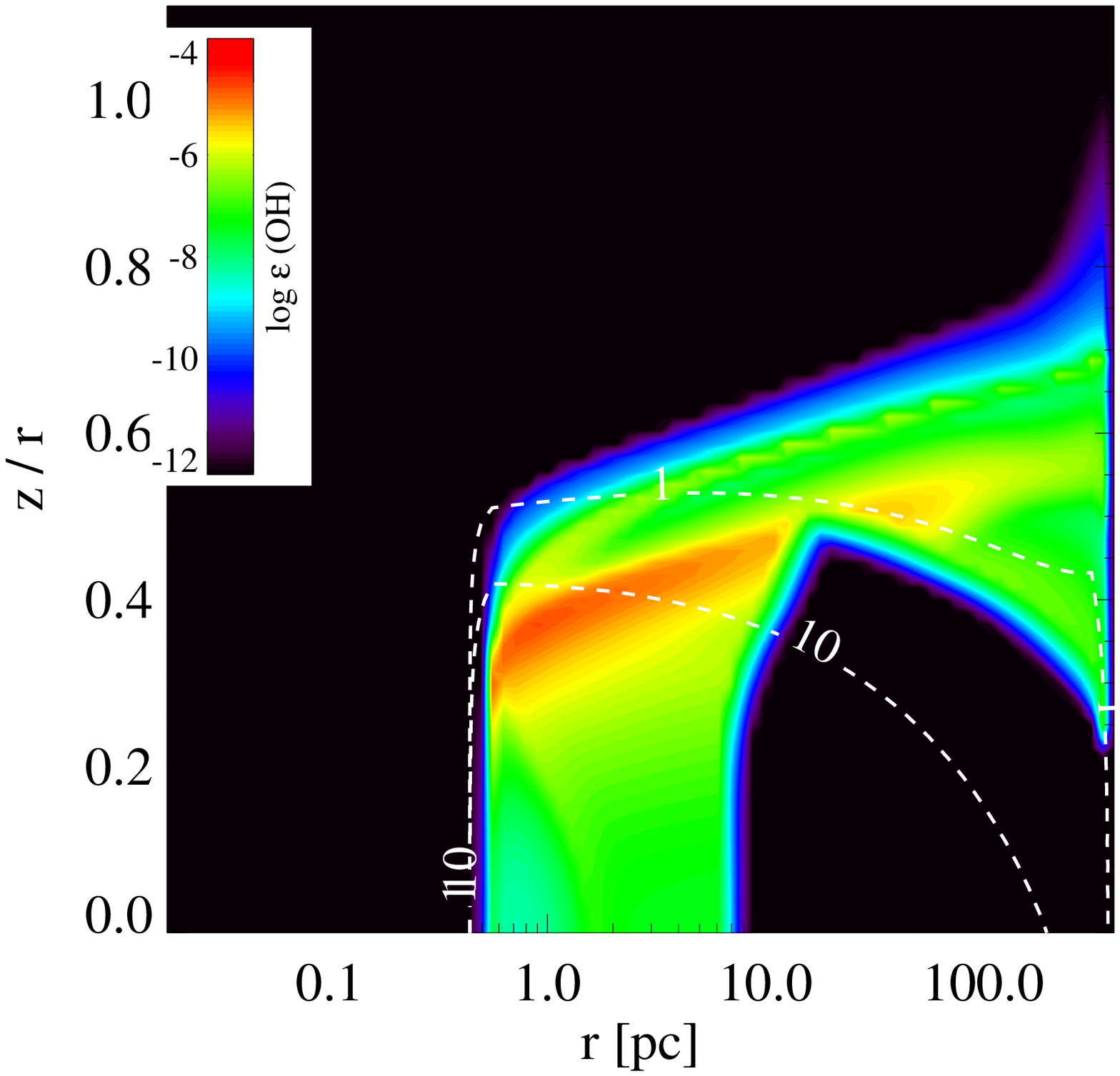}
  \includegraphics[width=5.4cm]{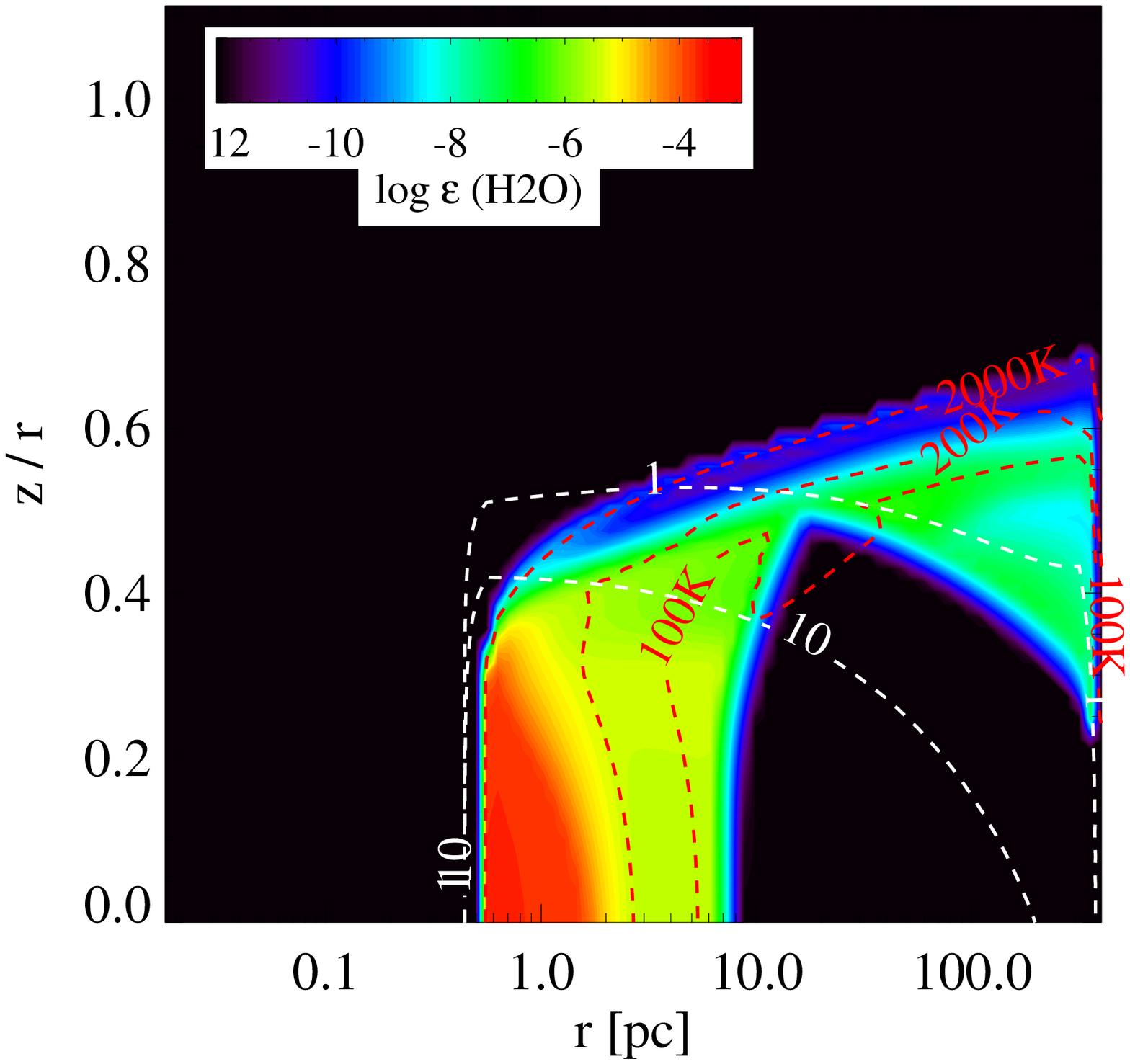}
  \includegraphics[width=5.4cm]{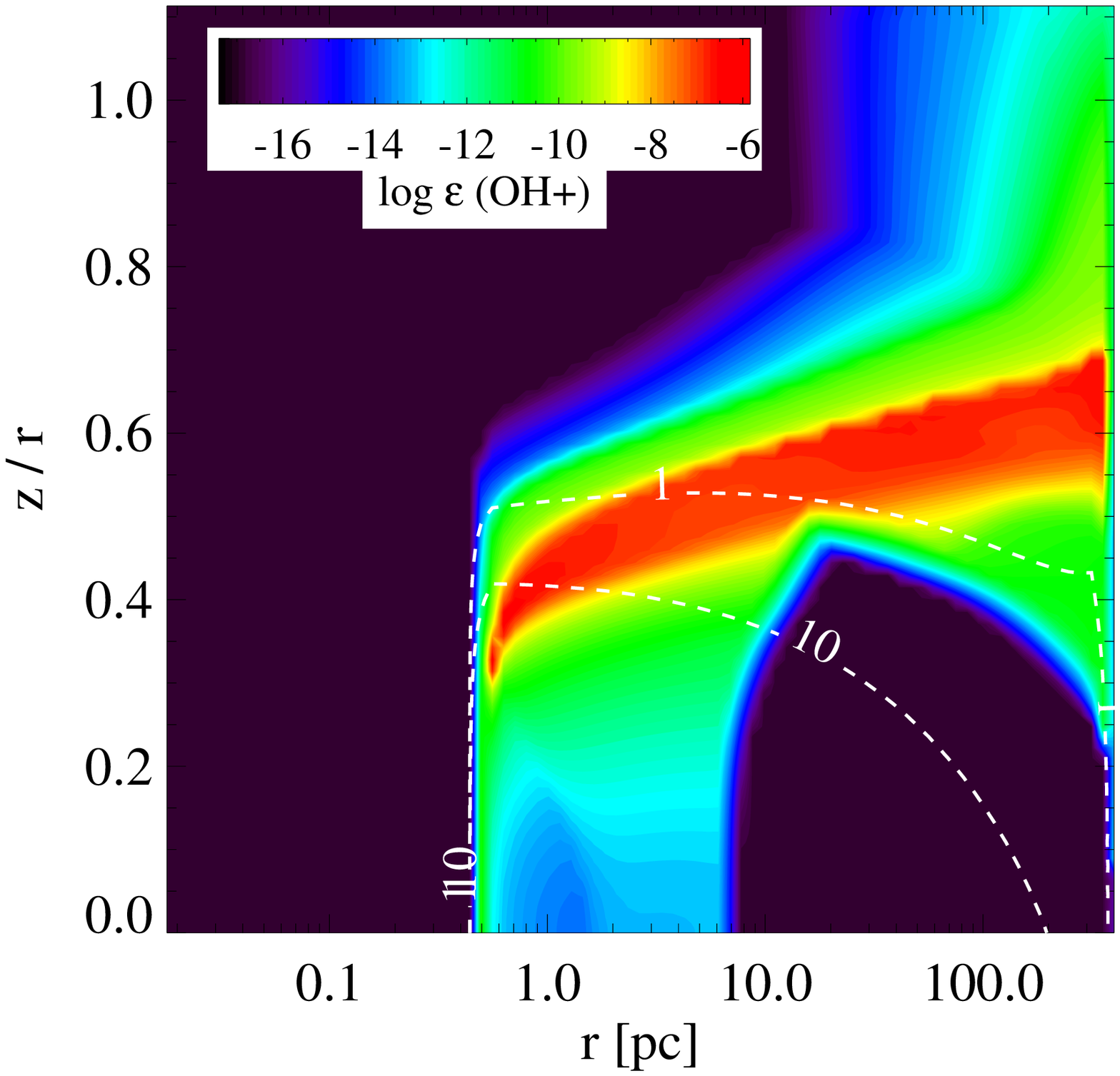}
  \includegraphics[width=5.4cm]{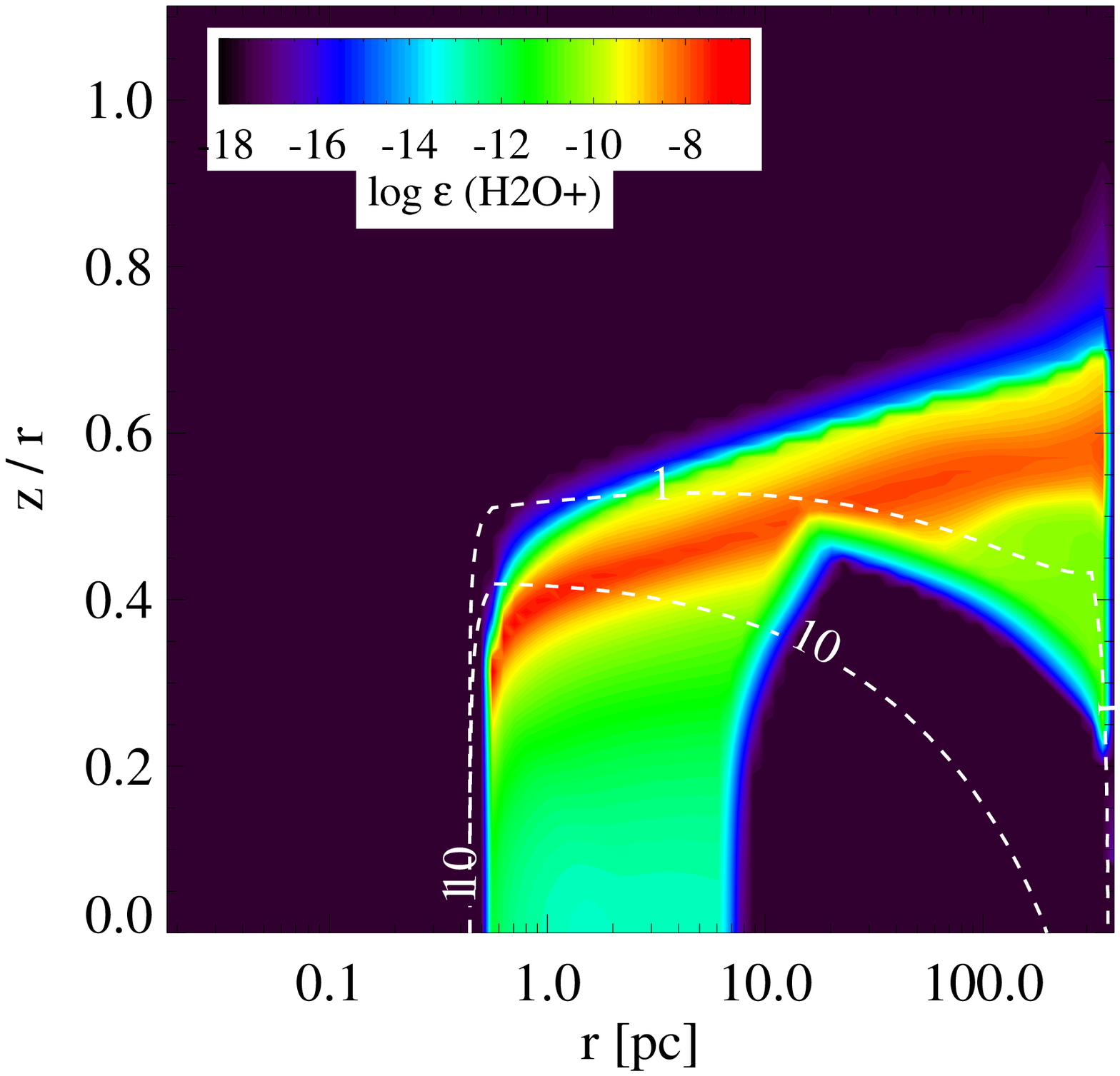}
  \includegraphics[width=5.4cm]{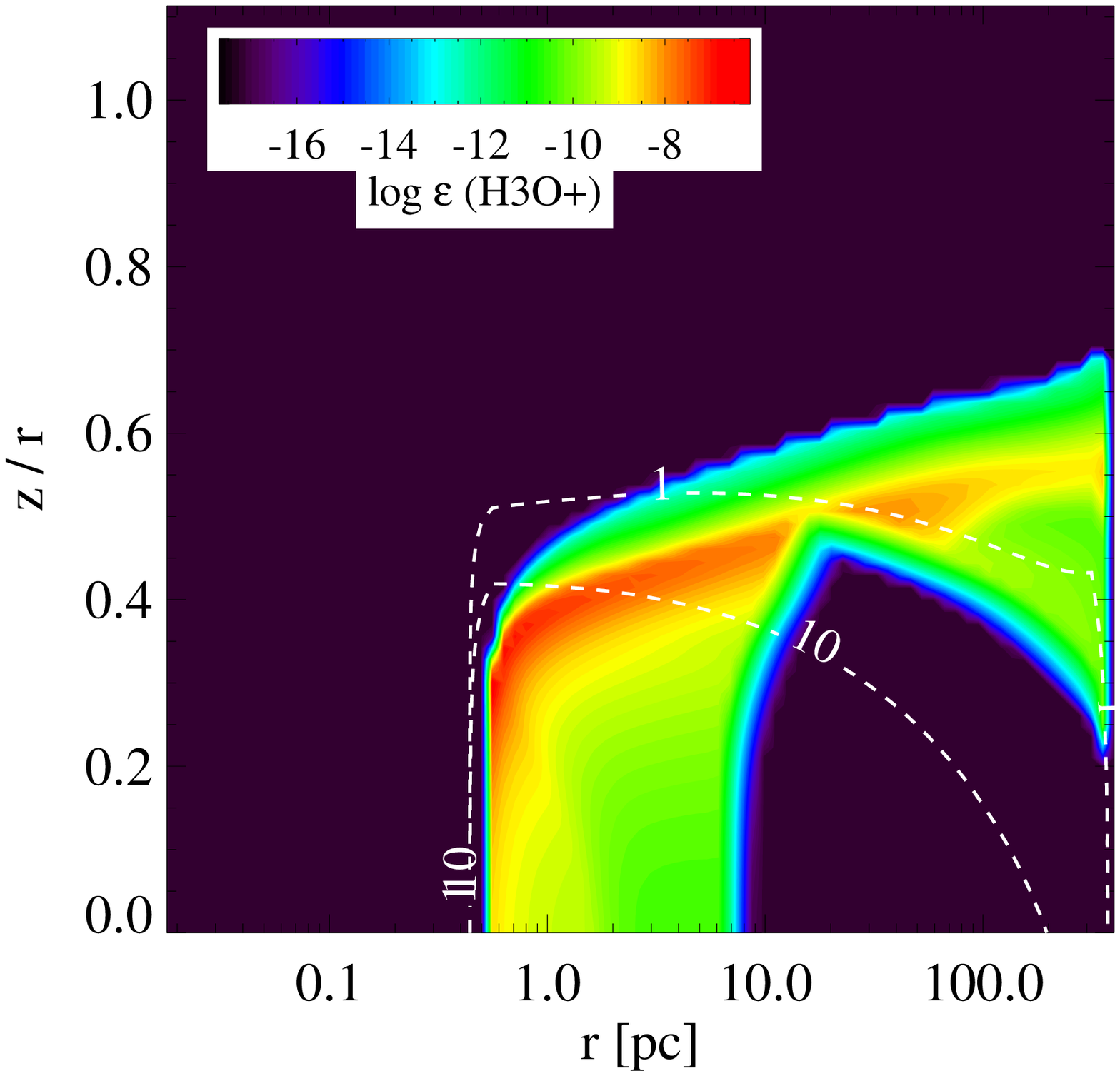}
  \caption{Abundances of O, OH, H$_2$O, OH$^+$, H$_2$O$^+$ and
    H$_3$O$^+$. Contours for $A_{\rm v}=1$ and 10 and $T_{\rm
      gas}=100$, 200, and 2000~K are overplotted.}
  \label{oxygen_species}
\end{figure*}

{\it O, OH, H$_2$O, OH$^+$, H$_2$O$^+$, and H$_3$O$^+$:}
Fig. \ref{oxygen_species} shows the oxygen related species. Neutral oxygen
is the most abundant in those regions of the disk where temperatures
are $T < 5000$~K,and above the temperature where H$_2$O freezes
out. H$_2$O freezes out on grain surfaces when dust temperatures are
below $T\sim 90-110$~K.  This process removes oxygen from the gas
phase, which also affects other species (e.g., CO and H$_2$). H$_2$O
has an abundance of $10^{-4}$ with respect to H$_2$ in the dense ($n_H
> 10^7$~cm$^{-3}$) warm ($T_{\rm gas} \sim 300-1000$~K) part of the
disk, between 0.5 and 2.0 pc, although water remains present at lower
abundances in the more shielded part of the disk, until it freezes
out. In the warm dense, neutral ($x_e < 10^{-8}$) parts of the disk,
H$_2$O is formed through neutral-neutral reactions, while in the
regions of the disk where densities are low, and ionization fractions
of order $x_e\sim 10^{-4}$, it is also formed through ion-molecule
reactions $\rm H^+ + O \rightarrow O^+ + H$, $\rm O^+ + H_2
\rightarrow OH^+ + H$, $\rm OH^+ + H_2 \rightarrow H_2O^+ + H_2$, $\rm
H_2O^+ + H_2 \rightarrow H_3O^+ + H_2$, followed by $\rm H_3O^+ + e^-
\rightarrow H_2O + H$. In that particular part of the disk, we see a
clear transition zone from OH$^+$ to H$_2$O$^+$ to H$_3$O$^+$.

\begin{figure*}
  \centering
  \includegraphics[width=5.4cm]{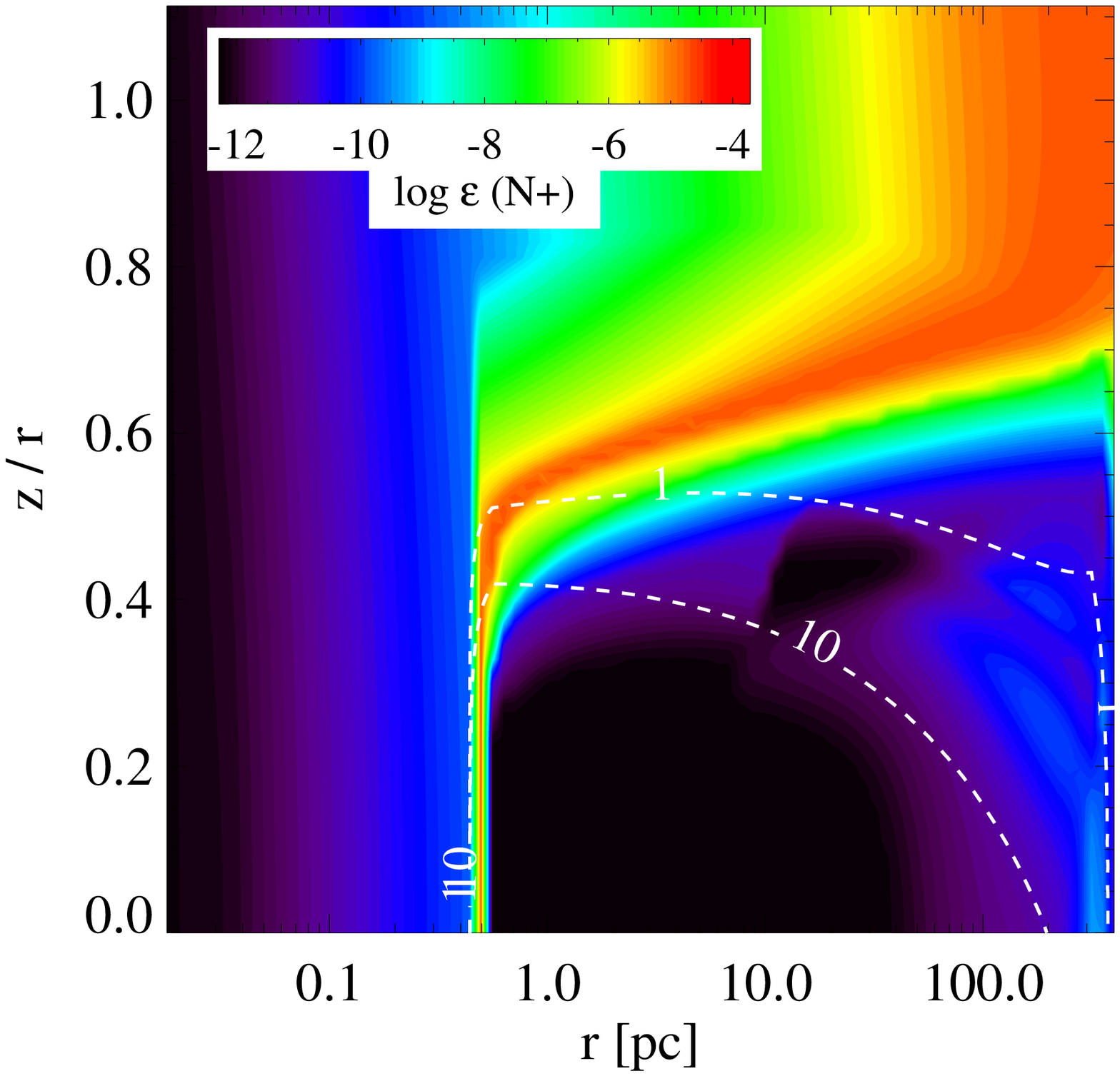}
  \includegraphics[width=5.4cm]{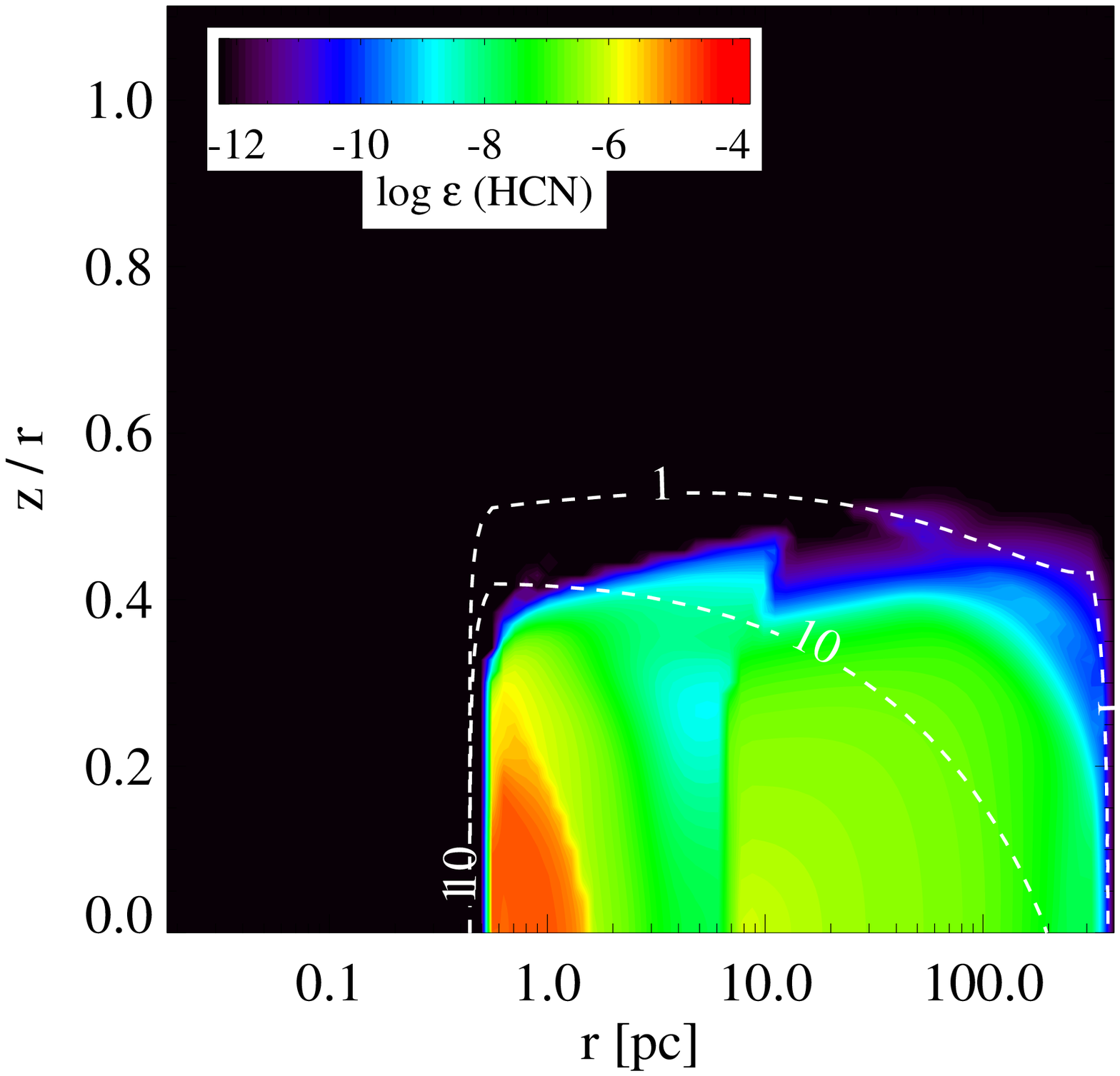}
  \includegraphics[width=5.4cm]{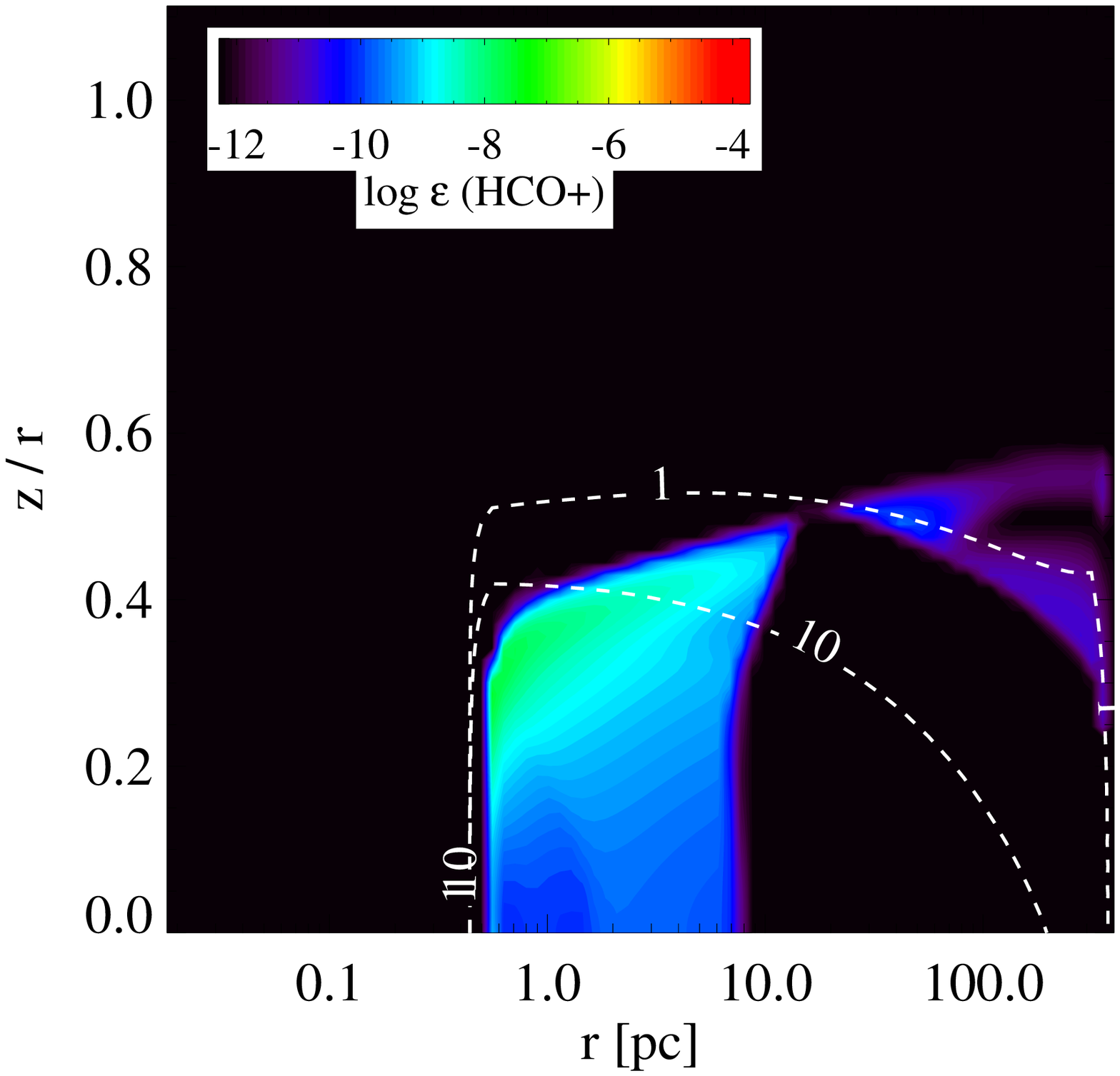}
  \caption{Abundances of N$^+$, HCN, and HCO$^+$. Contours for $A_{\rm
      v}=1$ and 10 are overplotted.}
  \label{nitrogen_species}
\end{figure*}

{\it N$^+$, HCN and HCO$^+$ (Fig. \ref{nitrogen_species}):} HCN and HCO$^+$
are commonly observed by ground-based telescopes and trace dense
($n_{\rm H} > 10^4-10^5$~cm$^{-3}$) gas. Extra-galactic N$^+$ has been
recently observed by the Herschel Space
Telescope \citep{VdWerf2010,Loenen2010} and also in the high-redshift
universe. N$^+$ is very abundant in the transition layer from ionized
to atomic hydrogen.  HCO$^+$ is entirely depleted in the water
freeze-out zone. HCN, on the contrary, has abundances in those regions
that are $x_{\rm HCN} \sim 10^{-8}- 10^{-6}$. HCN is most abundant in
the warm dense parts of the disk, following the abundance structure of
water. HNC, which is also commonly observed, follows the abundance
structure of HCN at $T < 100$~K. Interestingly, there is an
endothermic reaction $\rm H + HNC \rightarrow HCN + H$, that
efficiently converts HNC into HCN. Including additional lines from HNC
in the analysis would thus help constraining the temperature of the
dense gas.

\subsection{Chemical relaxation timescales and the need for a
  time-dependent solution}

\begin{figure}
\centering
  \includegraphics[width=5.4cm]{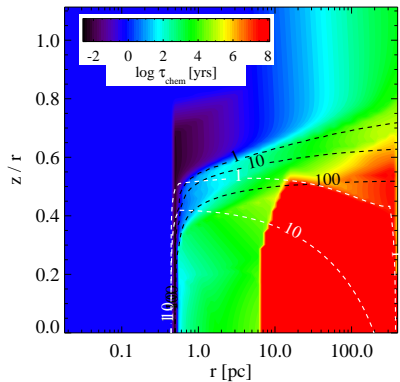}
  \caption{Chemical relaxation timescale for a disk in thermal and
    chemical equilibrium. Contours for $A_{\rm v}=1$ and 10, as
      well as $\tau(1 {\rm keV})=1$, 10 and 100 are overplotted.}
  \label{chem_time_scale}
\end{figure}

The chemical relation time \citep{Woitke2009} is defined as:

\begin{equation}
  \tau_{\rm chem}= \max_{\rm valid\ n}|Re\{\lambda_n\}^{-1}|,
\end{equation}

\noindent where $\lambda_n$ are the valid eigenvalues of the chemical
Jacobian $\delta F_i / \delta n_j$.  Chemical relaxation timescales in
the regions of the disk where freeze out occurs are of the order of
$\tau_{\rm chem} \sim 1$ to 100 Myrs (see Fig. \ref{chem_time_scale}). This
was already pointed in earlier work \citep{Bergin2000}, where it was
shown that CO can be depleted from the gas phase on a timescale
$\tau_{\rm depl} > 10^6$~yrs. This happens through CO destruction by
He$^+$, after which oxygen is locked into H$_2$O. They included in
this work also the formation of water on grains ($\rm O(grain)
\rightarrow \rm H_2O(grain)$), without allowing it to return into the
gas-phase. This is true for grains that have already an icy layer, but
for bare grains, a significant (a fraction of 0.5 to 0.6 for H$_2$O)
amount will be desorb from the grain into the gas phase upon
formation \citep{Meijerink2012a}. Accreted oxygen will not necessarily
remain on the grain. As mentioned in the introduction, the chemical
relaxation timescales are larger than the dynamical and supernova
timescales, which are typically $\tau_{\rm dyn} < 0.1 - 1$~Myr, and
are removing molecules from the grain.

In order to study the effect of time-dependent freeze-out on the
chemistry, we take the model described in the previous section, but
without any molecules frozen out onto grains. The initial conditions
are thus fully molecular, which we expect for ULIRGs. One exception is
the gas that is directly exposed to radiation. Then the gas is partly
atomic/ionized or even fully ionized (see equilibrium solution
described in previous section). We consequently add the possibility
that the molecules CO, H$_2$O, CH$_4$, CO$_2$, and NH$_3$ can
freeze-out onto grains. No surface reactions are involved into the
current calculation, and therefore the oxygen depletion timescale is
possibly overestimated, as oxygen is not accreted onto the grain and
reacting to OH and H$_2$O. We expect however that most of these
species are entering the gas-phase upon formation, since they are
forming on mostly bare grains. The current version of the ProDiMo code
allows the evolution of the chemistry for a given gas and dust
temperature structure. In the results we show below, we take the gas
and dust temperature from the equilibrium solution, and evolve the
chemistry for 100 Myrs. We checked the differences in the thermal
balance between the equilibrium model and the starting conditions for
the time-dependent model and the changes are marginal, and only
present in a small fraction of the disk.

\subsection{Effect on the chemical structure of commonly observed molecules}

We concentrate on the species that are affected by time-dependent
chemistry. The effect of oxygen depletion on the hydrogen species are
marginal, and the abundance structures do not significantly differ
from the equilibrium solution, and are therefore not discussed here.

\begin{figure*}
  \centering
  \includegraphics[width=4.0cm]{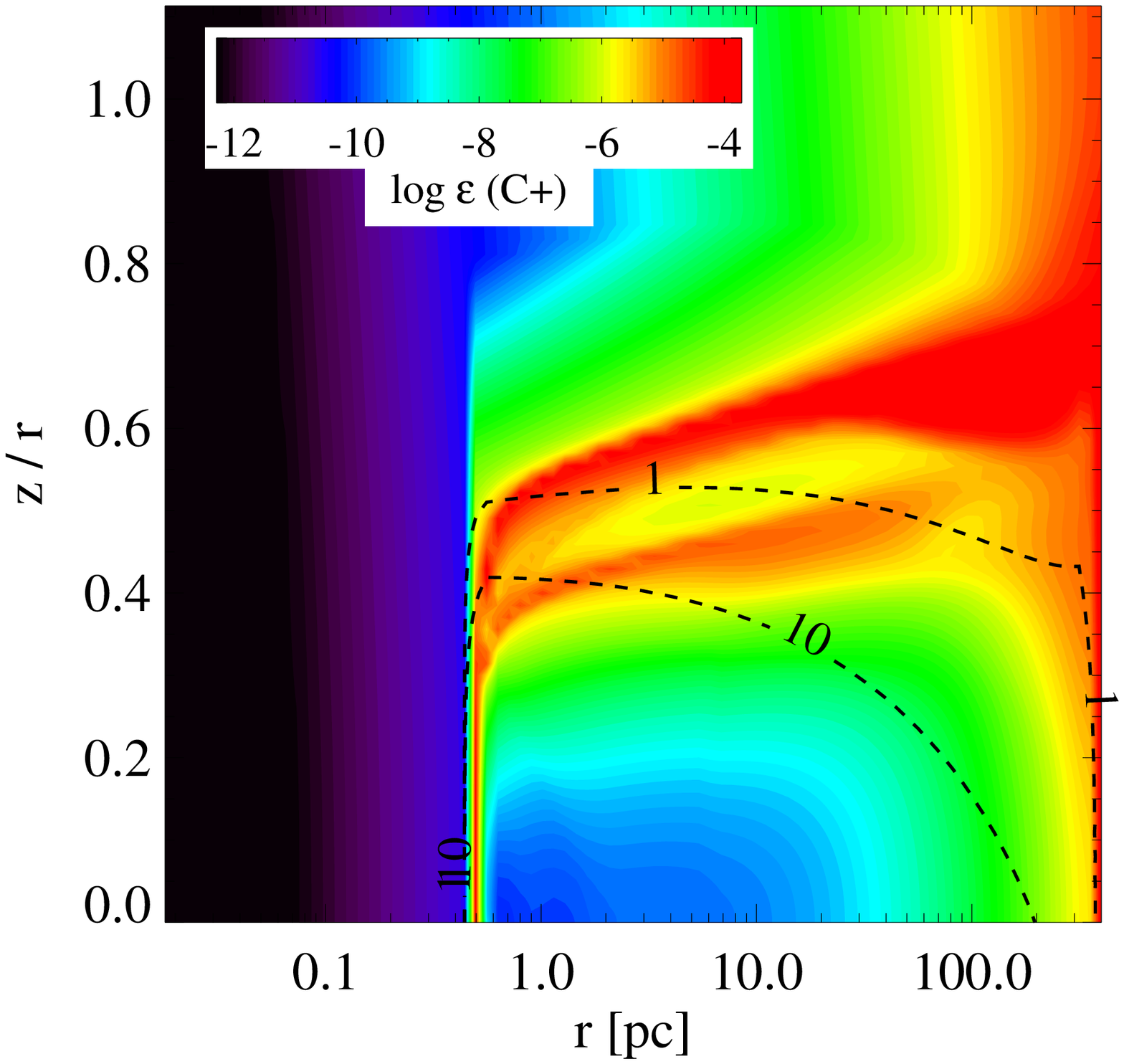}
  \includegraphics[width=4.0cm]{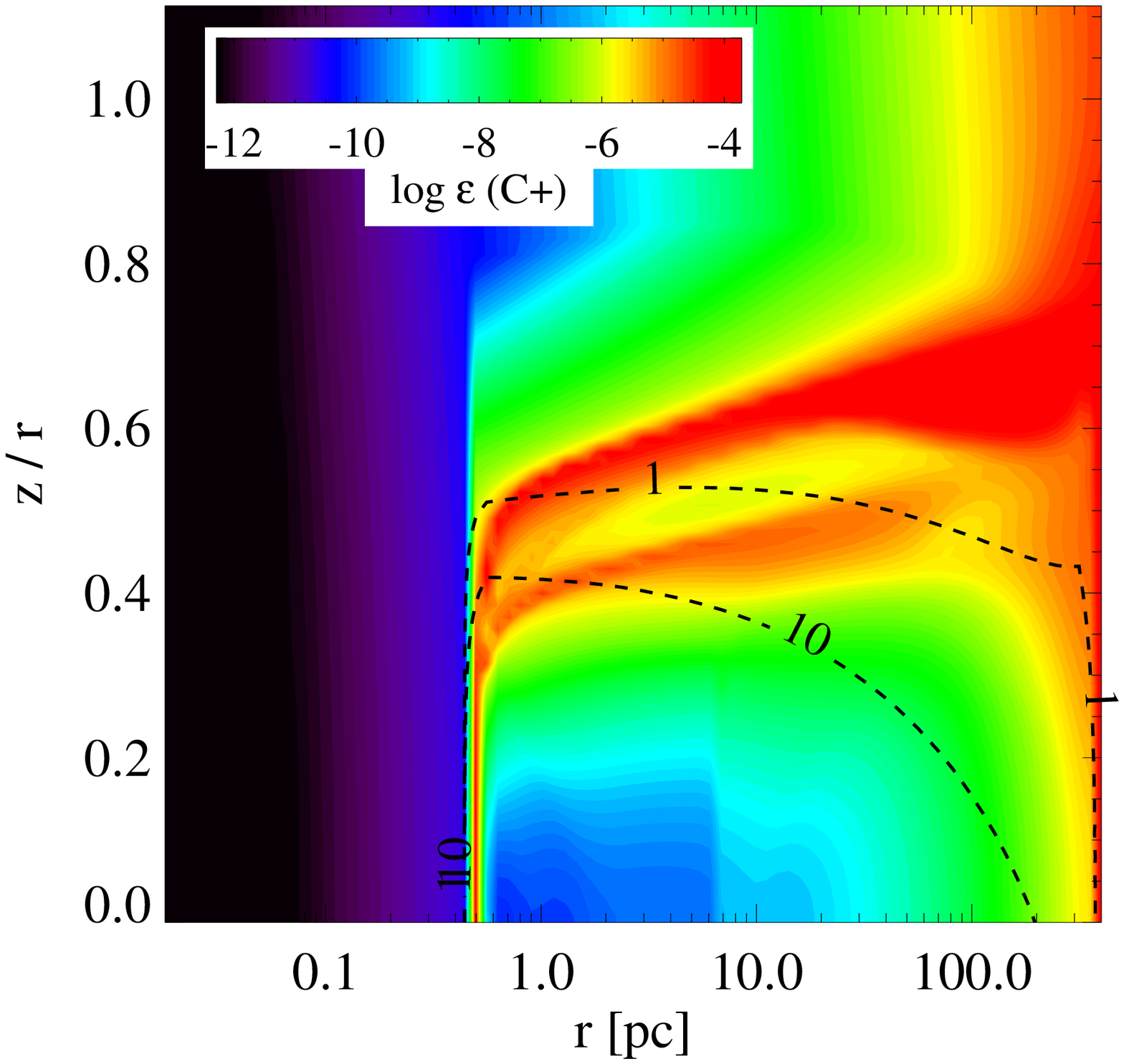}
  \includegraphics[width=4.0cm]{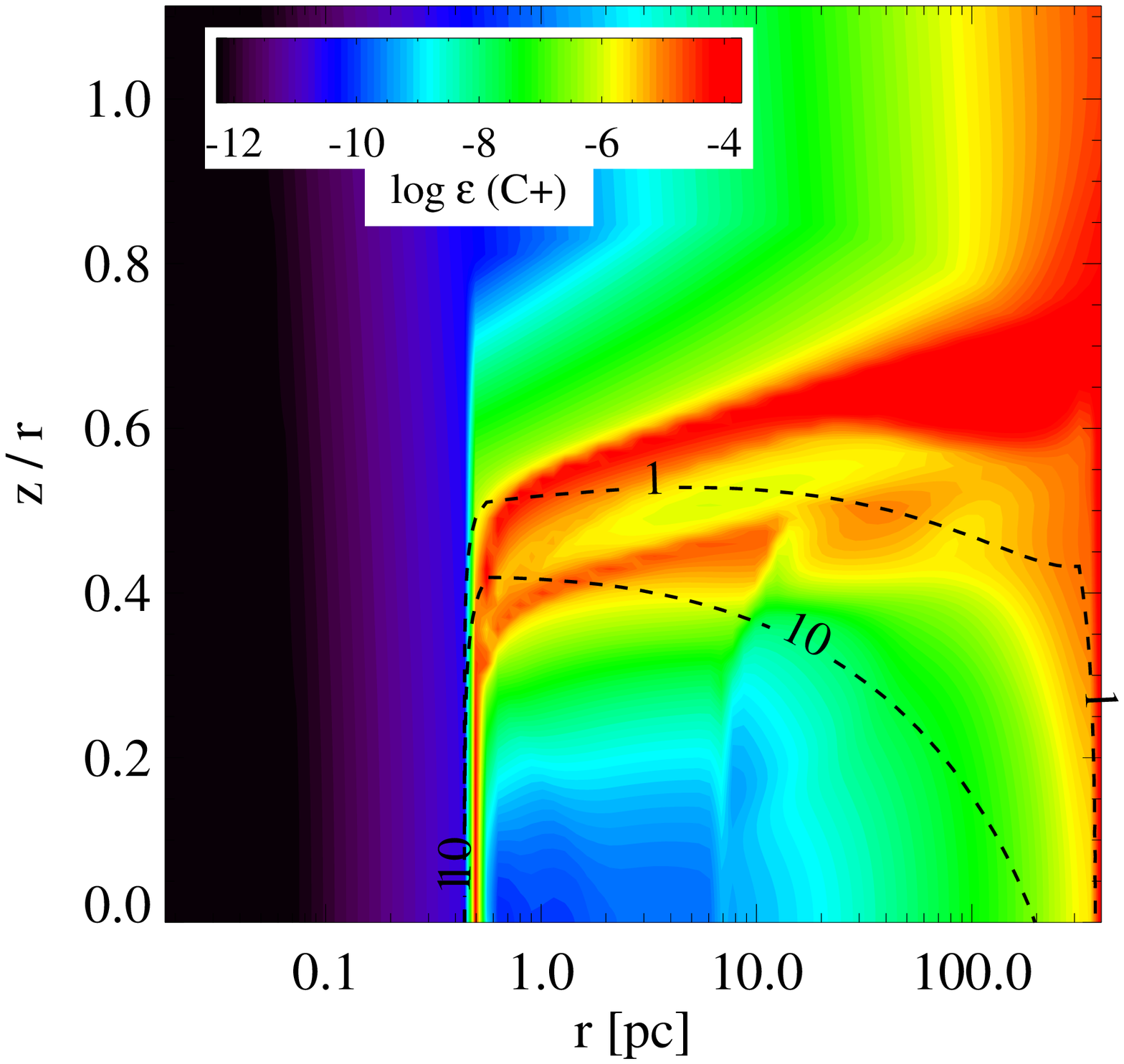}
  \includegraphics[width=4.0cm]{Cpabundance.ps}
  \includegraphics[width=4.0cm]{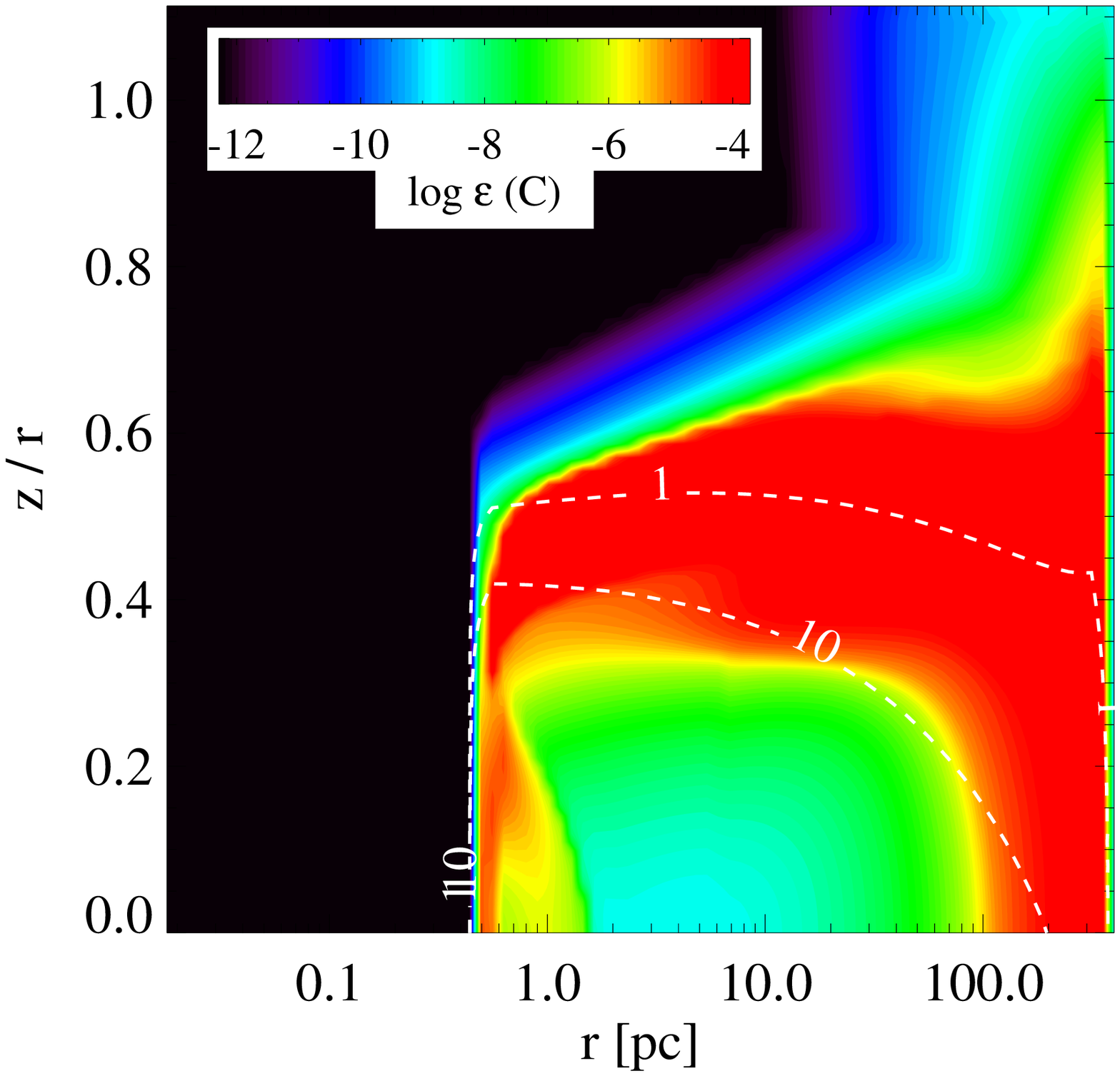}
  \includegraphics[width=4.0cm]{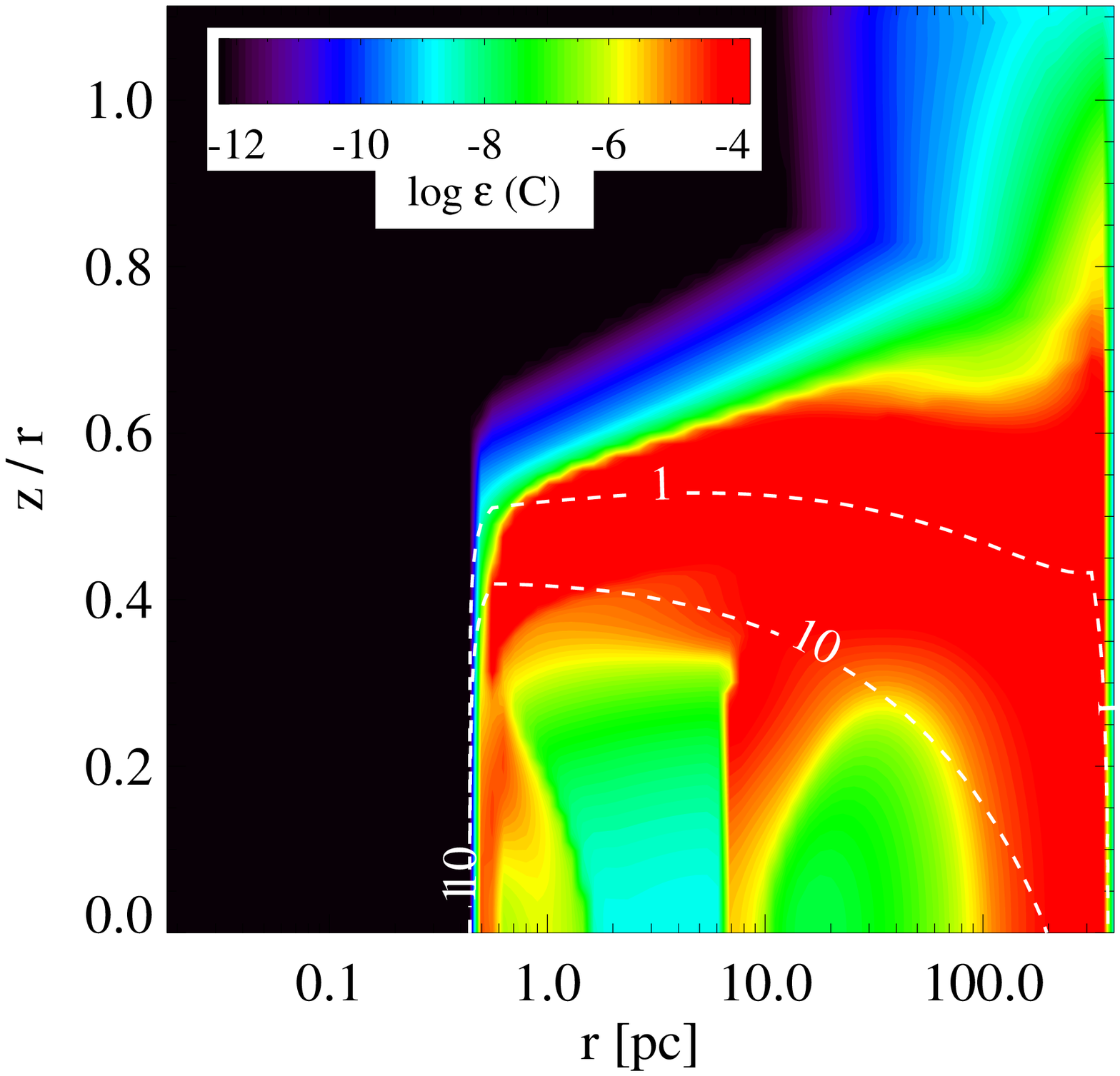}
  \includegraphics[width=4.0cm]{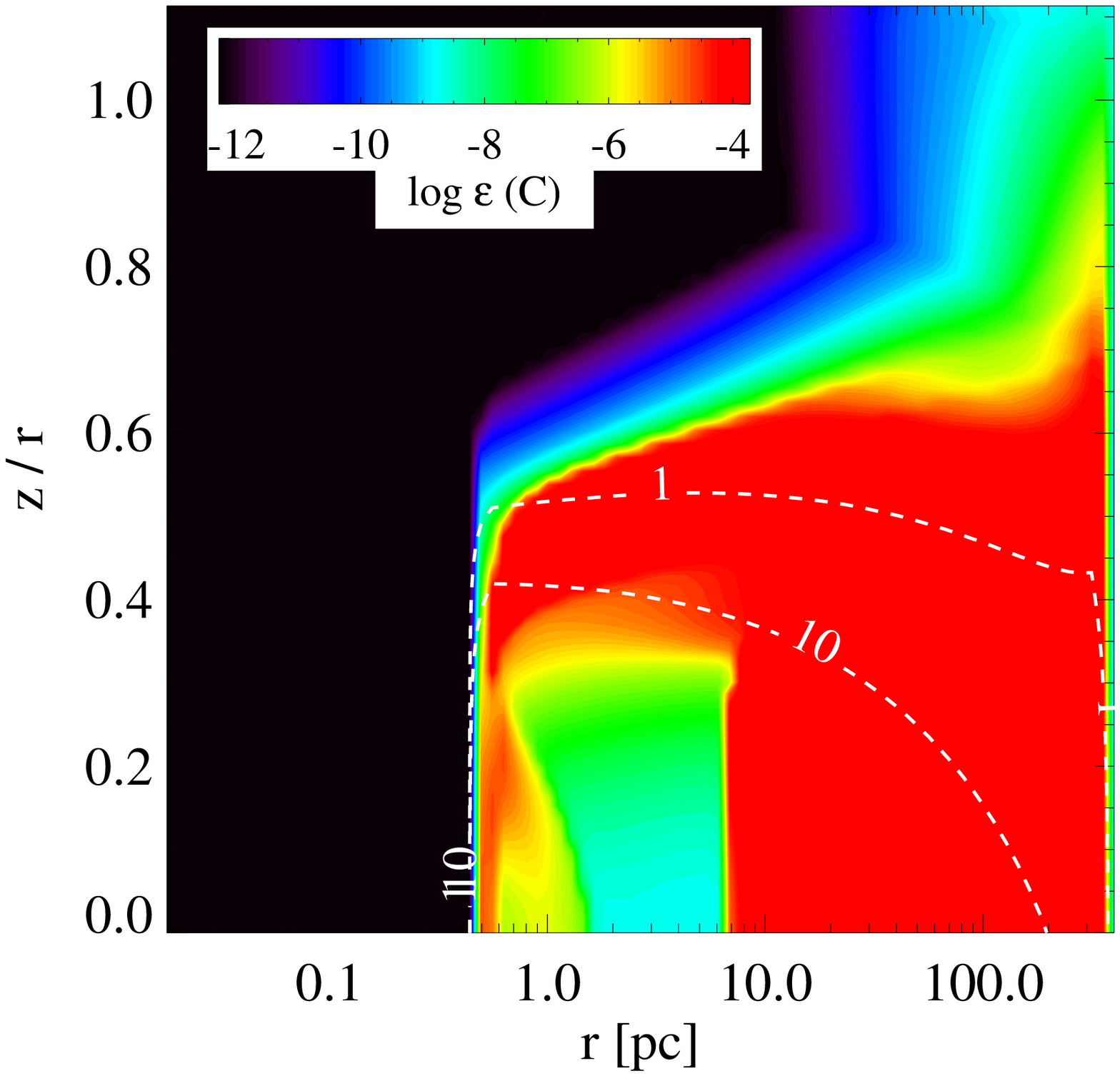}
  \includegraphics[width=4.0cm]{Cabundance.ps}
  \includegraphics[width=4.0cm]{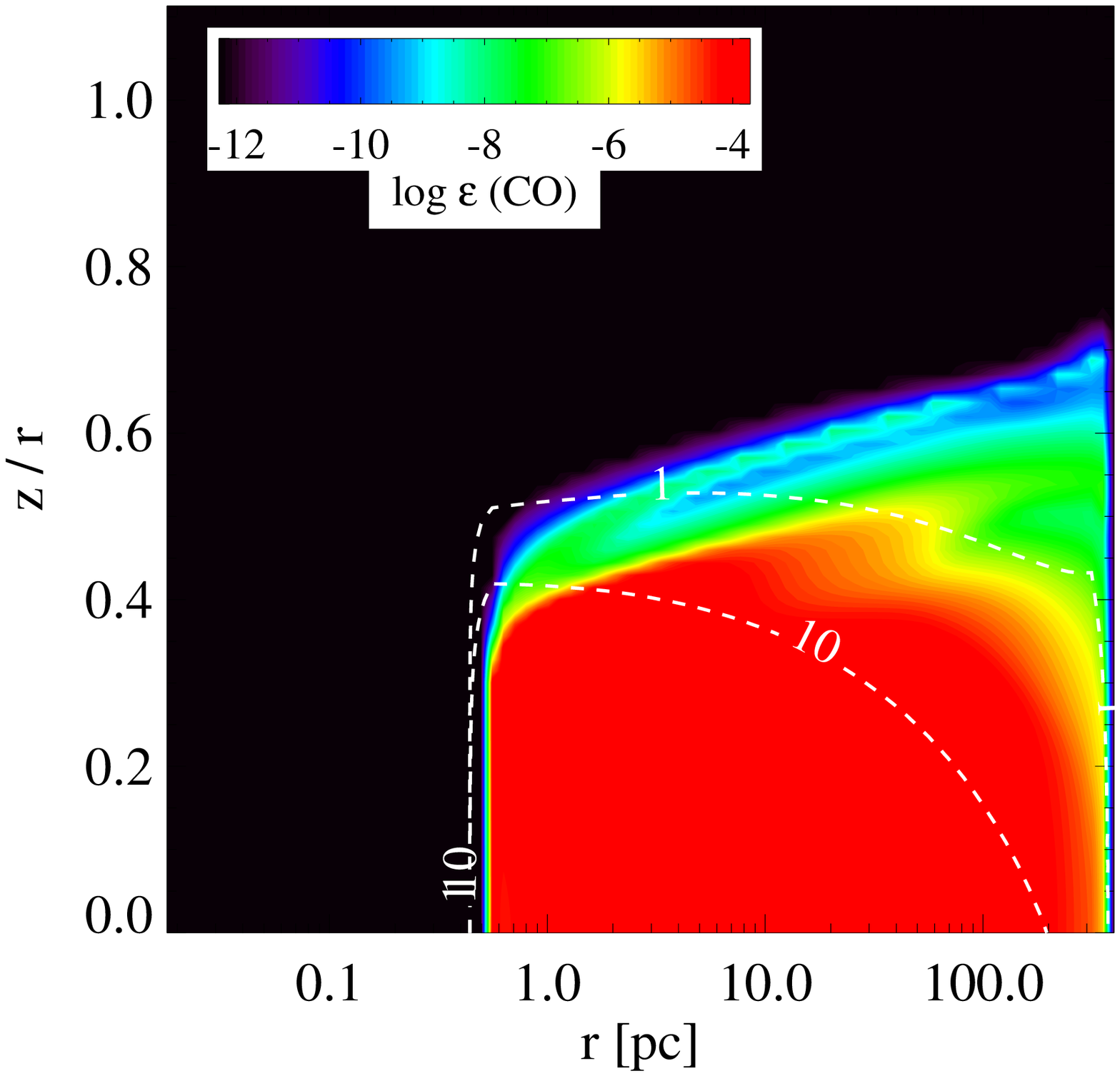}
  \includegraphics[width=4.0cm]{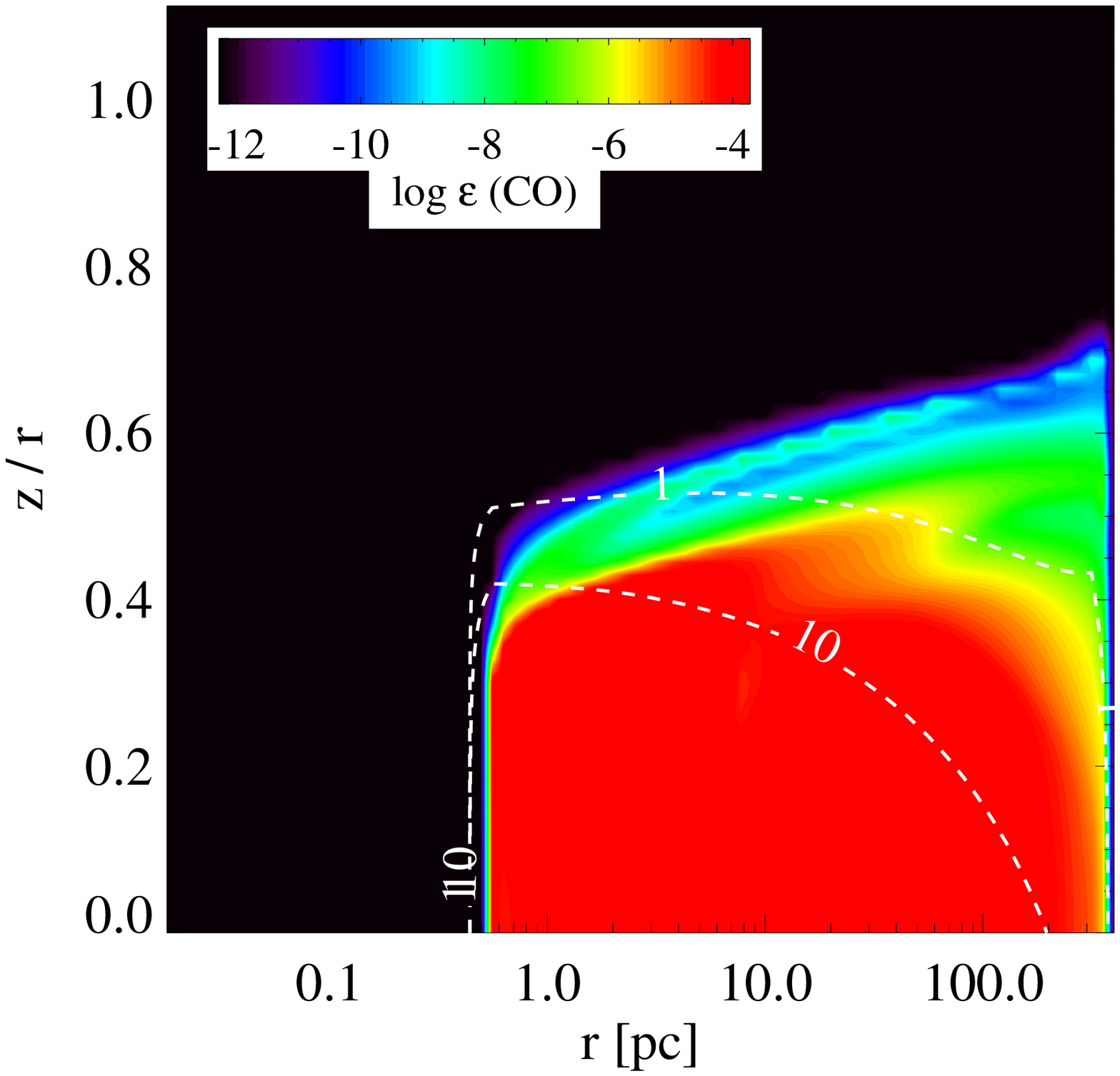}
  \includegraphics[width=4.0cm]{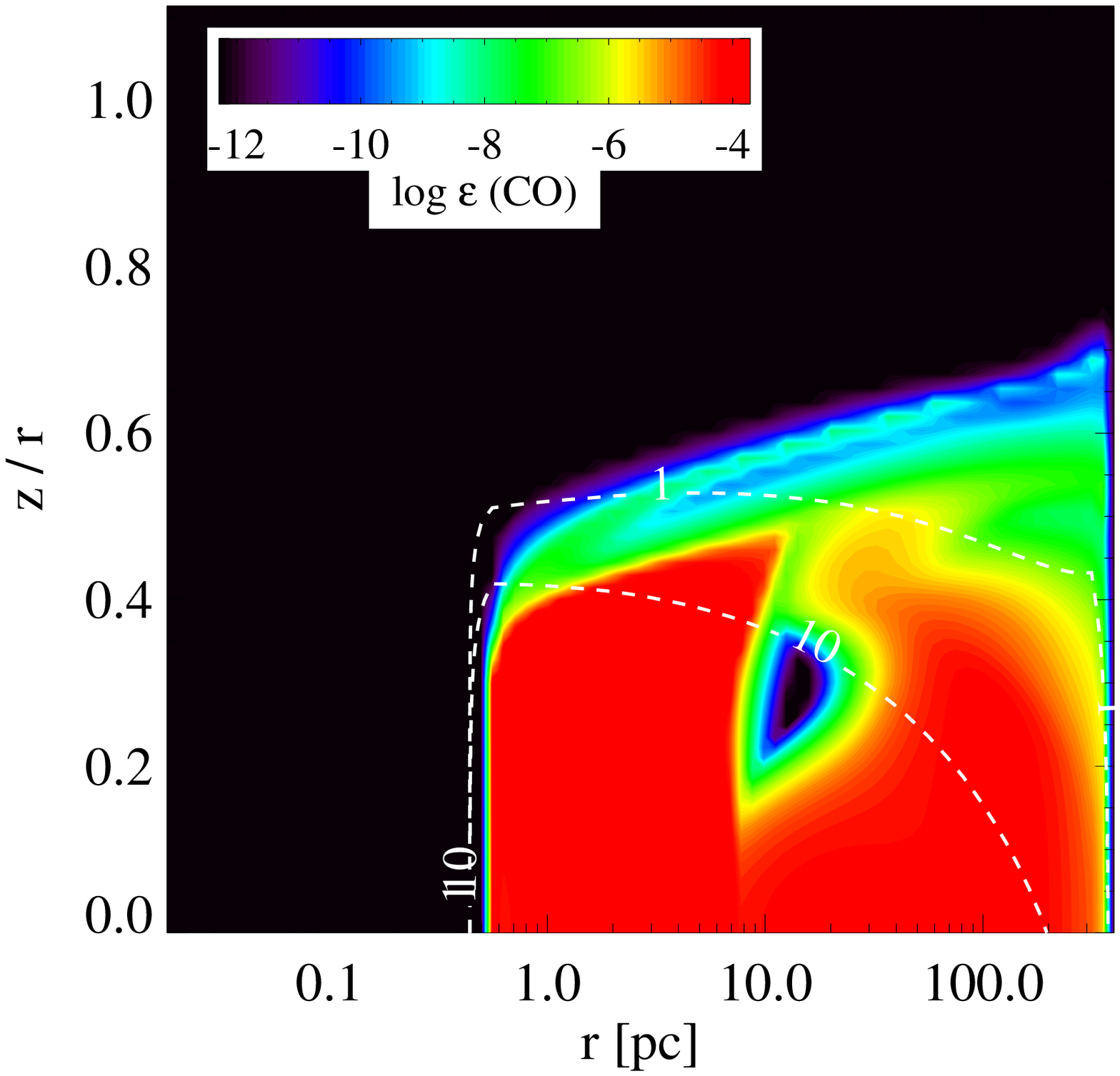}
  \includegraphics[width=4.0cm]{COabundance.ps}
  \includegraphics[width=4.0cm]{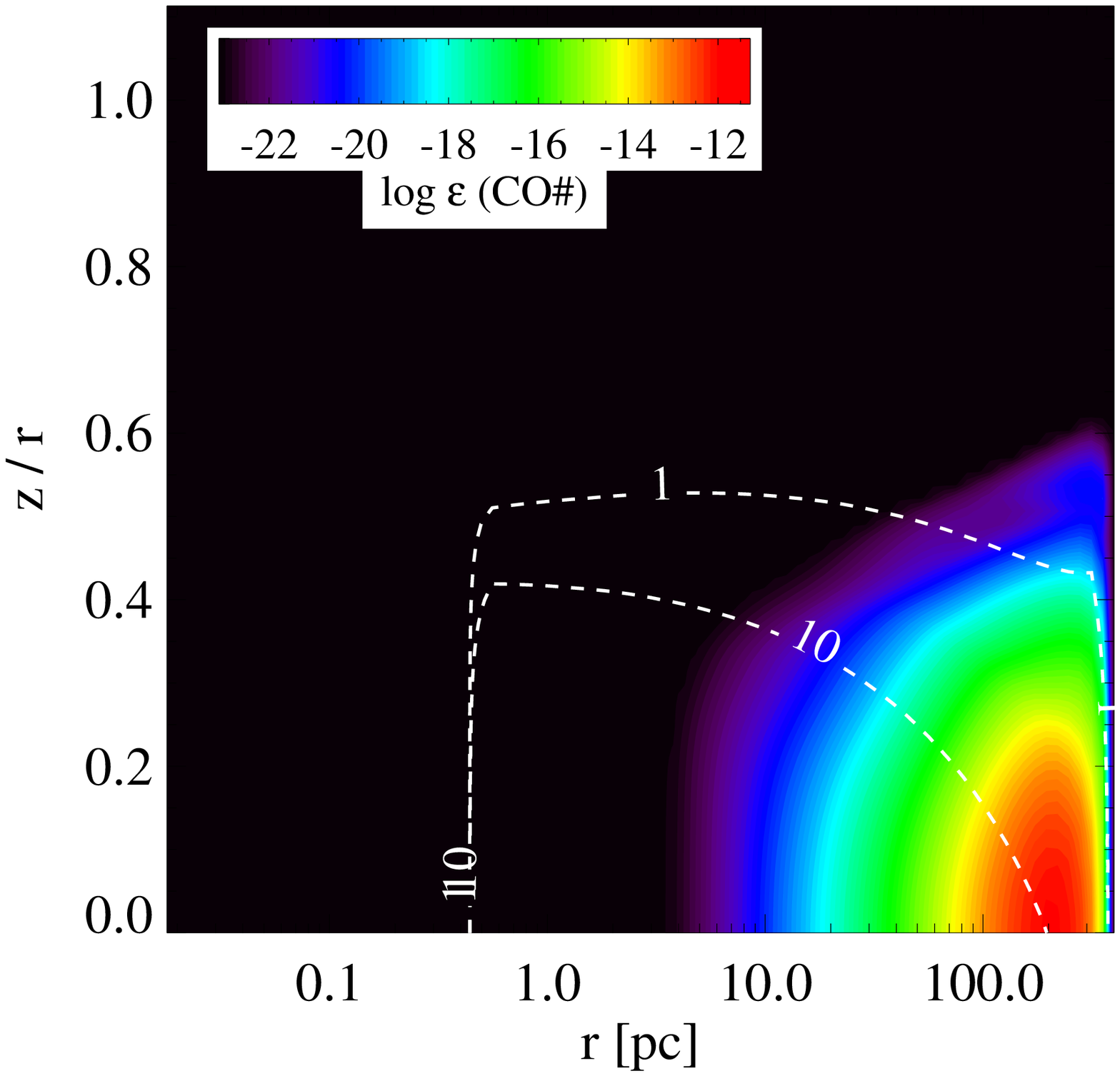}
  \includegraphics[width=4.0cm]{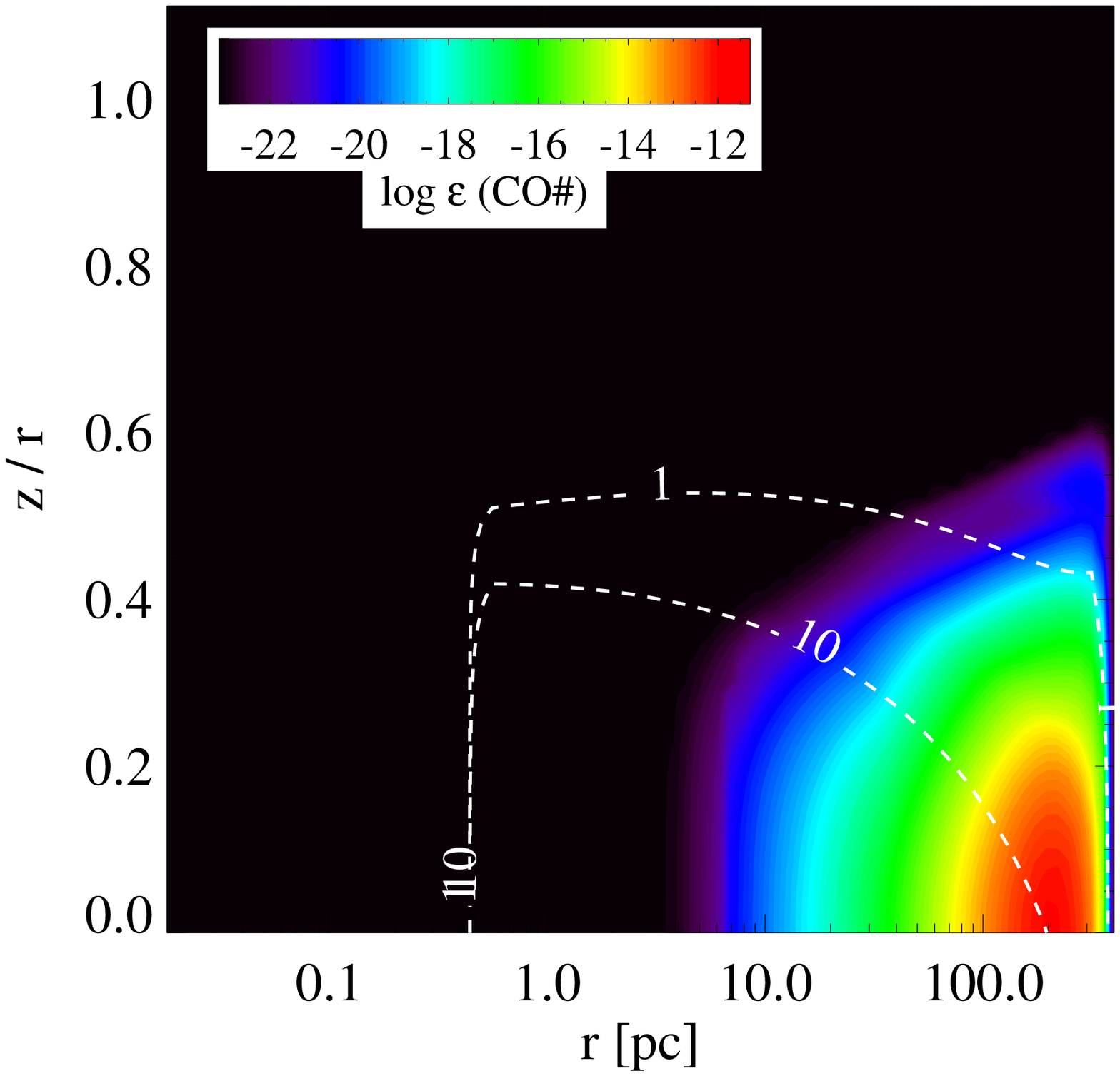}
  \includegraphics[width=4.0cm]{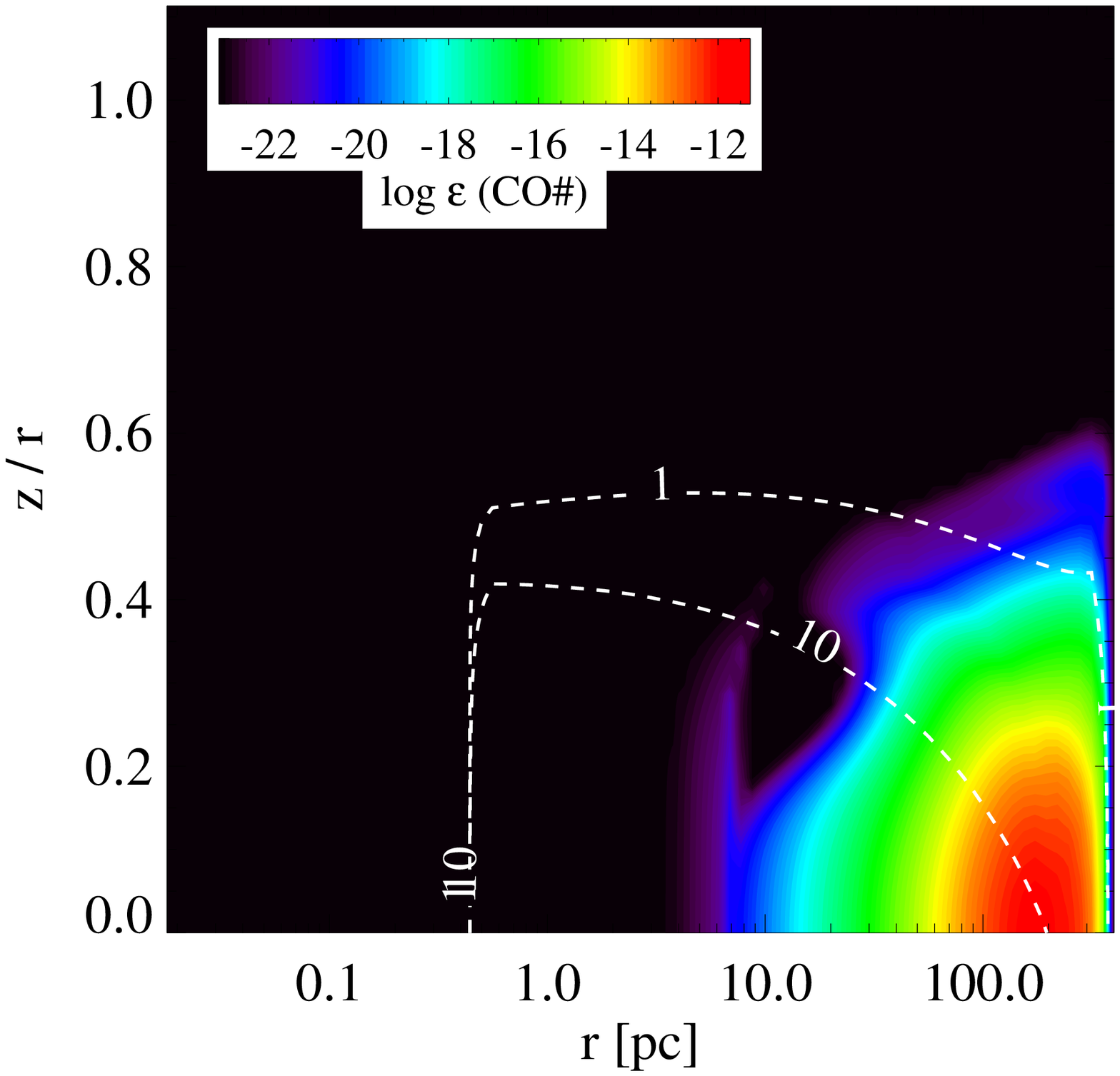}
  \includegraphics[width=4.0cm]{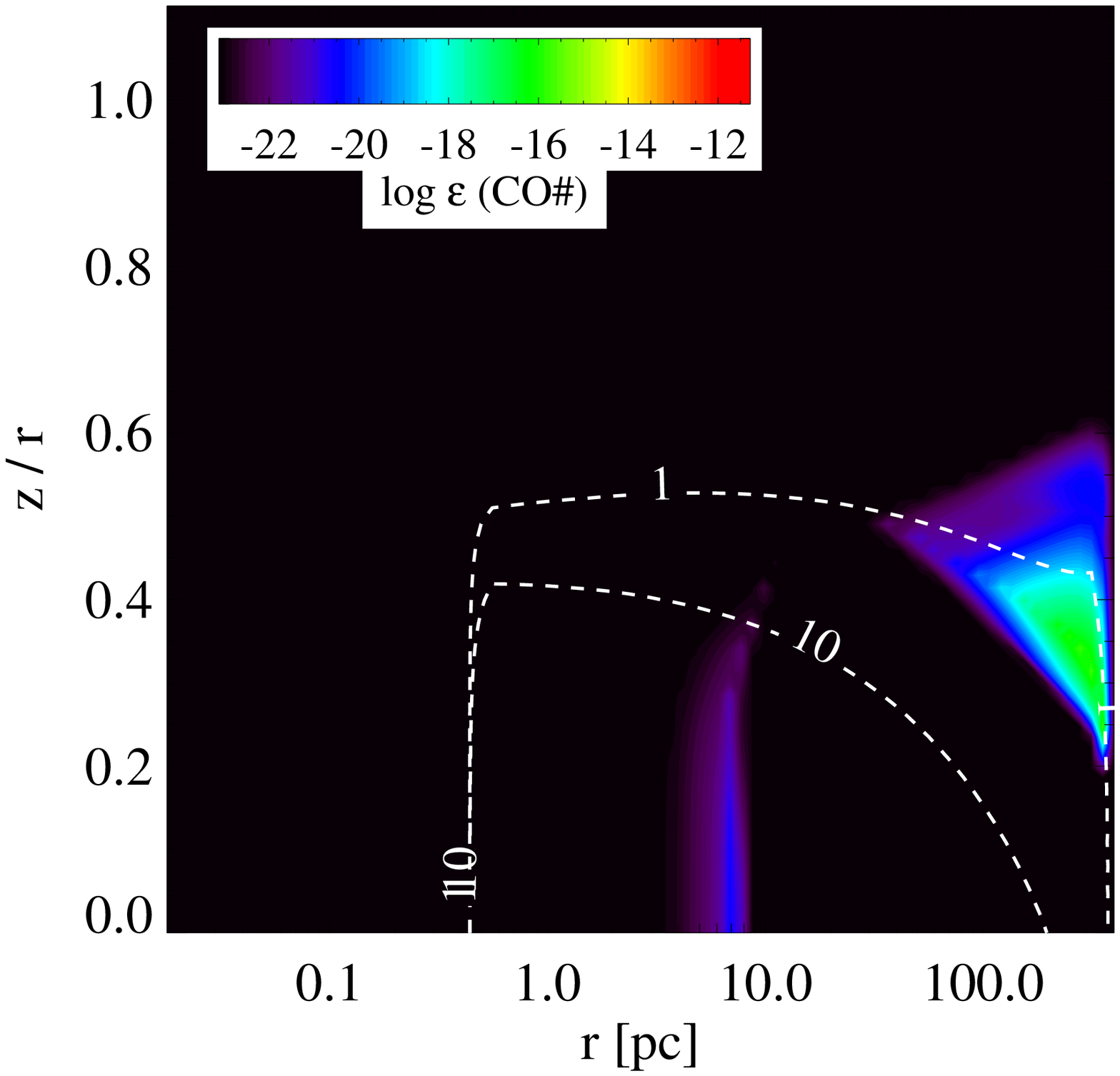}
  \caption{Chemical abundances of C$^+$ (top), C, CO and CO ice
    (bottom) from the time-dependent solution at time $t=10^4$,
    $10^6$, and $10^8$ years (left), and the equilibrium solution
    (right). Contours for $A_{\rm v}=1$ and 10 are overplotted.}
  \label{time_dep_chem2}
\end{figure*}

{\it C$^+$, C, CO, and CO ice:} Fig. \ref{time_dep_chem2} shows the
abundance structure of C$^+$, C, CO, and CO ice abundances at time
$t=10^4$, $10^6$, and $10^8$~yrs compared to the equilibrium
solution. The C$^+$ abundance structure shows a thick layer with
$x_{\rm C^+} \sim 10^{-4}$ at all times, similar to the equilibrium
solution. The second deeper-lying layer, however, is not interrupted
at early times. The interruption of this second layer, that occurs in
the equilibrium solution due to the freeze-out effects, starts to show
at $t>10^6$~yrs. At $t=10^4$~yrs, the neutral carbon abundance
structure shows high ($x_{\rm C} \sim 10^{-4}$) abundances down to the
mid-plane at both the inner edge of the disk and at $R > 100$ pc,
while CO has abundance $x_{\rm CO} \sim 10^{-4}$ in the entire disk up
to $R \sim 300$~pc, and $z/R \sim 0.4$. Over time, neutral carbon
becomes more abundant, and at $t=10^8$~yrs, the radius beyond which
carbon is dominant at the mid-plane moved from $R\sim 100$~pc to 5~pc,
while the CO abundance slowly drops. Compared to the equilibrium
model, the CO is not completely depleted from the gas-phase, and still
has significant abundances out to $R\sim 300$~pc at $t=10^8$~yrs. We
also find that a small abundance ($x_{\rm CO ice} < 10^{-12}$) of CO
ice is developing on the grains. The fraction of CO ice compared to
gas phase CO is orders of magnitude smaller.

\begin{figure*}
  \centering
  \includegraphics[width=4.0cm]{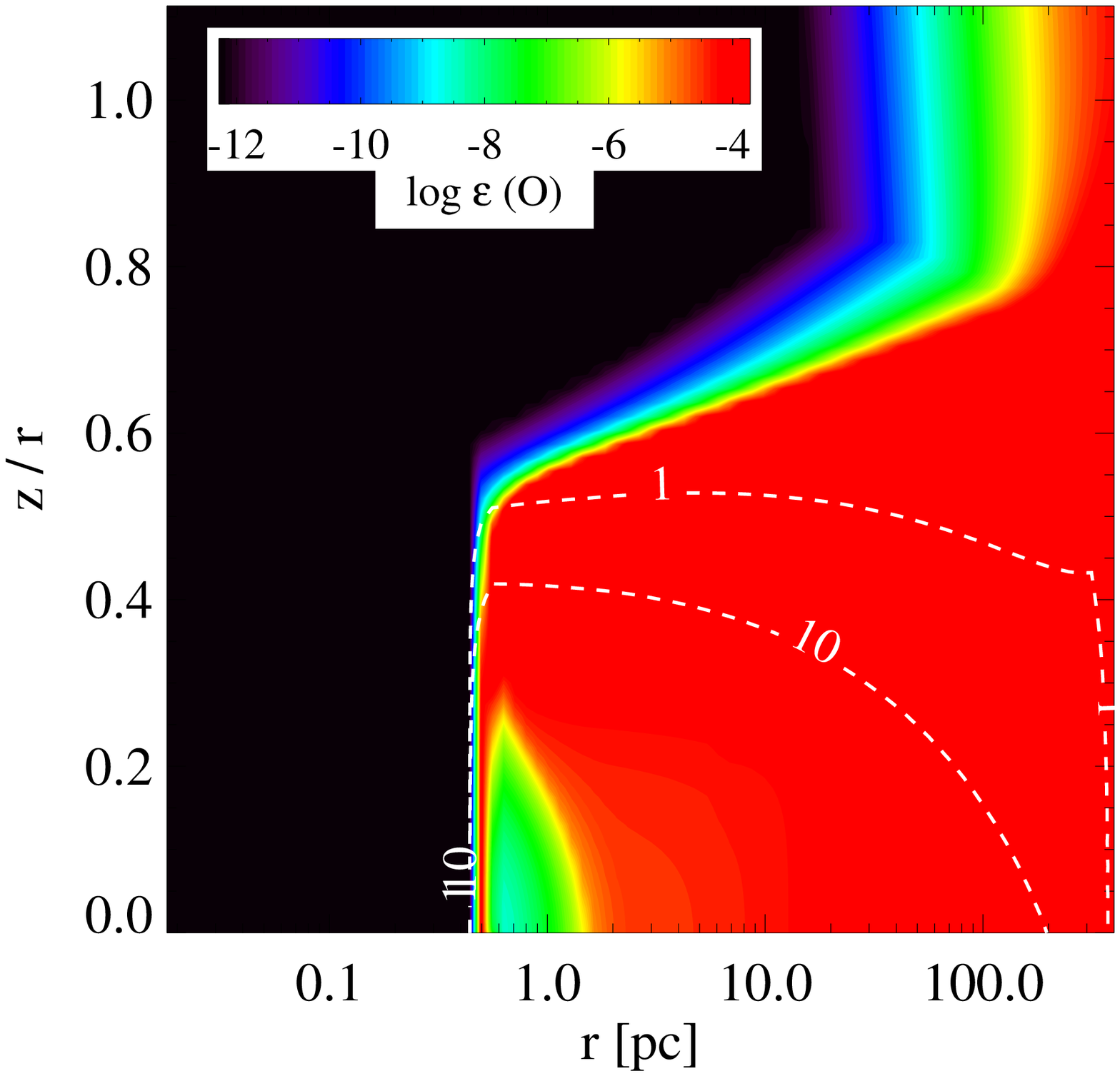}
  \includegraphics[width=4.0cm]{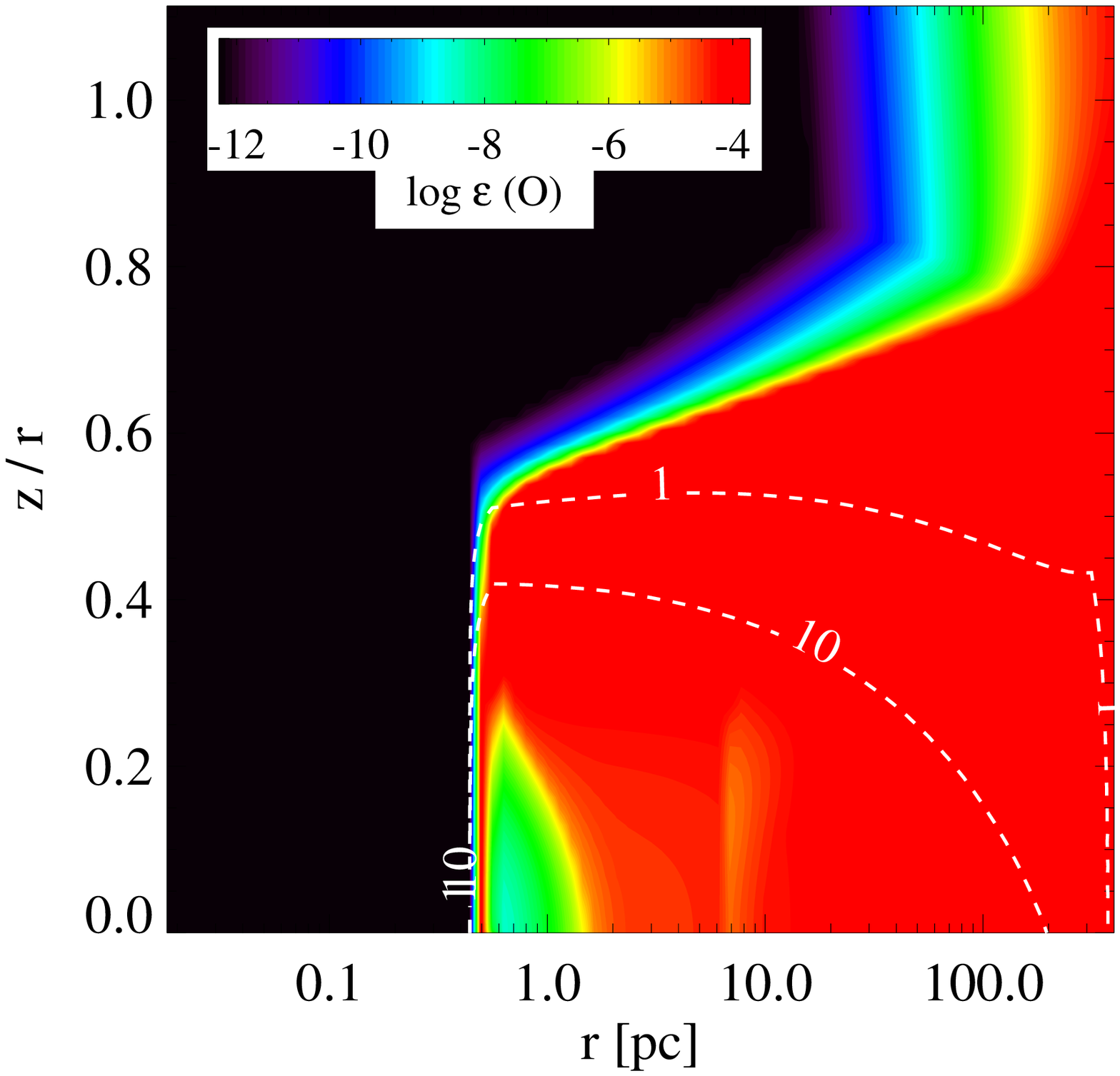}
  \includegraphics[width=4.0cm]{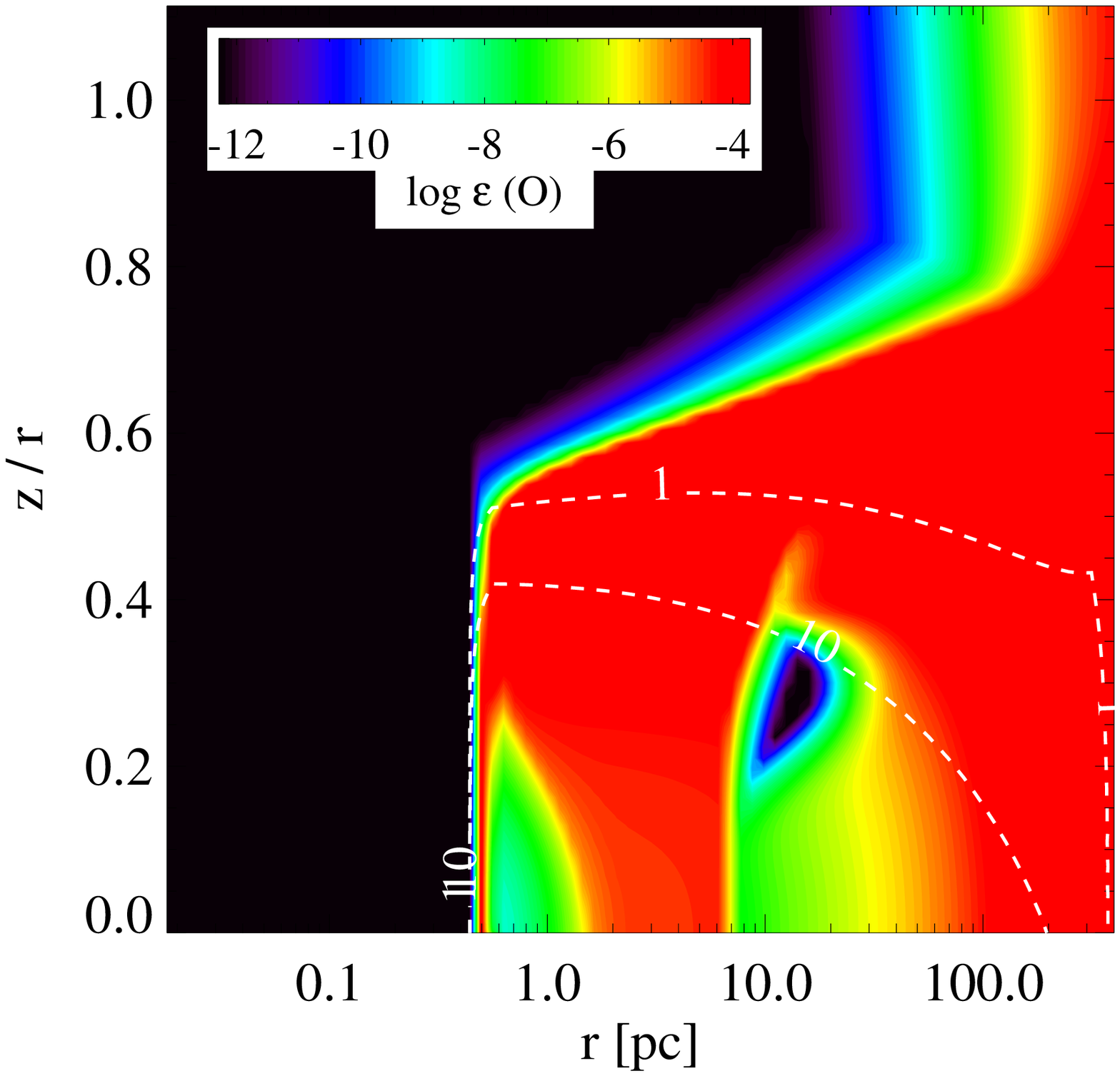}
  \includegraphics[width=4.0cm]{Oabundance.ps}
  \includegraphics[width=4.0cm]{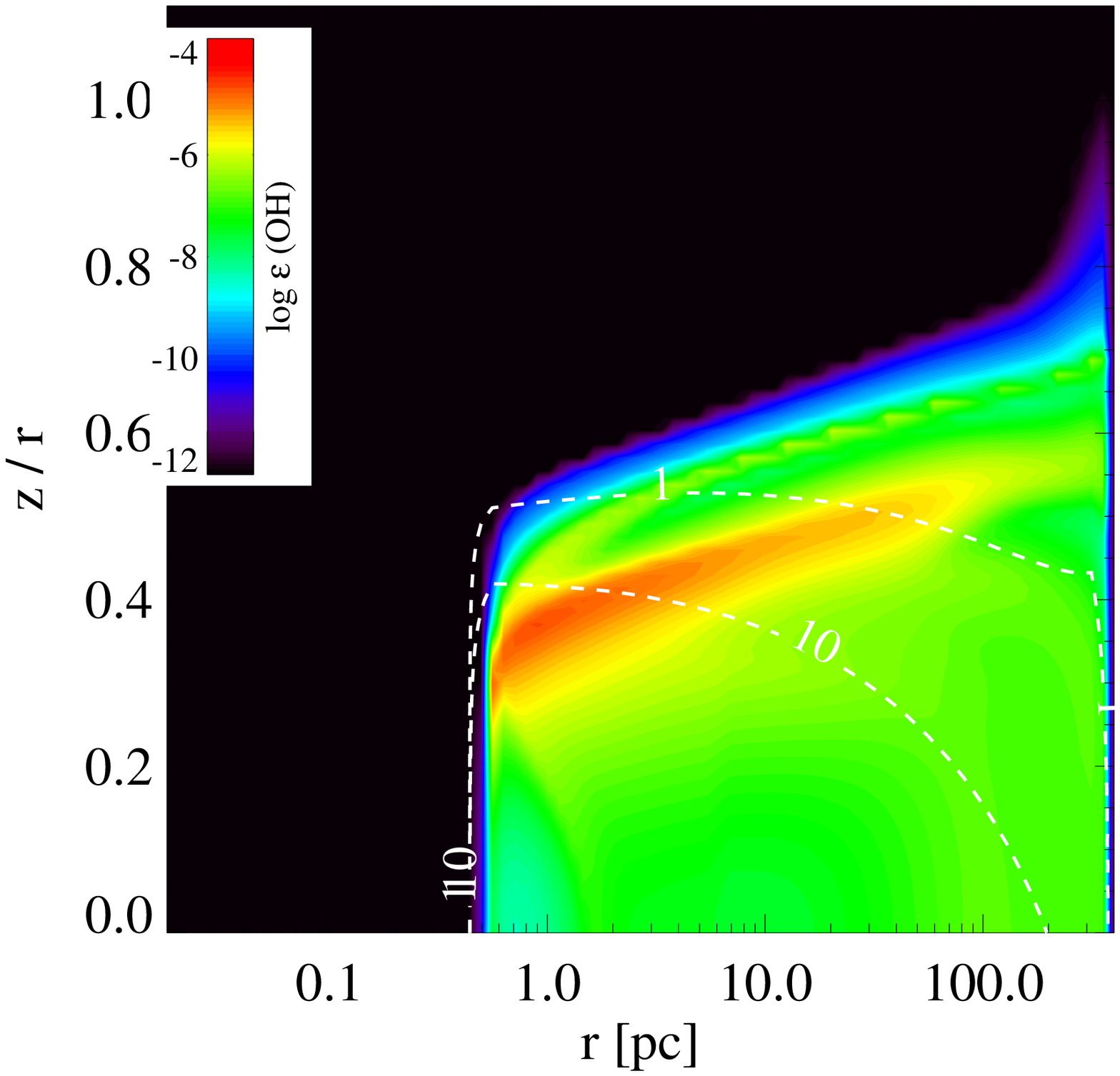}
  \includegraphics[width=4.0cm]{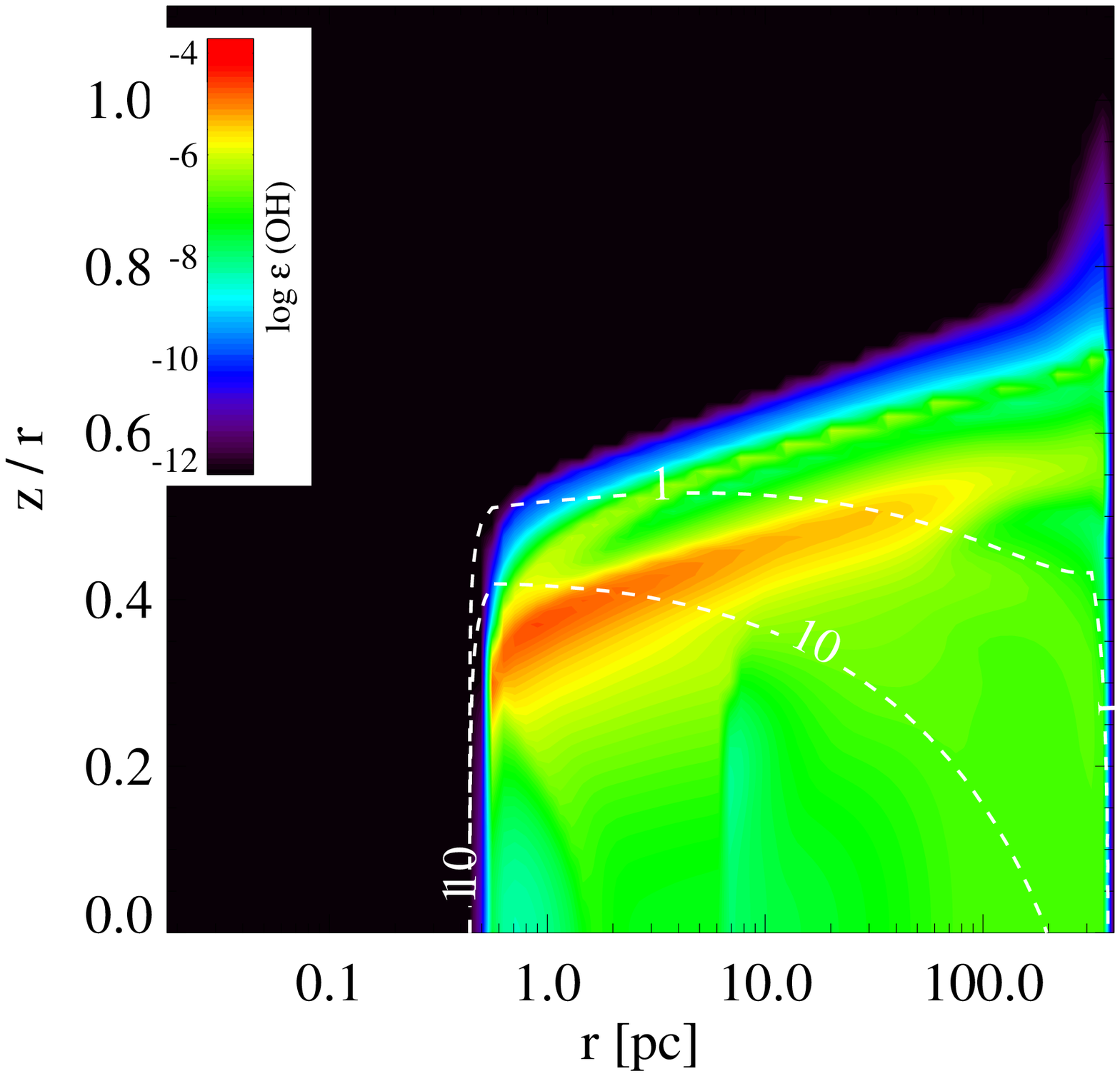}
  \includegraphics[width=4.0cm]{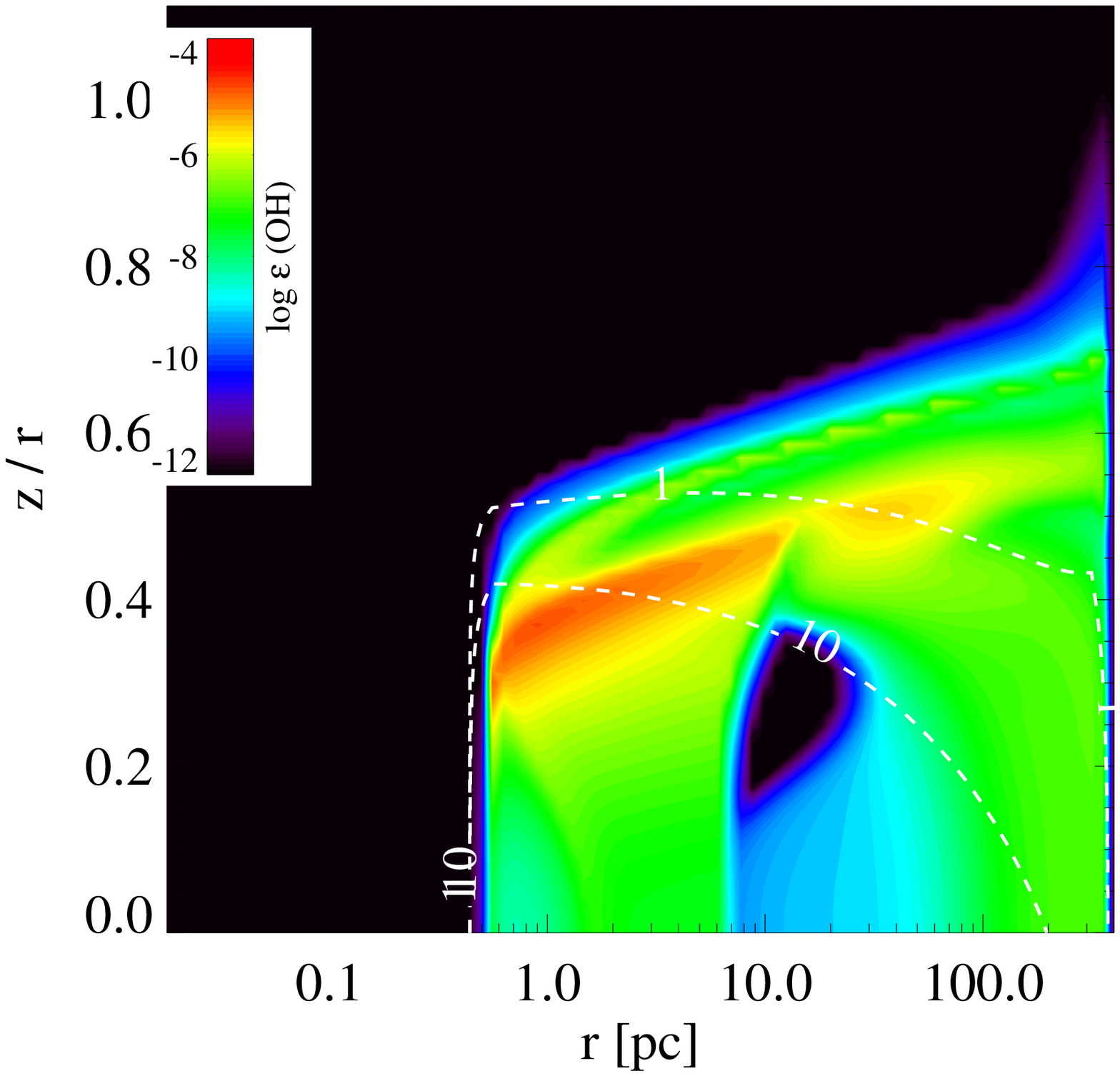}
  \includegraphics[width=4.0cm]{OHabundance.ps}
  \includegraphics[width=4.0cm]{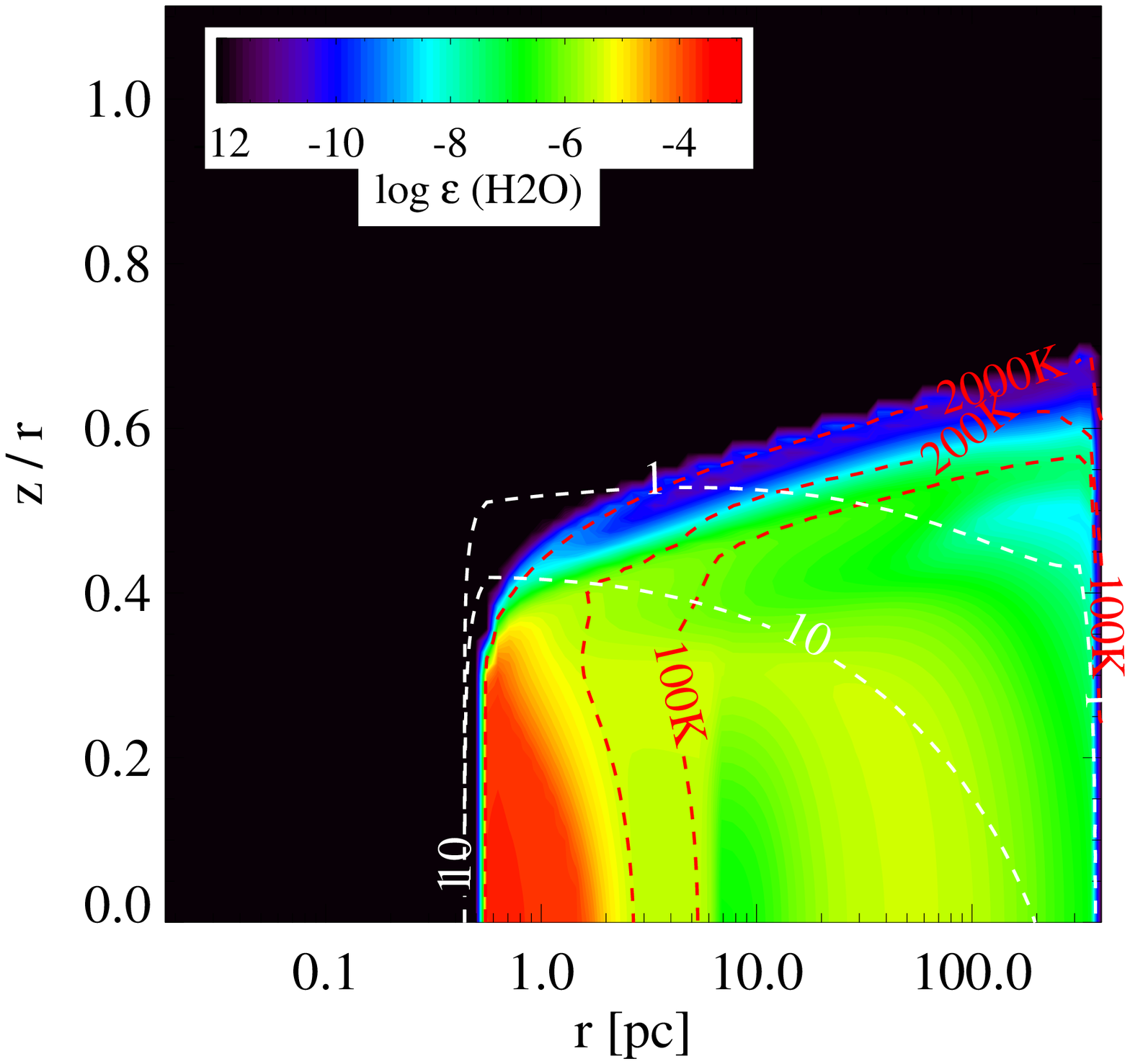}
  \includegraphics[width=4.0cm]{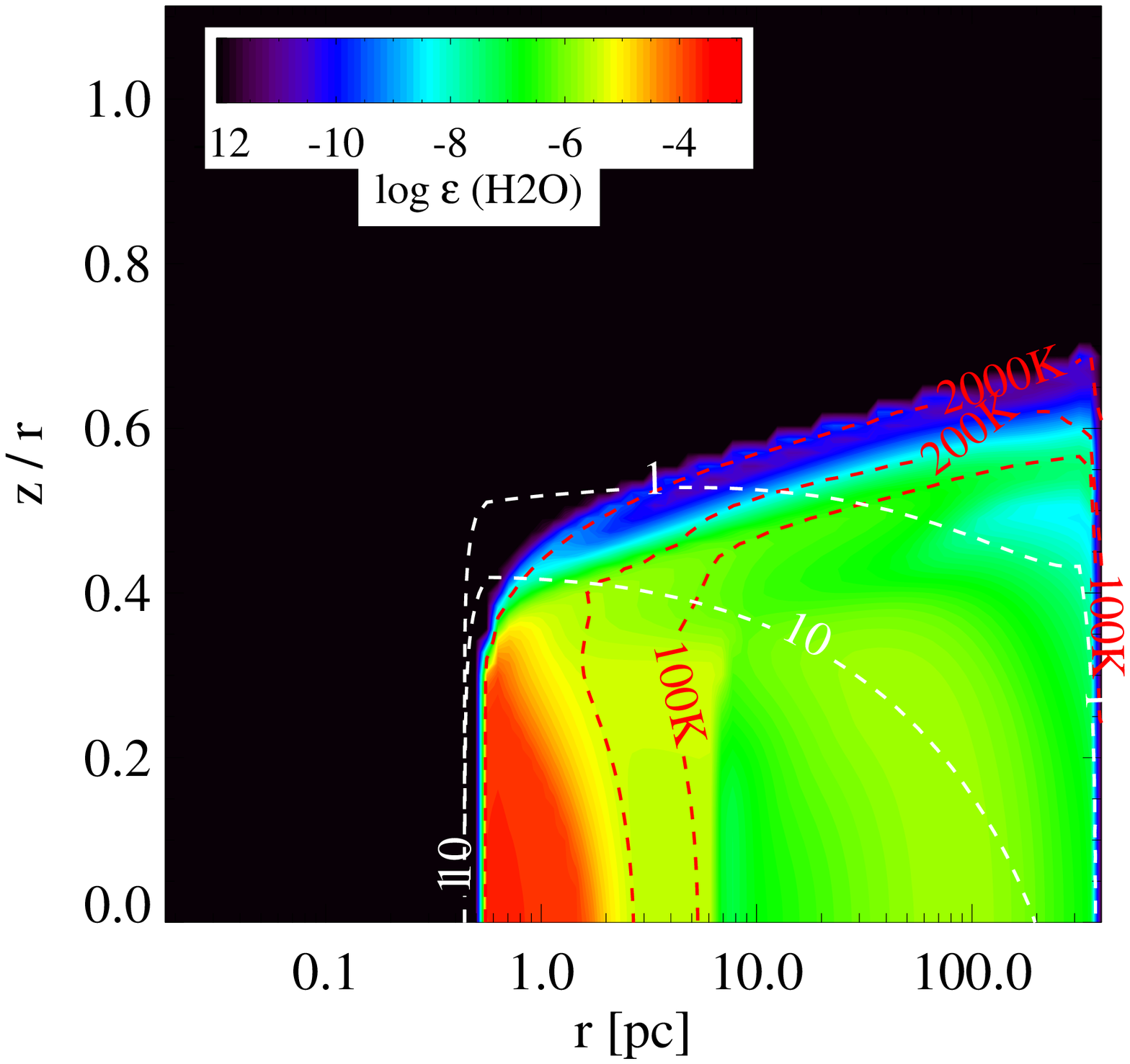}
  \includegraphics[width=4.0cm]{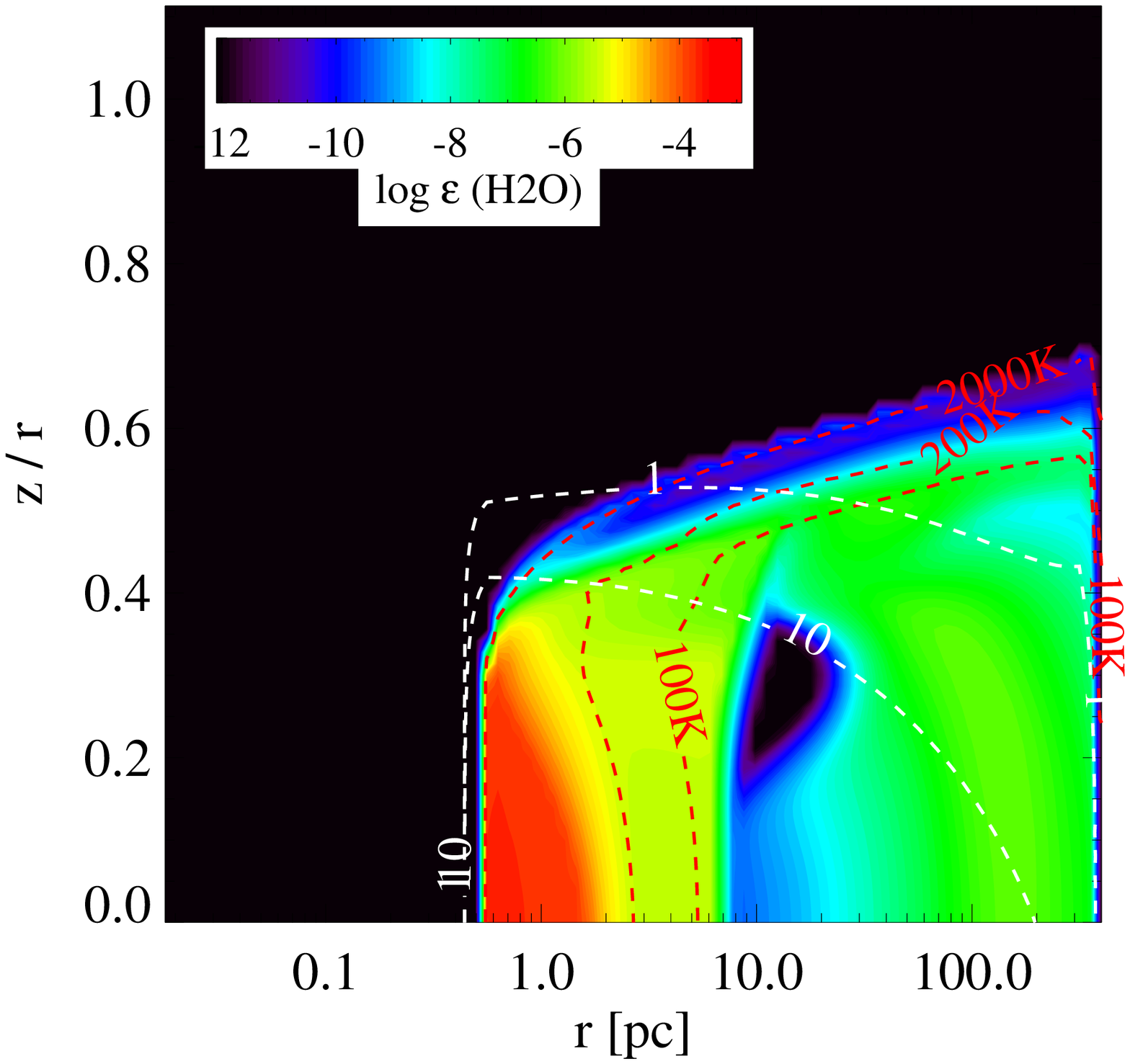}
  \includegraphics[width=4.0cm]{H2Oabundance.ps}
  \includegraphics[width=4.0cm]{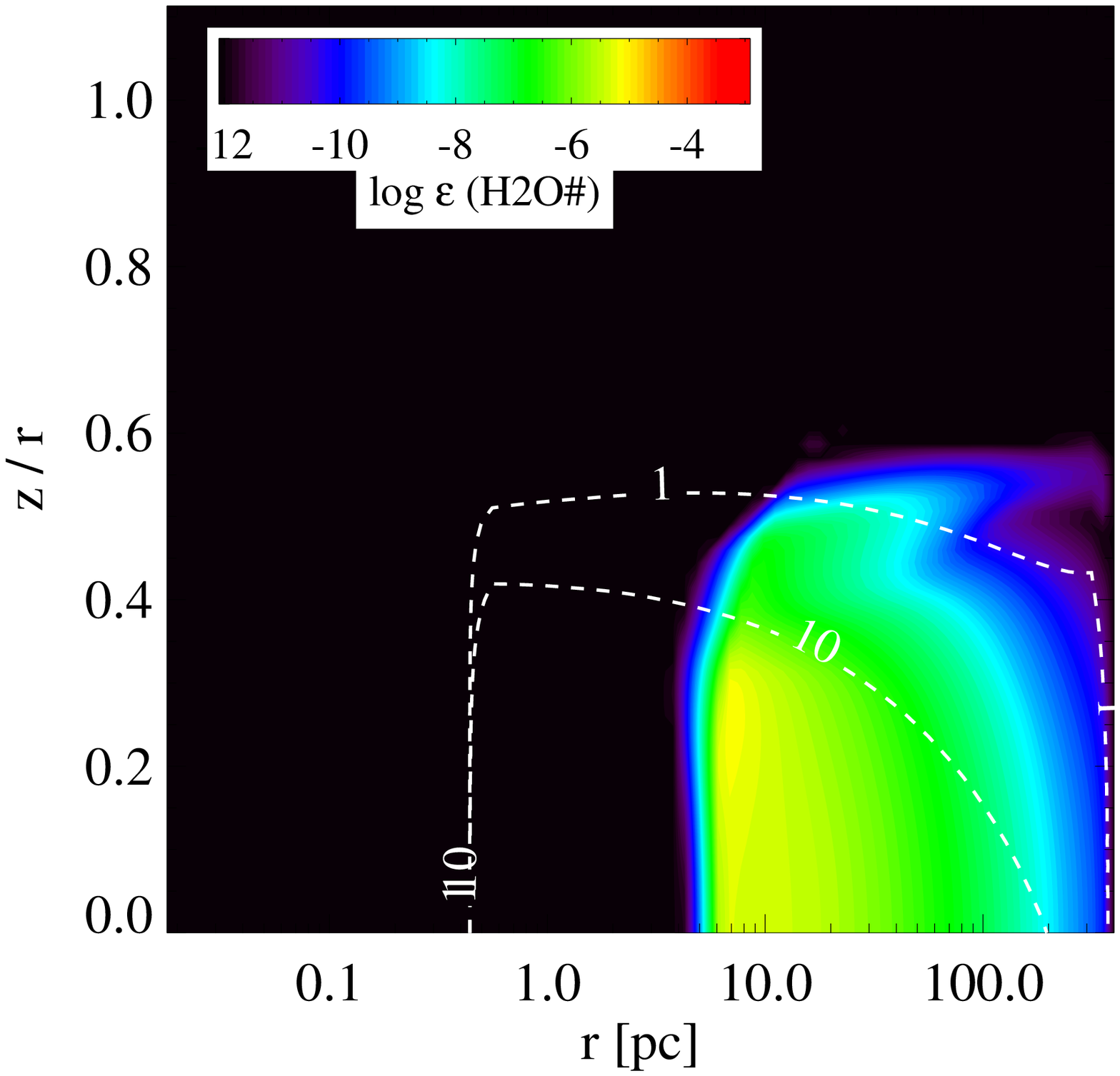}
  \includegraphics[width=4.0cm]{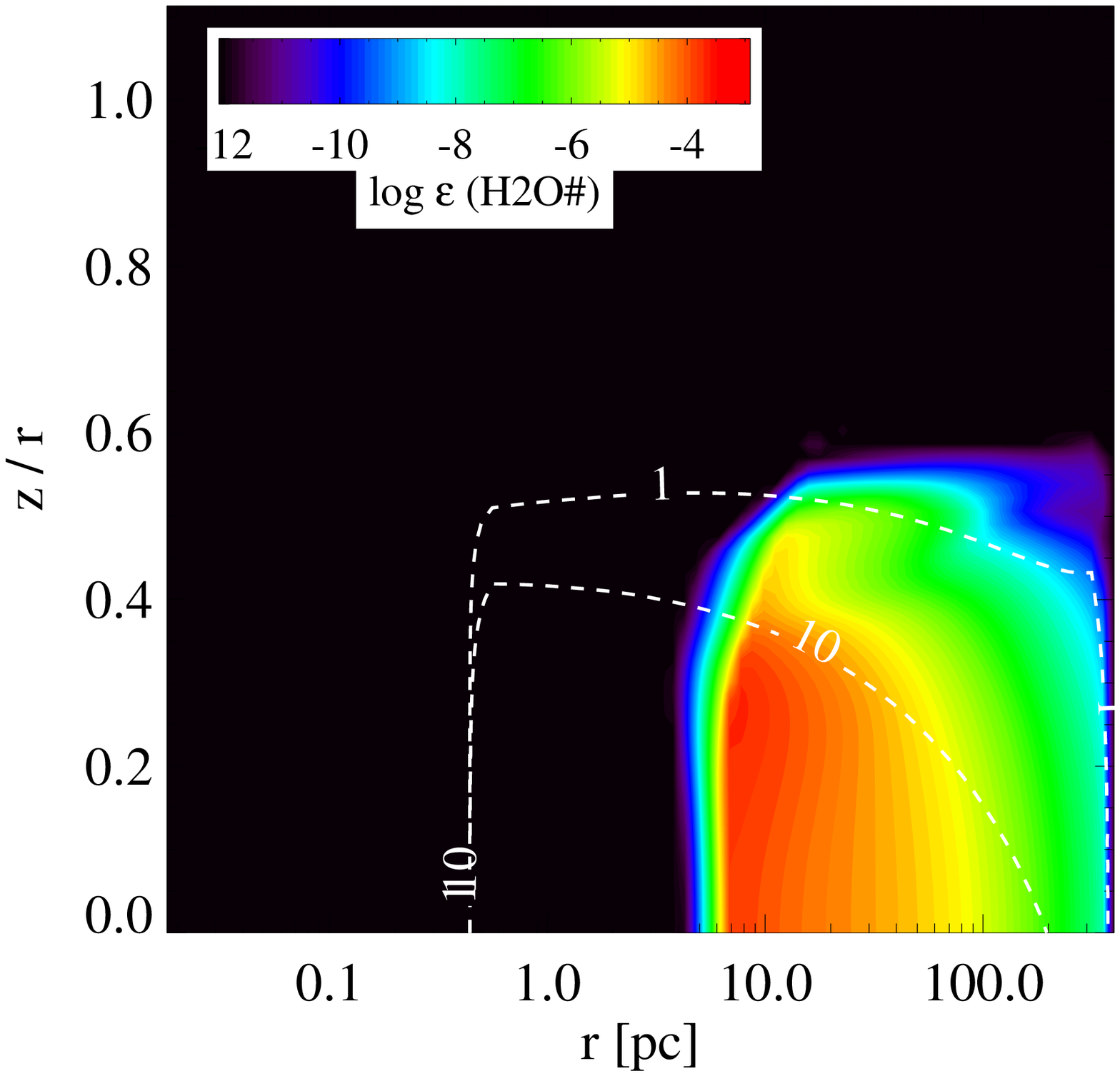}
  \includegraphics[width=4.0cm]{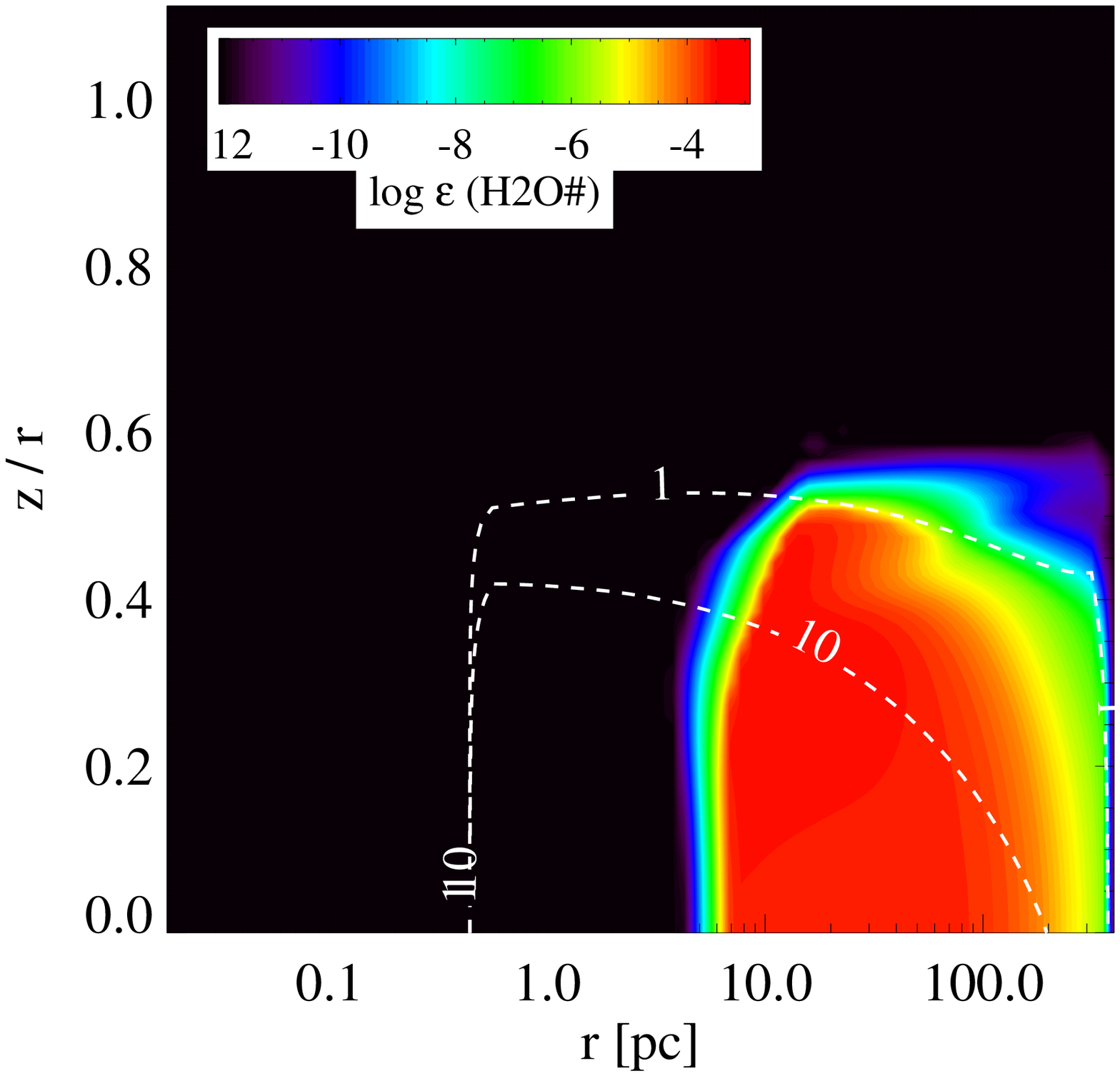}
  \includegraphics[width=4.0cm]{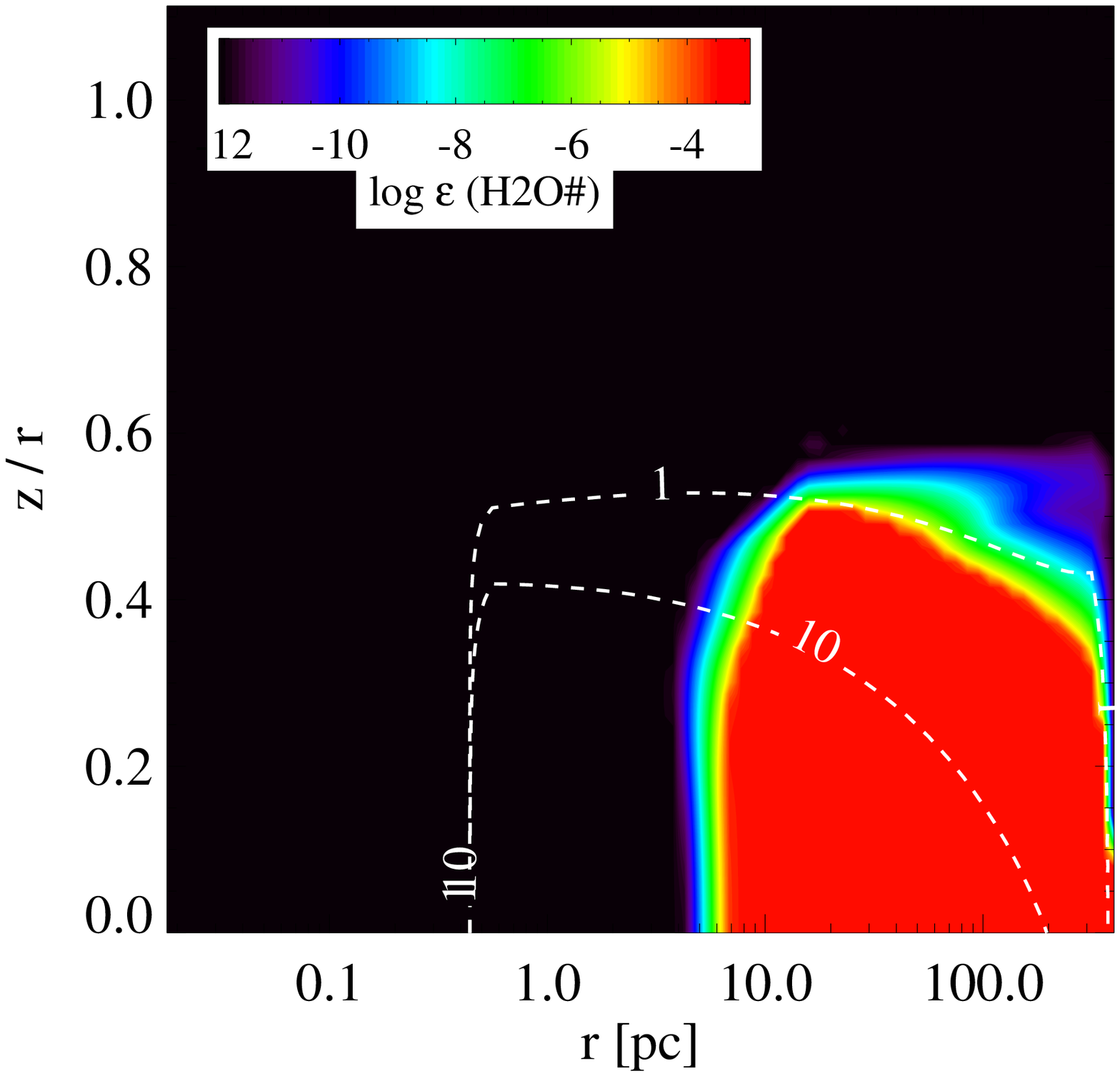}
  \caption{Chemical abundances of O (top), OH, H$_2$O, and H$_2$O ice
    (bottom) from the time-dependent solution at time $t=10^4$,
    $10^6$, and $10^8$ years (left), and the equilibrium solution
    (right). Contours for $A_{\rm v}=1$ and 10 and temperature
    $T=100$, 200 and 2000~K are overplotted.}
  \label{time_dep_chem3}
\end{figure*}

{\it O, OH, and H$_2$O (Fig. \ref{time_dep_chem3}):} Oxygen starts to get
affected at $t > 10^6$~years. It only drops below an abundance $x_{O}
< 10^{-12}$ in a small part of the disk at $t=10^{8}$~ yrs, while the
region where oxygen is below $x < 10^{-6}$, spans from $R=5-30$~pc and
$z/R=0-0.4$. Overall, the time-dependent oxygen abundance structure
does not show the prominent freeze-out region as in the case of the
equilibrium model. The OH abundance shows an interrupted layer with a
maximum abundance $x_{\rm OH} \sim 10^{-5}$ in the equilibrium model,
which is due to freeze-out. This interruption is not present at early
times $t < 10^6$~yrs. Although it starts to show at $t=10^8$~yrs, it
is not as prominent as in the equilibrium model. The OH is slowly
being depleted from the gas phase, but it only drops below $x_{\rm OH}
< 10^{-12}$ in a small fraction of the molecular part of the disk,
similar to oxygen. H$_2$O is very abundant at the inner rim ($x_{\rm
  H_2O} \sim 10^{-4}$), but also shows abundances $x_{\rm} > 10^{-6}$
in large parts of the disk, where gas temperatures are less favorable
for water formation. Similar to oxygen and OH, H$_2$O is slowly
depleted from the gas phase, but there are still significant
abundances out to $R\sim 300$~pc after $t=10^8$~yrs. The water ice
fraction is slowly increasing over time at $R > 5$~pc. However, after
$t=10^8$~yrs, the region where H$_2$O ice abundance exceeds $x_{\rm
  H_2O ice} > 10^{-4}$ is not as extended as in the equilibrium model
($R\sim 100-200$~pc compared to $R=400$~pc).

\begin{figure*}
  \centering
  \includegraphics[width=4.0cm]{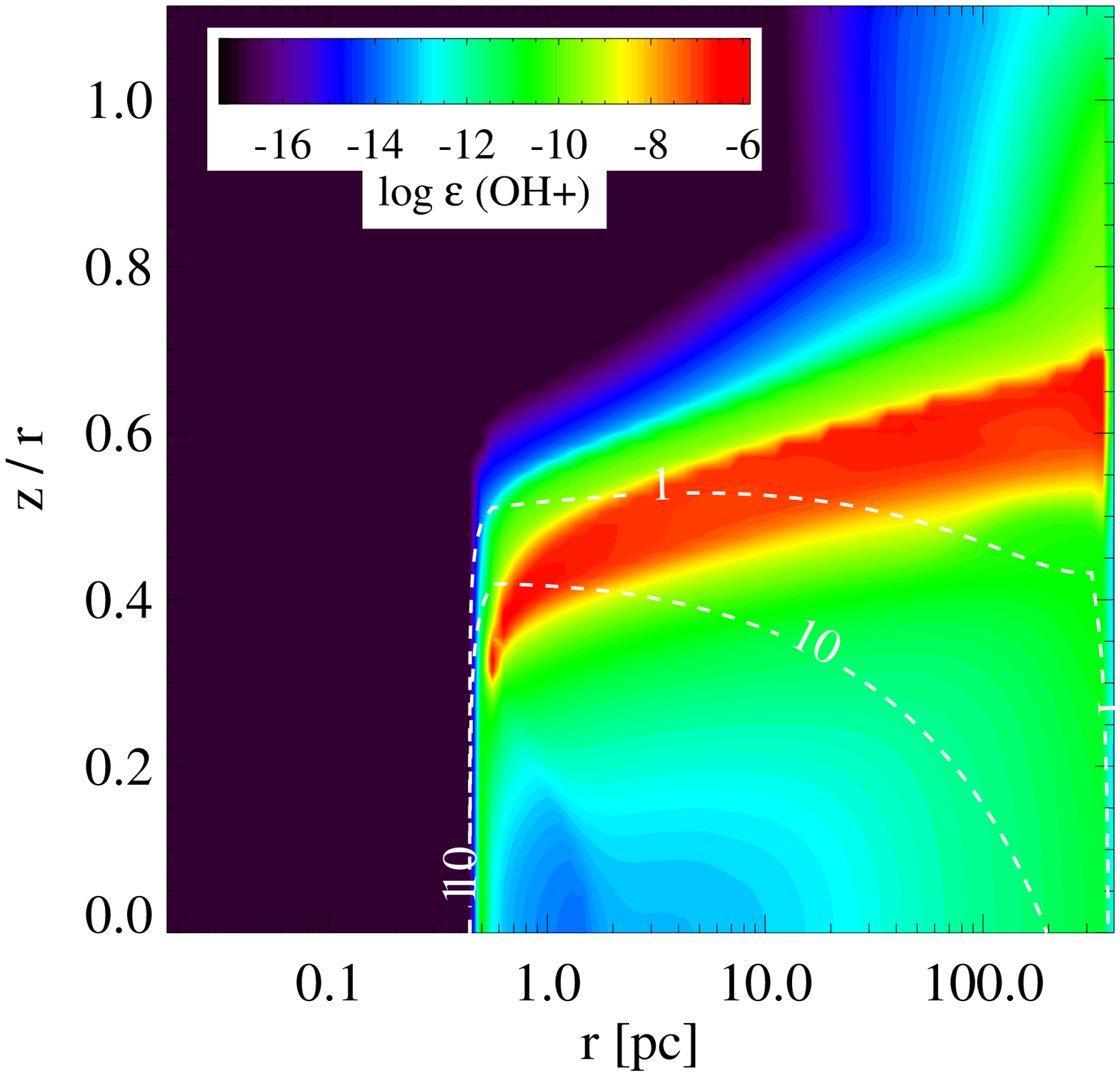}
  \includegraphics[width=4.0cm]{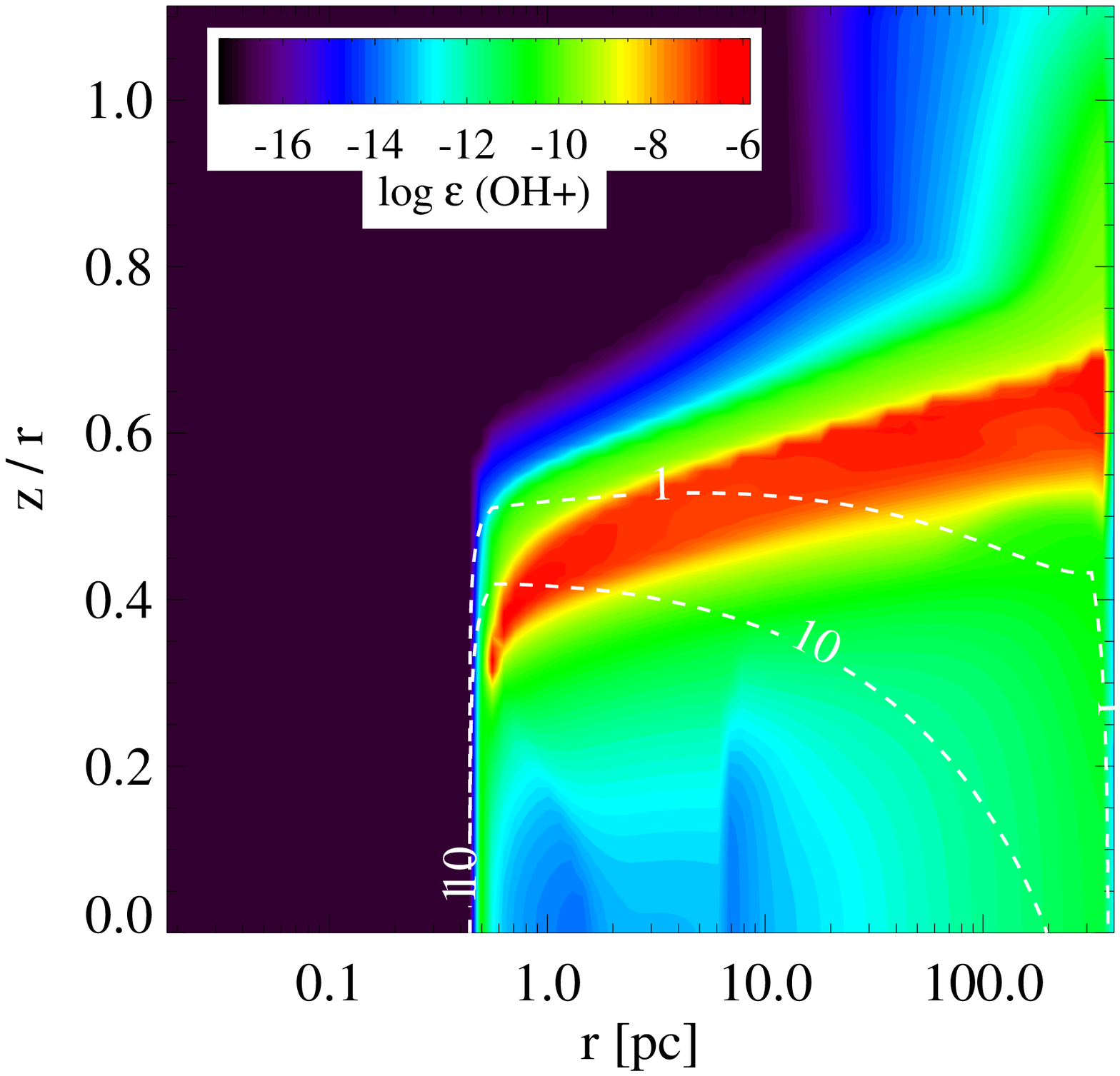}
  \includegraphics[width=4.0cm]{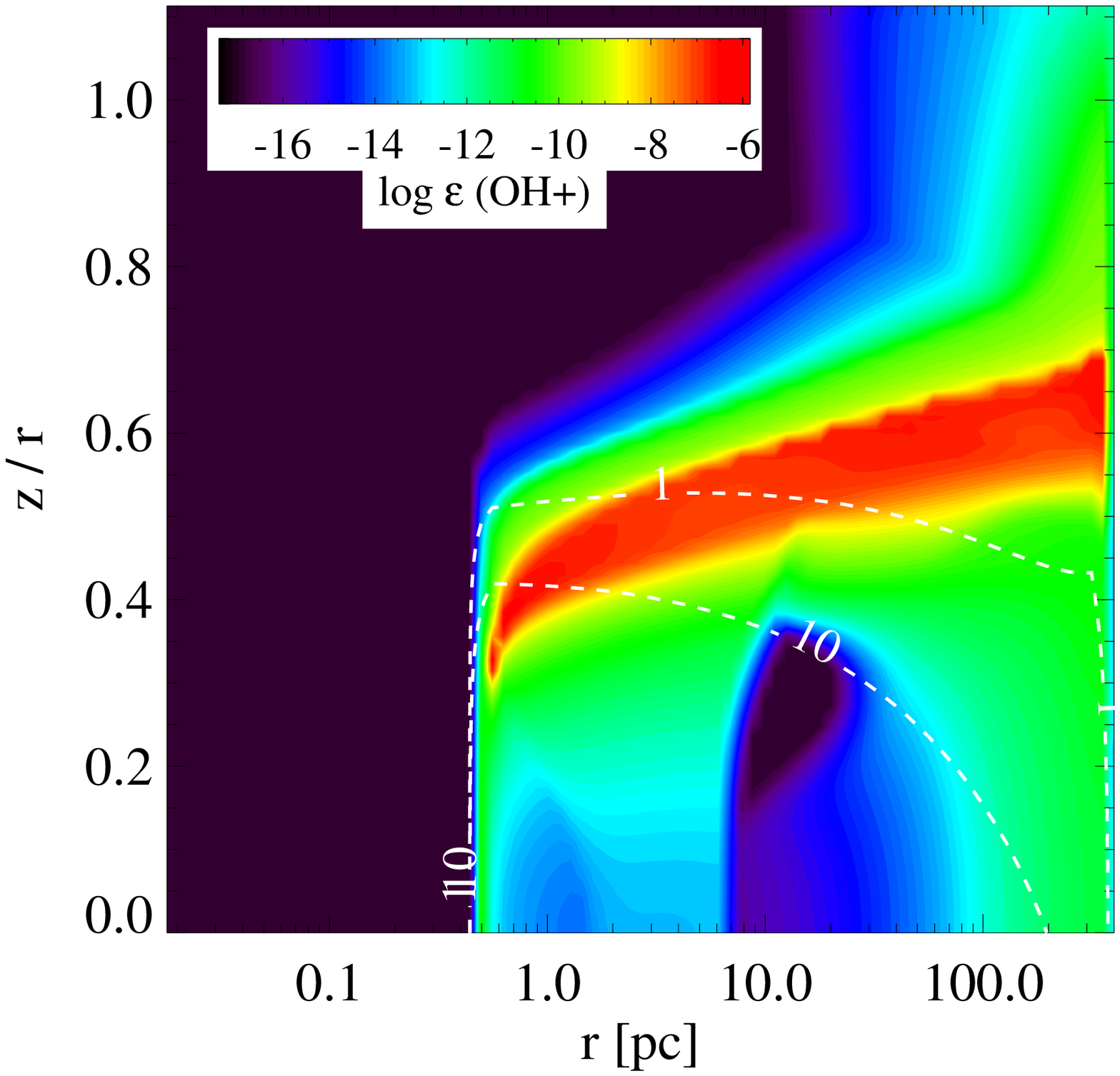}
  \includegraphics[width=4.0cm]{OHpabundance.ps}
  \includegraphics[width=4.0cm]{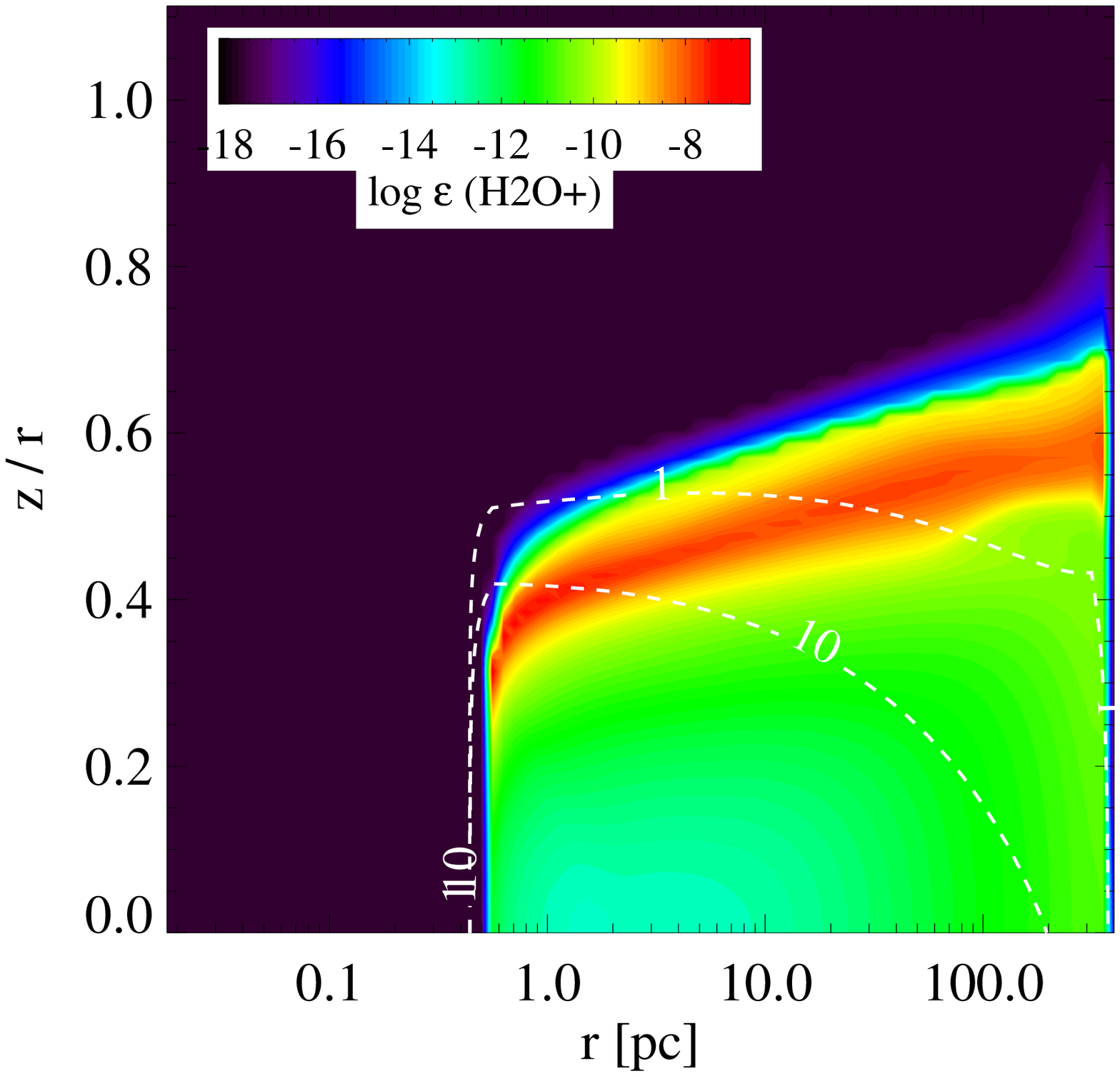}
  \includegraphics[width=4.0cm]{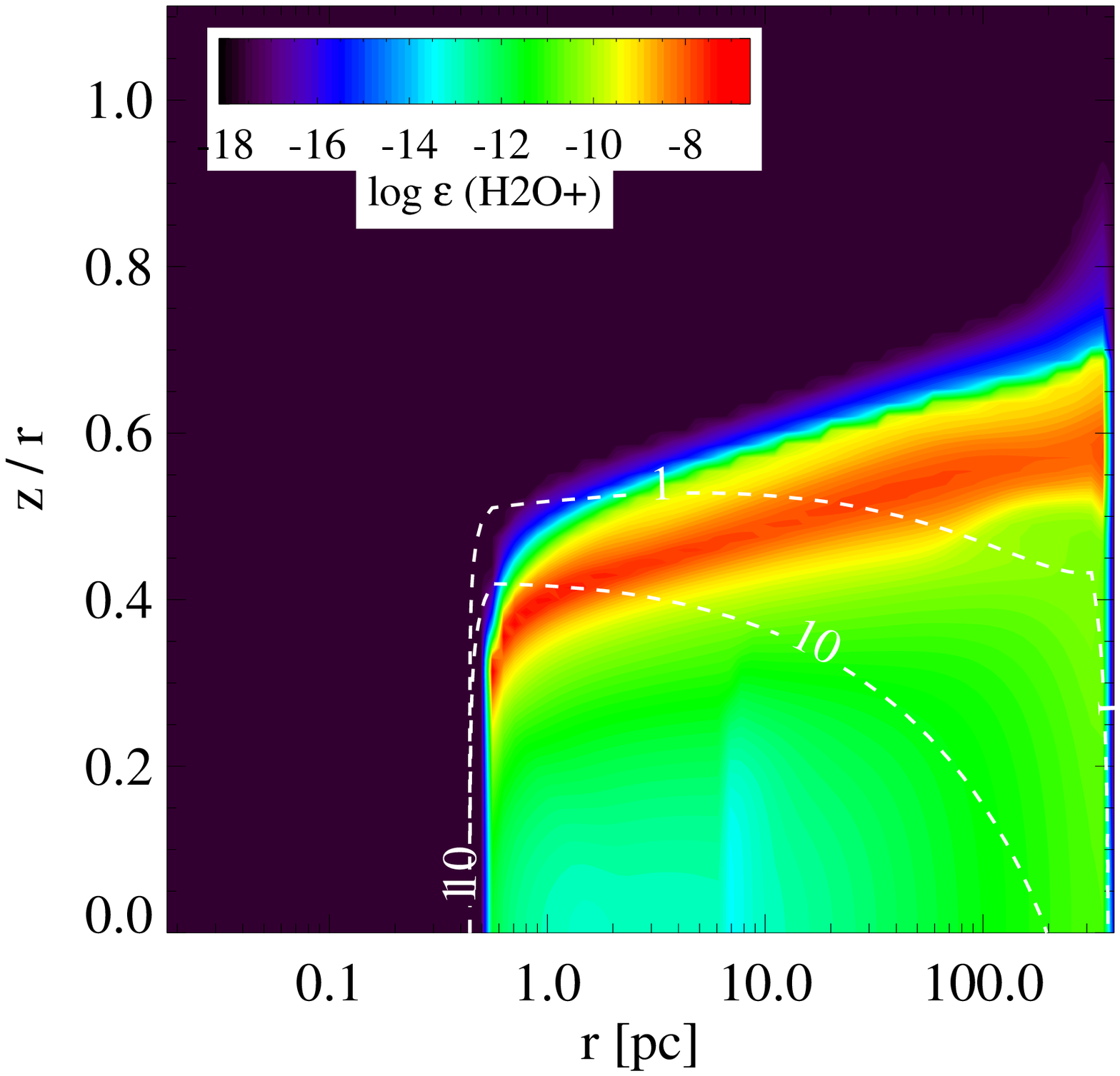}
  \includegraphics[width=4.0cm]{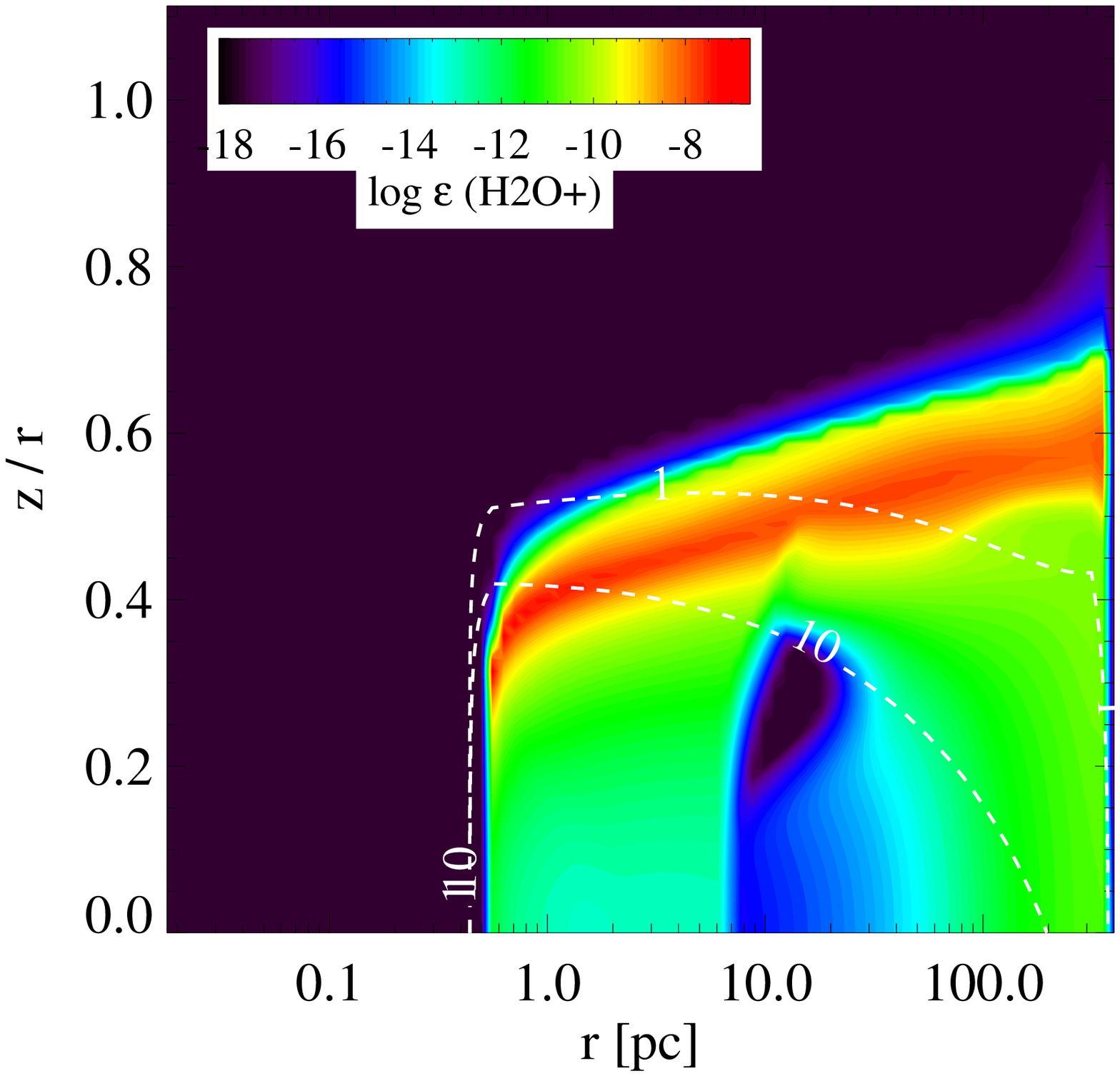}
  \includegraphics[width=4.0cm]{H2Opabundance.ps}
  \includegraphics[width=4.0cm]{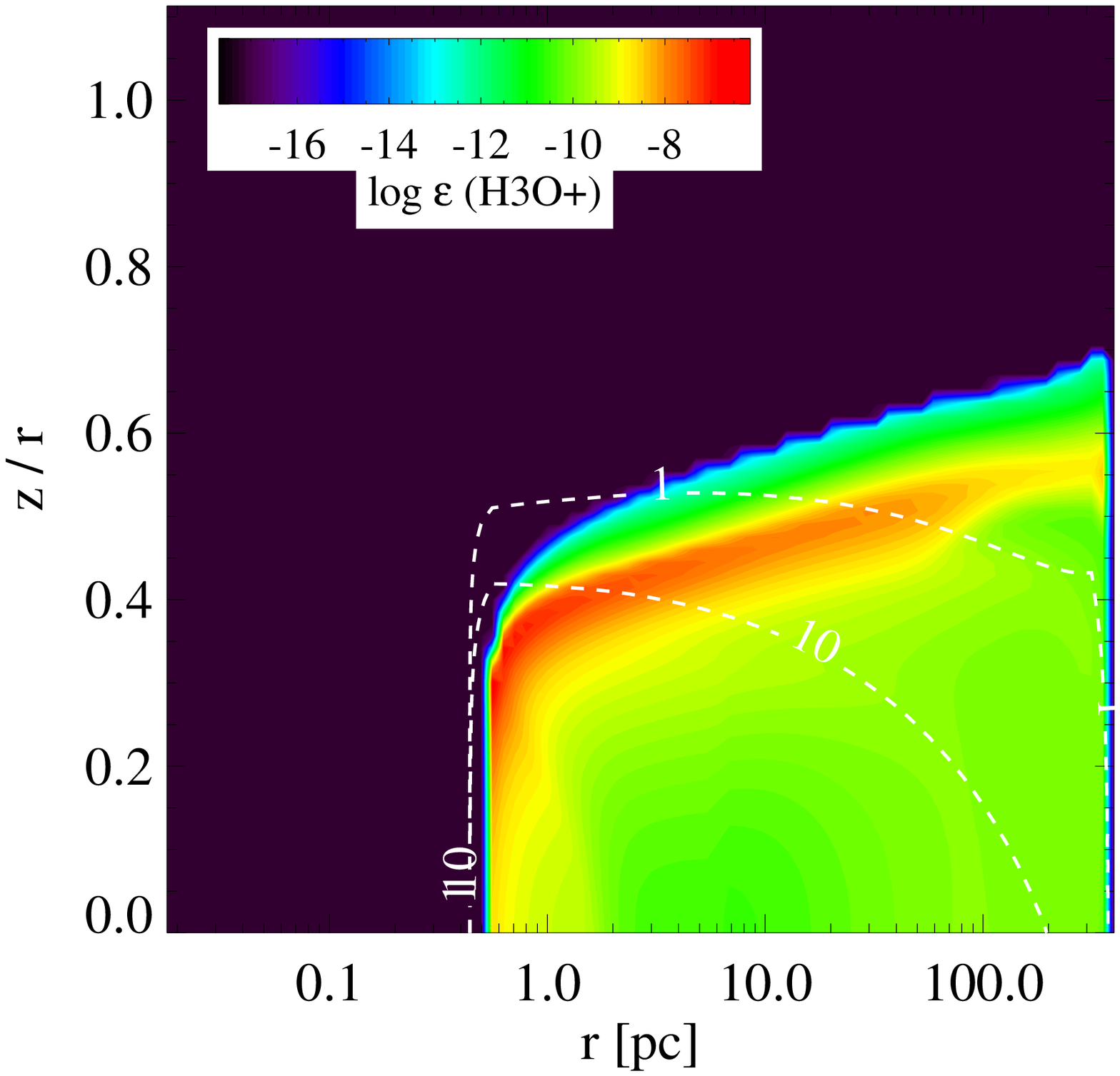}
  \includegraphics[width=4.0cm]{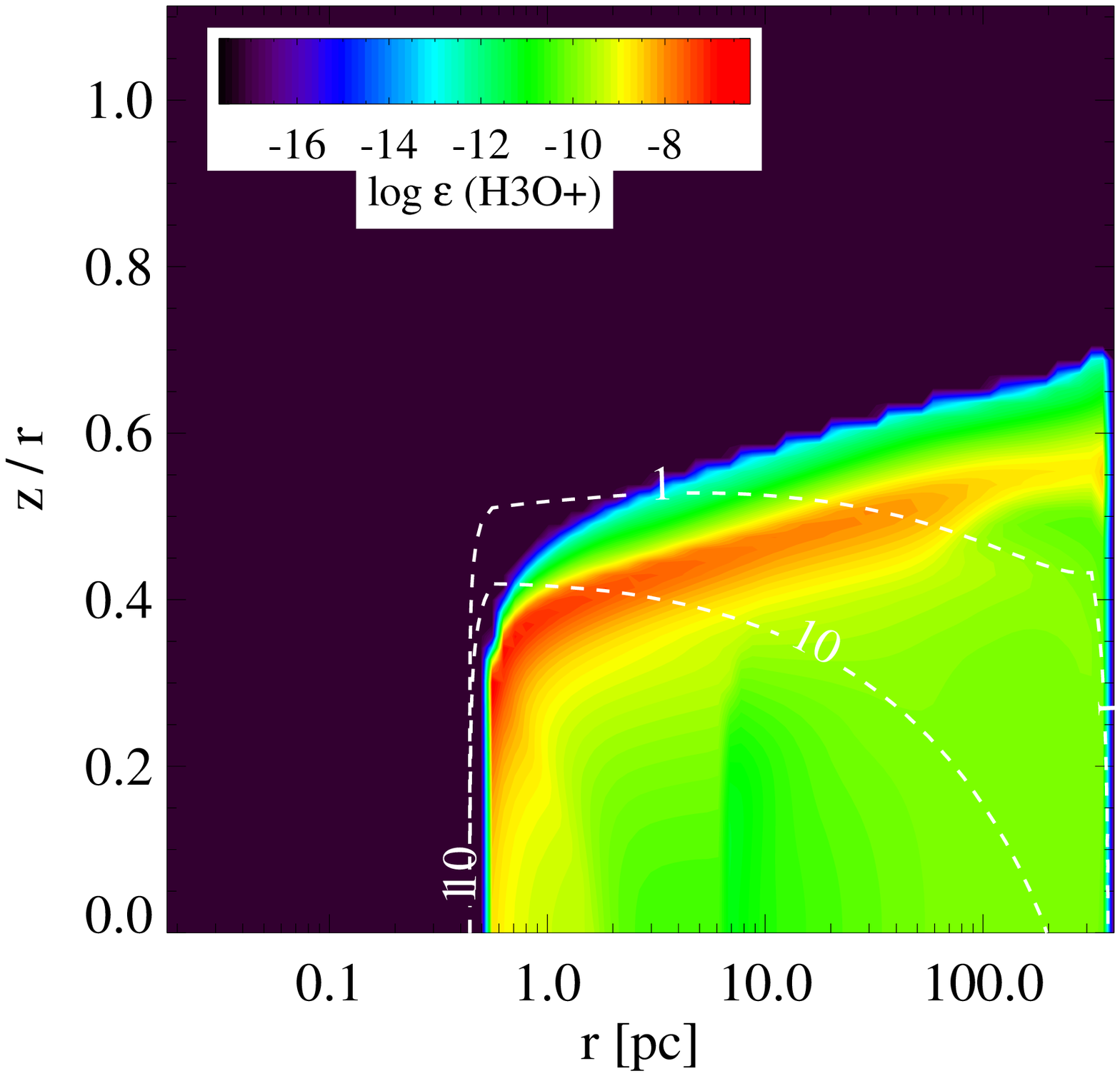}
  \includegraphics[width=4.0cm]{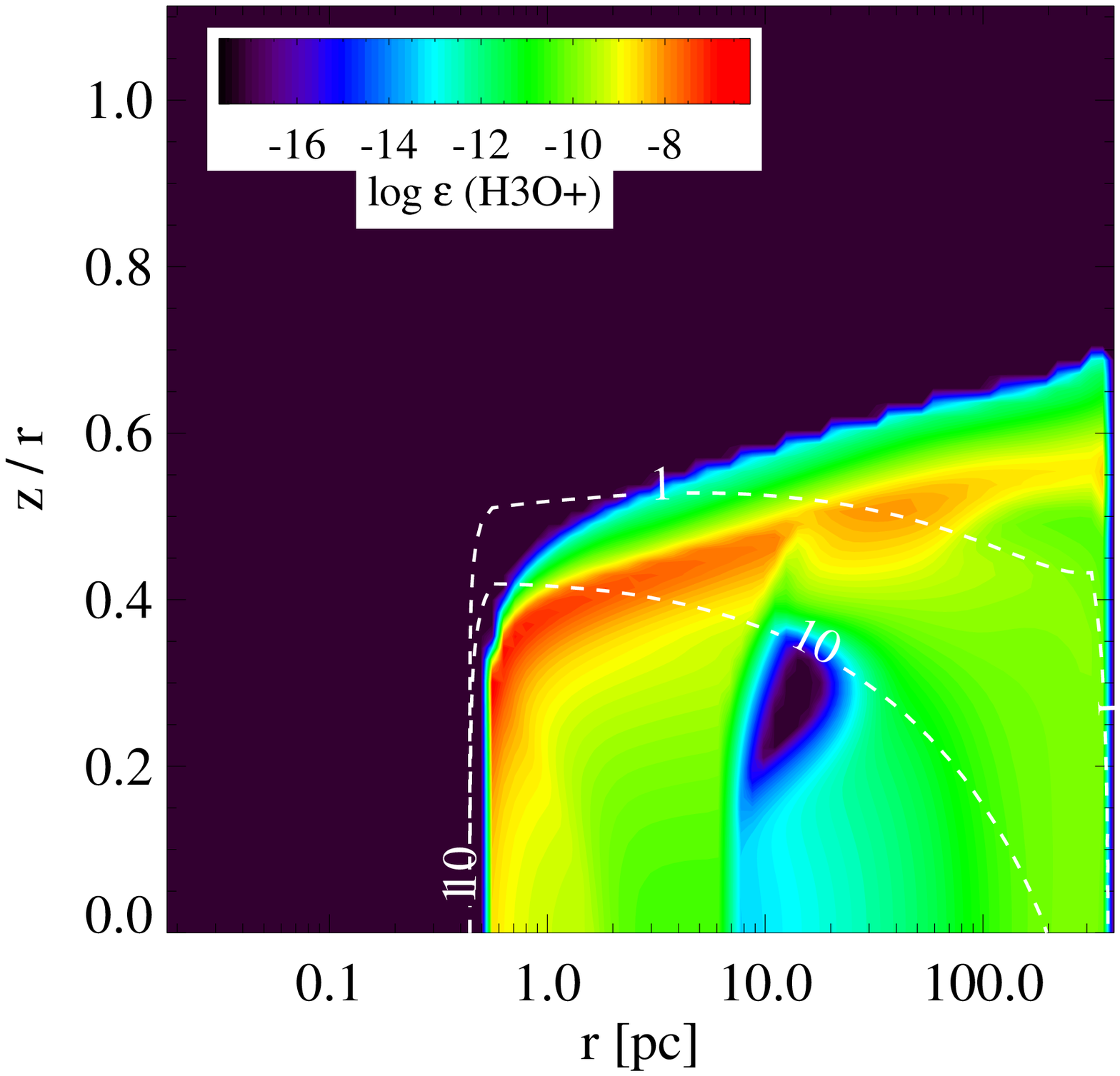}
  \includegraphics[width=4.0cm]{H3Opabundance.ps}
  \caption{Chemical abundances of OH$^+$ (top), H$_2$O$^+$ (middle),
    and H$_3$O$^+$ (bottom) from the time-dependent solution at time
    $t=10^4$, $10^6$, and $10^8$ years (left), and the equilibrium
    solution (right). Contours for $A_{\rm v}=1$ and 10 are
    overplotted.}
  \label{time_dep_chem4}
\end{figure*}

{\it OH$^+$, H$_2$O$^+$, and H$_3$O$^+$:} 
Similar to OH, OH$^+$, H$_2$O$^+$ and H$_3$O$^+$ have a continuous
layer of maximum abundance in the upper part of the disk at
$t=10^4$~yrs: The interruption of these layers by freeze-out effects
is not present at $t=10^4$~yrs as is the case in the equilibrium
model. When time is progressing and H$_2$O is slowly being depleted
onto grains, the freeze-out also affects the ionic species and, as a
result, the high-abundance layer is slowly being affected (OH$^+$ and
H$_2$O$^+$) or even interrupted (H$_3$O$^+$) at $t=10^8$~yrs. Since
the water is not completely frozen-out in the regions with $T <
90-110$~K, the ionized water-related species still have very
significant abundances at radii $R > 7$~pc.

\begin{figure*}
  \centering
  \includegraphics[width=4.0cm]{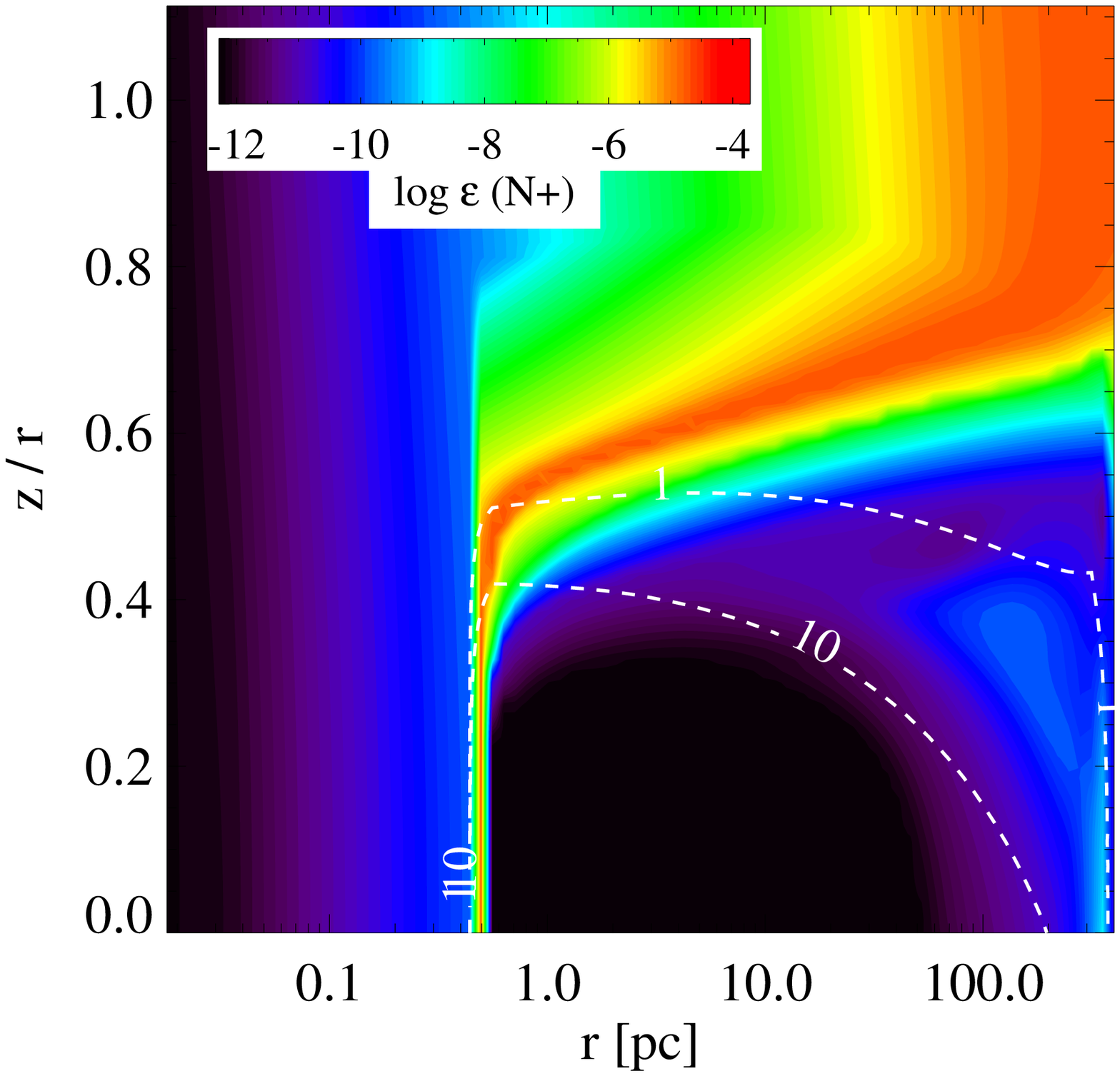}
  \includegraphics[width=4.0cm]{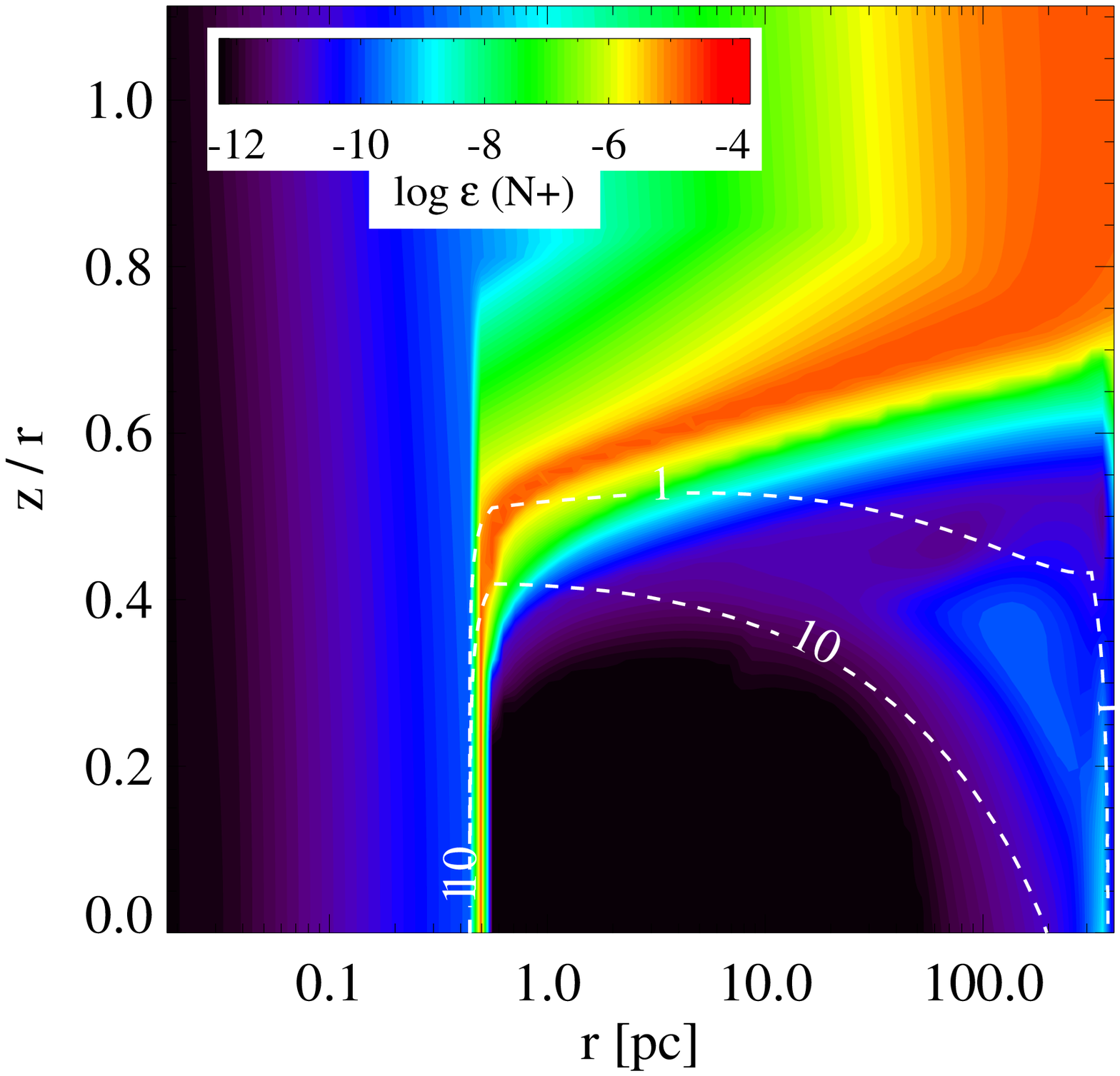}
  \includegraphics[width=4.0cm]{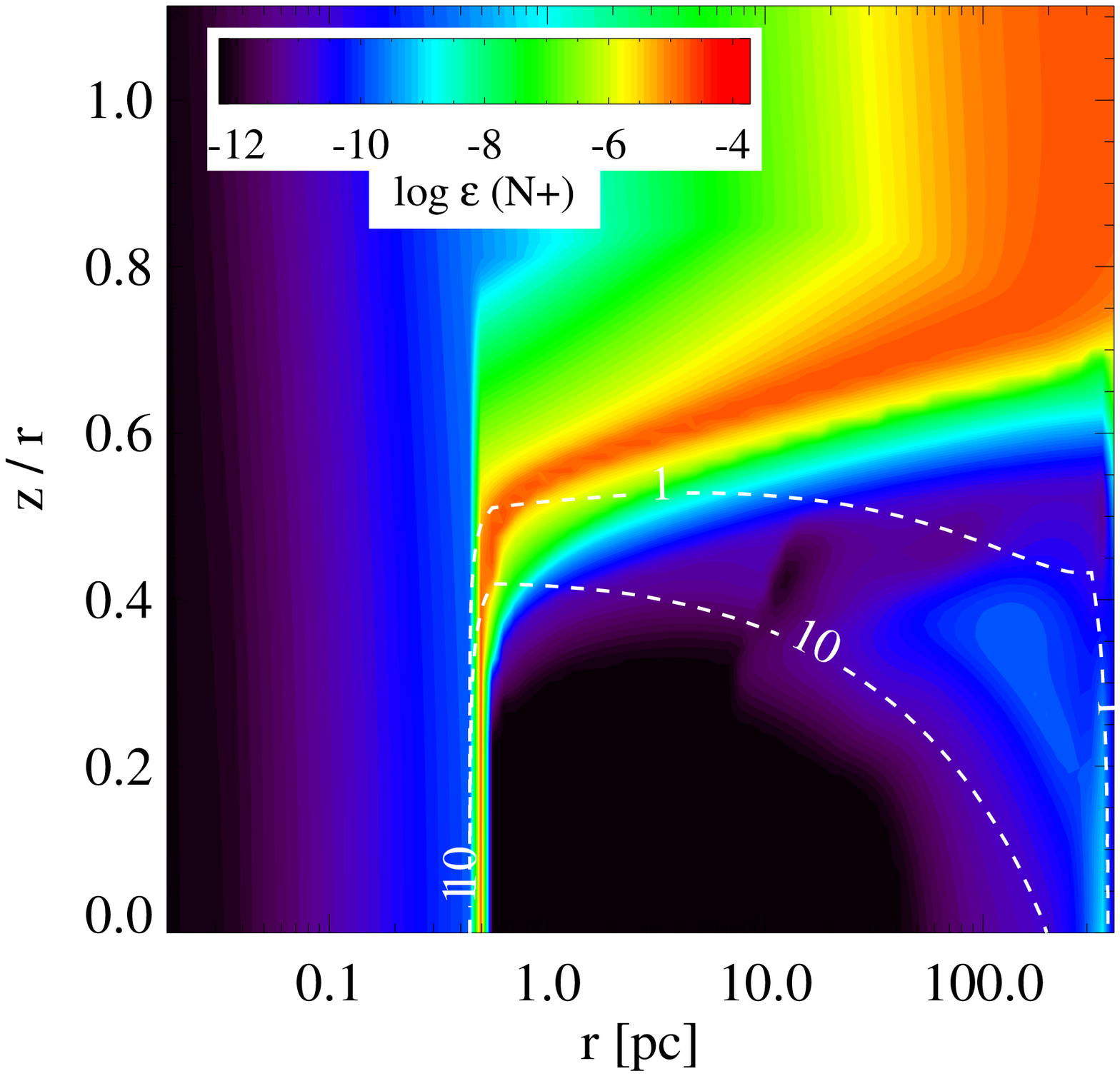}
  \includegraphics[width=4.0cm]{Npabundance.ps}
  \includegraphics[width=4.0cm]{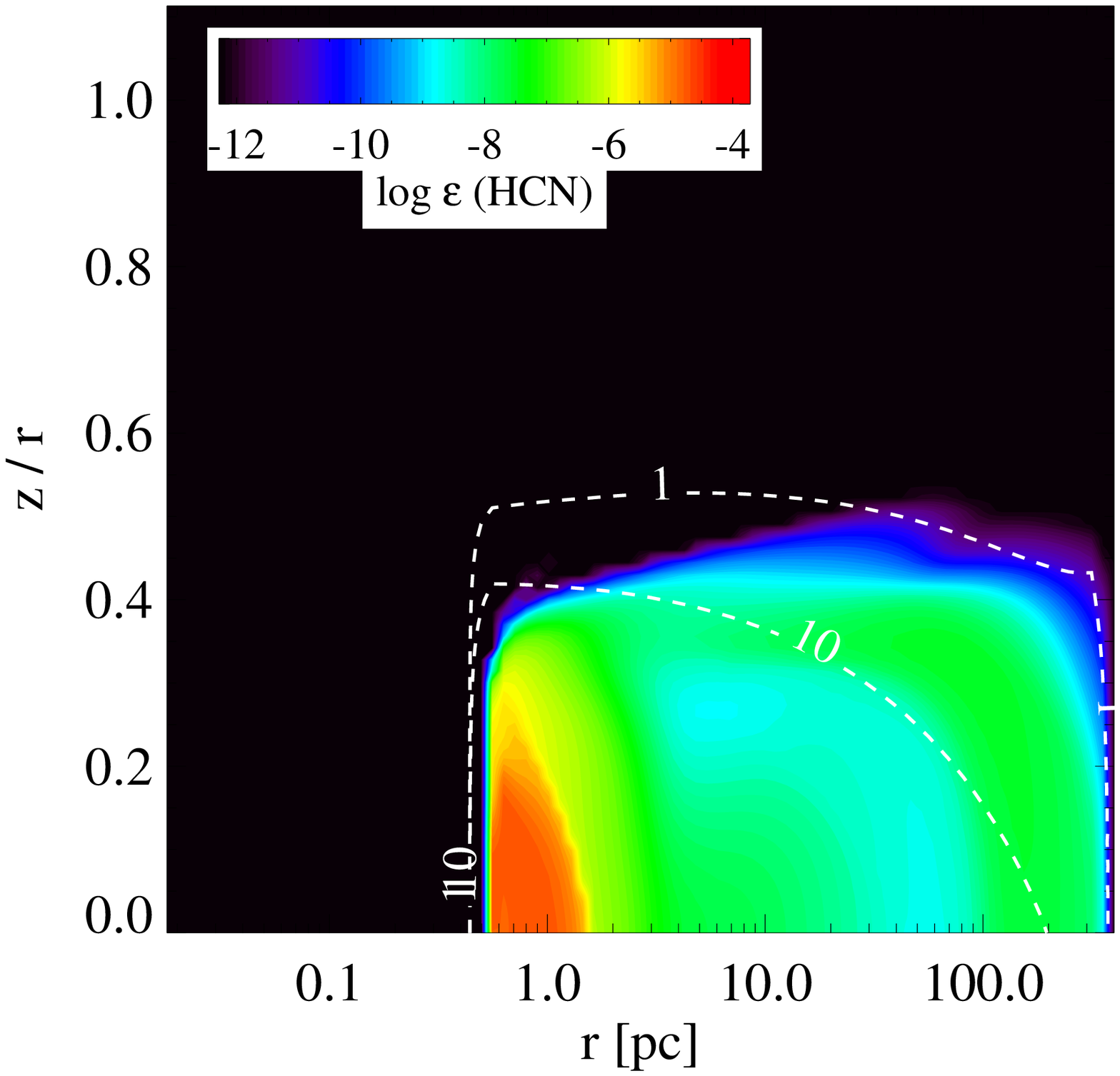}
  \includegraphics[width=4.0cm]{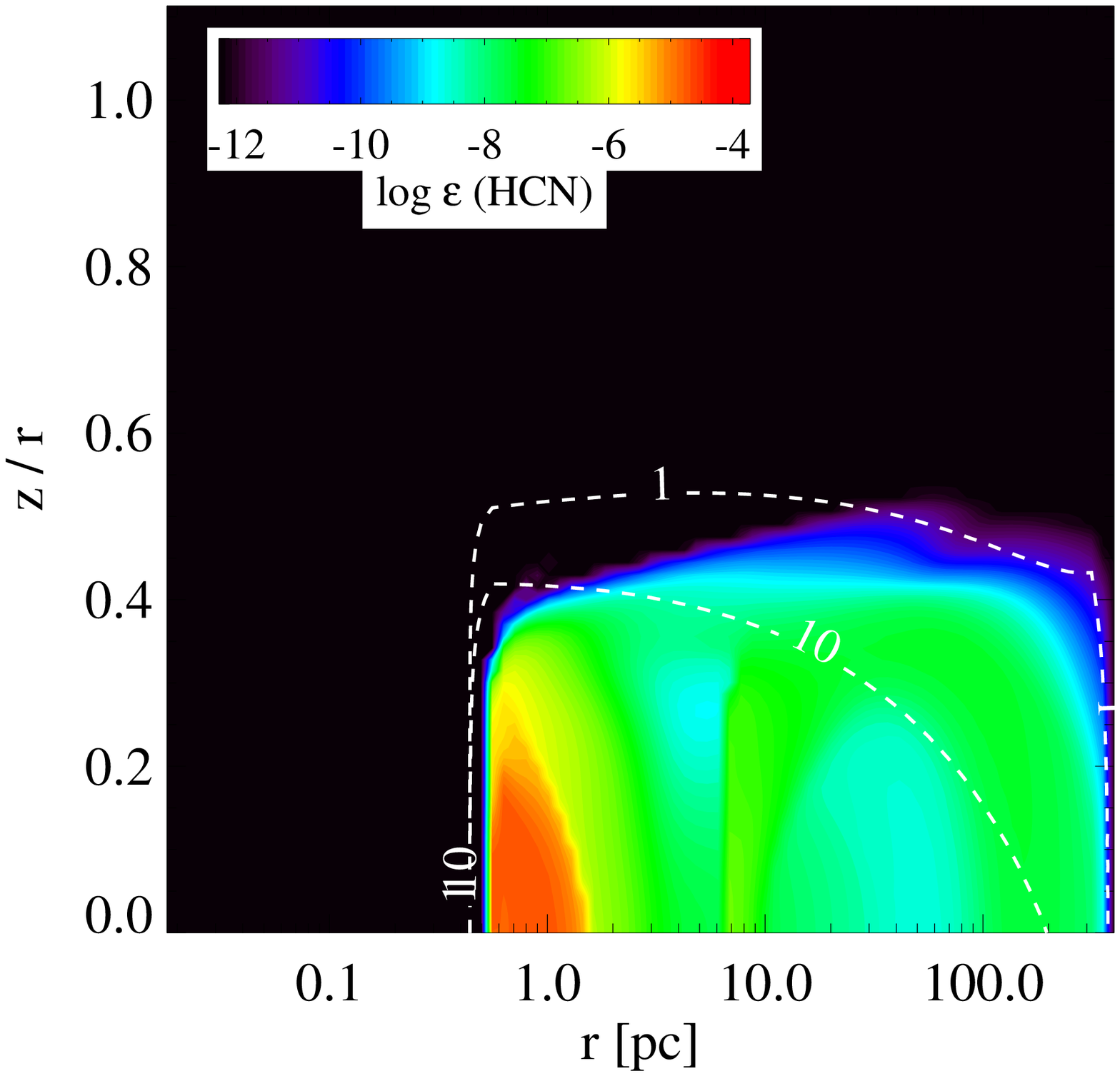}
  \includegraphics[width=4.0cm]{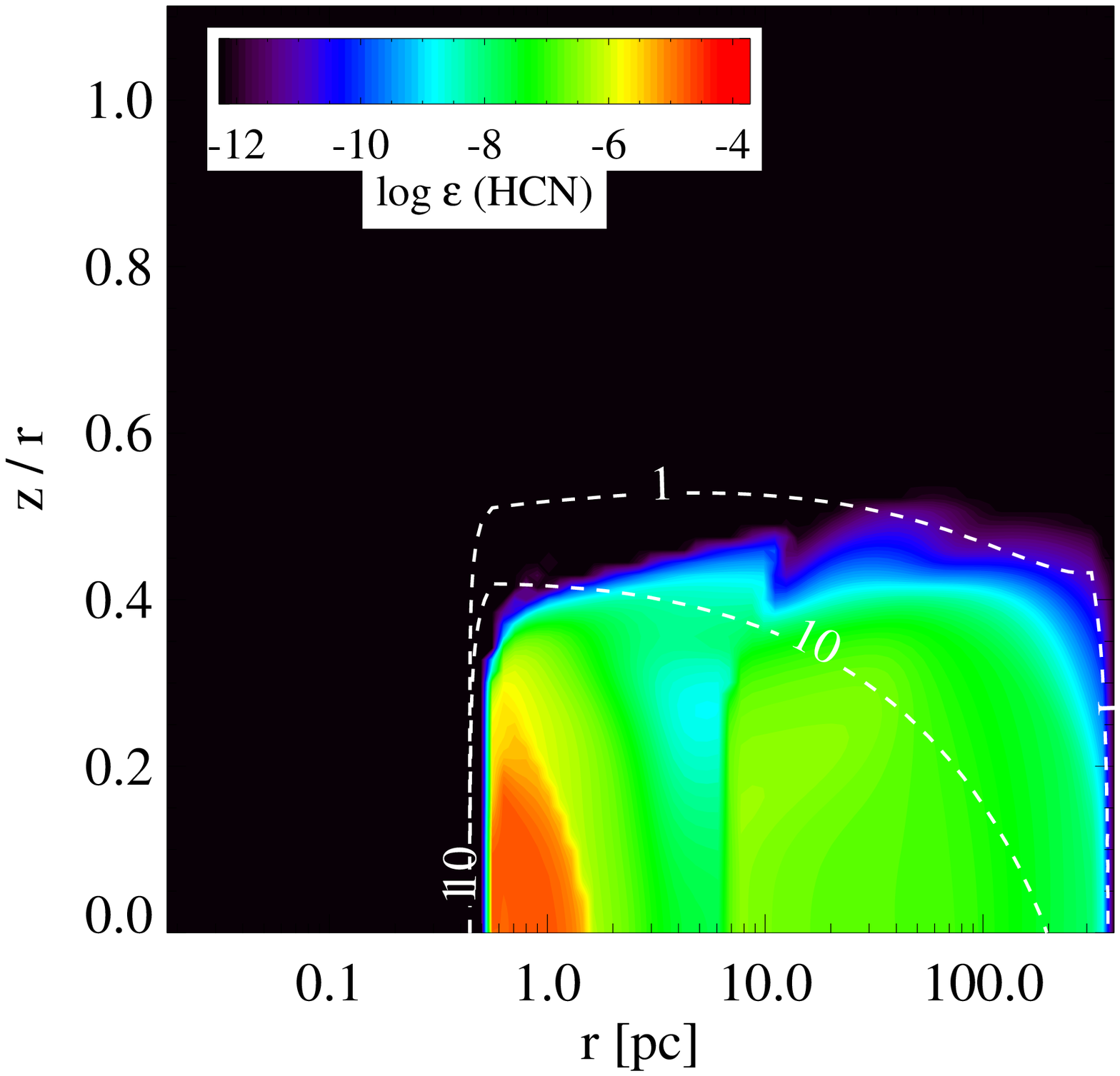}
  \includegraphics[width=4.0cm]{HCNabundance.ps}
  \includegraphics[width=4.0cm]{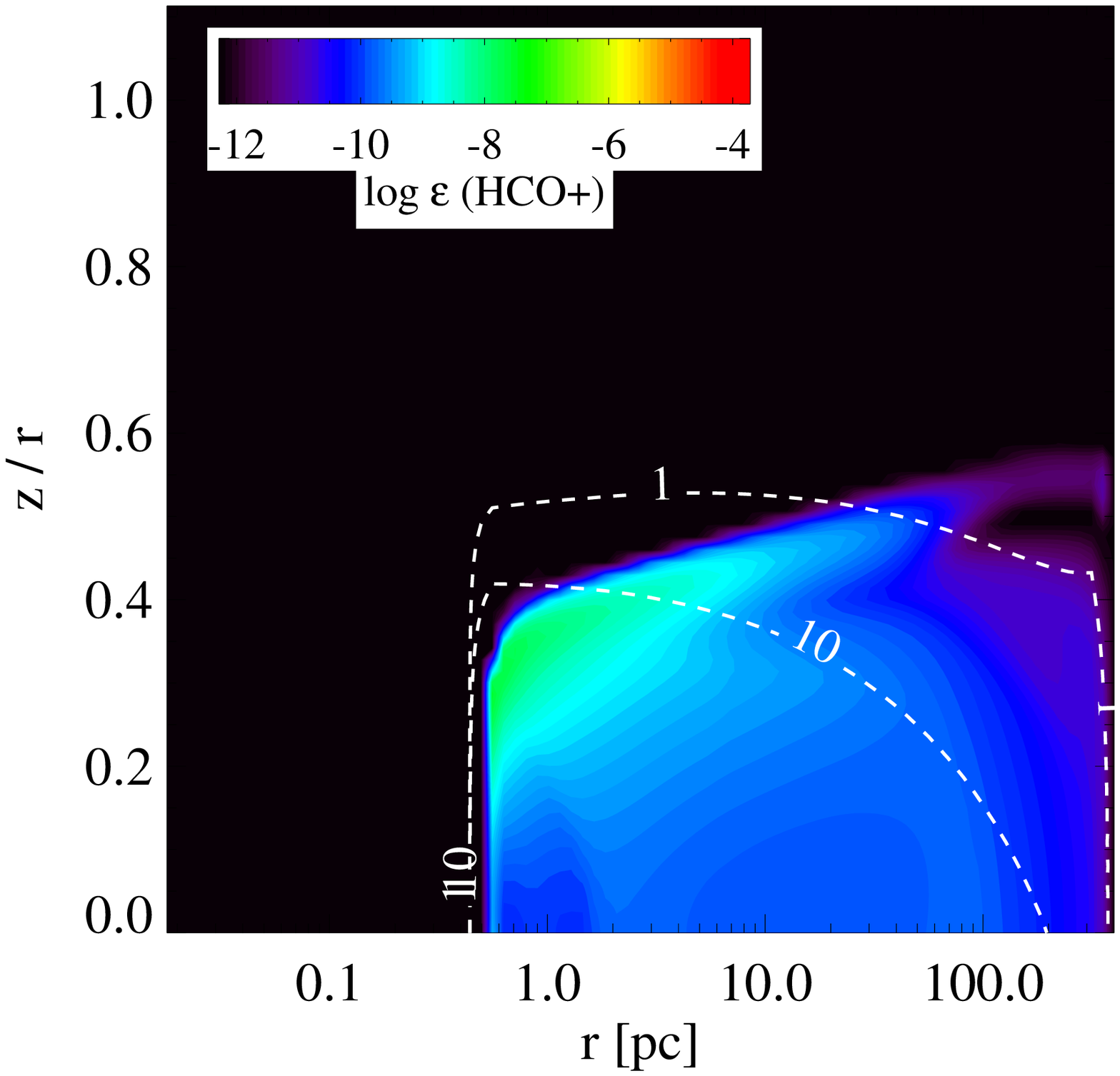}
  \includegraphics[width=4.0cm]{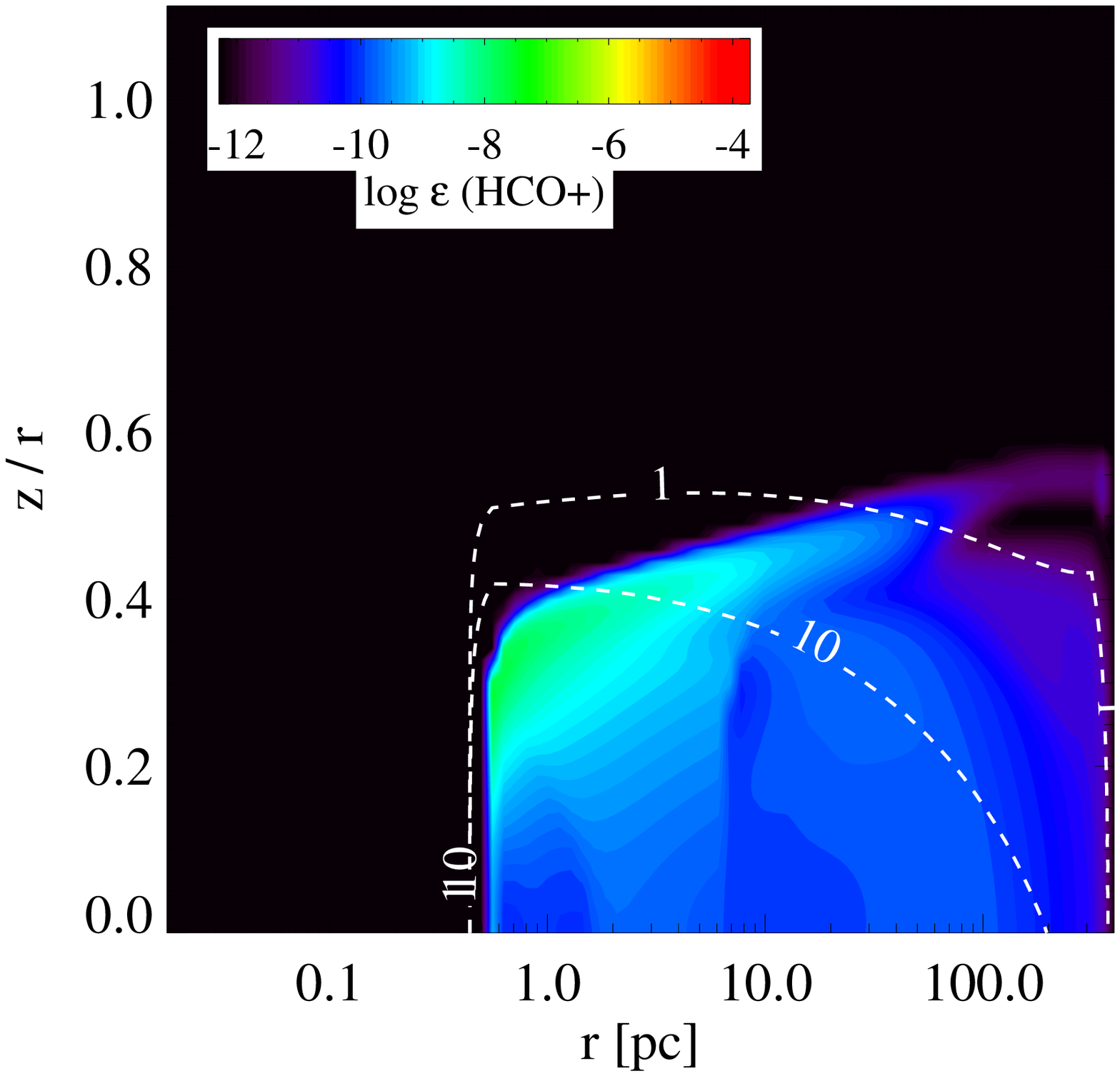}
  \includegraphics[width=4.0cm]{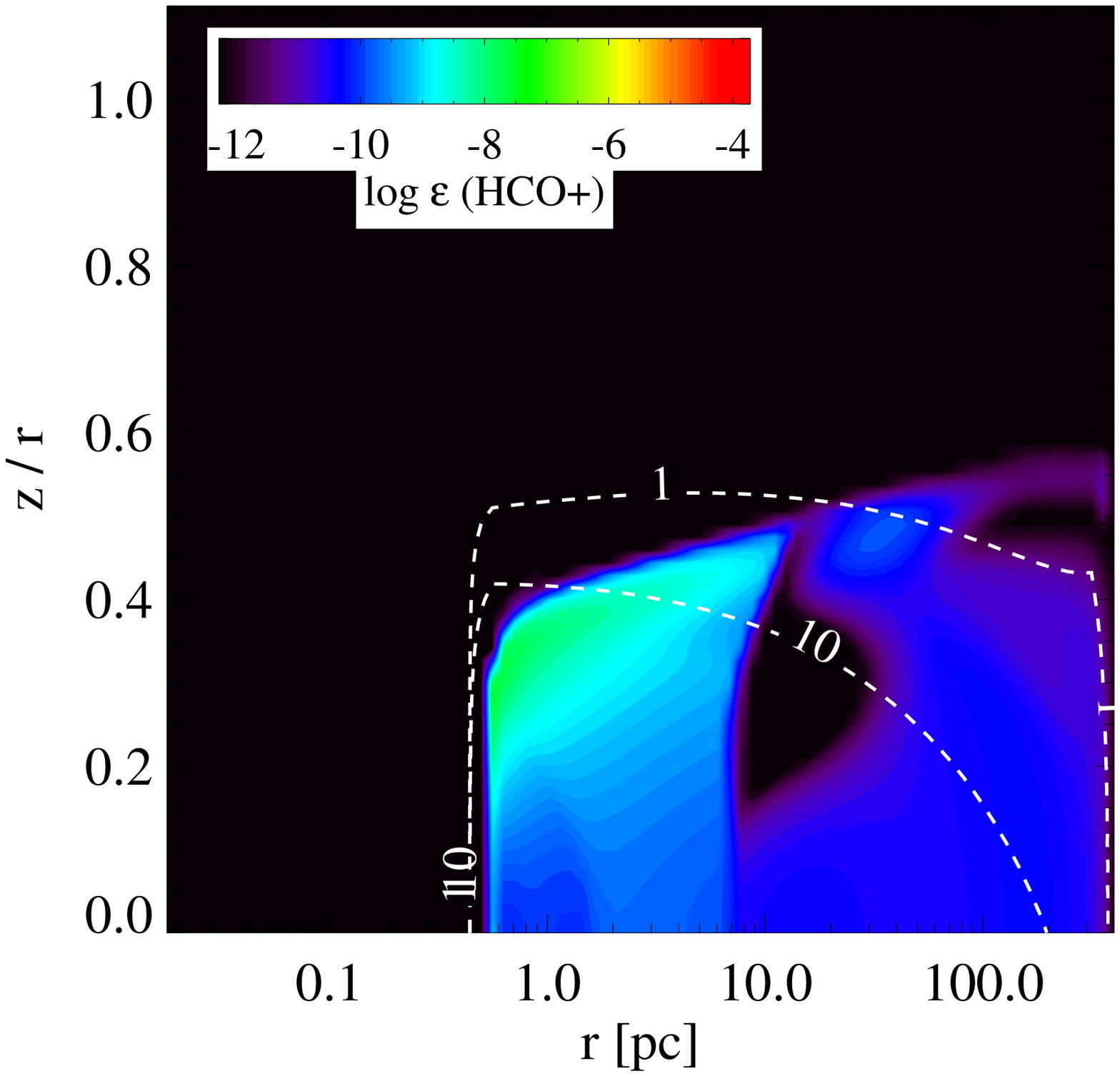}
  \includegraphics[width=4.0cm]{HCOpabundance.ps}
  \caption{Chemical abundances of N$^+$ (top), HCN (middle), and
    HCO$^+$ (bottom) from the time-dependent solution at time
    $t=10^4$, $10^6$, and $10^8$ years (left), and the equilibrium
    solution (right). Contours for $A_{\rm v}=1$ and 10 are
    overplotted.}
  \label{time_dep_chem5}
\end{figure*}

{\it N$^+$, HCN, and HCO$^+$:} N$^+$ settles very quickly to a stable
abundance structure, contrary to HCN and HCO$^+$. Only where
abundances are relatively small ($x_{\rm N^+} \sim 10^{-12}$) in the
outer region of the AGN torus and below $z/R < 0.4$, the N$^+$ shows
some fluctuations over time, which are a consequence of freeze-out
effects. Where oxygen is slowly being depleted from the gas phase
chemistry due to H$_2$O freeze out, the HCN abundance is slowly
increasing from $x_{\rm HCN} \sim 10^{-9}$ ($t=10^4$~yrs) to $10^{-7}$
($t=10^8$~yrs). The solution is very similar to the equilibrium
solution, although the HCN abundance in the region where water freezes
out is lower by a factor 3-10. HCO$^+$ is abundant throughout the
disk, when no water is frozen out, and is slowly depleted with
time. Contrary to the equilibrium model, HCO$^+$ is still present at
the level $x_{\rm HCO^+} \sim 10^{-10}$ in the freeze-out zone after
100 million years.

\section{Effects of time dependent chemistry}

\subsection{Total masses of atomic and molecular species}

\begin{table}

  \centering
  \caption{Total masses in the disk of atomic and molecular species}
  \label{atomic_molecular_masses}
  \begin{tabular}{lrrrrrrrr}

    \hline
    Species    & \multicolumn{4}{c}{Mass [M$_{\odot}$]} \\
               &  $10^4$~yrs & $10^6$~yrs & $10^8$~yrs & Equil. \\
    \hline
    H          & 8.0(7) & 8.0(7) & 8.0(7) & 8.0(7) \\
    H$_2$      & 6.9(8) & 6.9(8) & 6.9(8) & 6.9(8) \\
    H$^-$      & 2.4(-3) & 2.4(-3) & 2.4(-3) & 2.4(-3) \\
    H$^+$      & 1.9(7) & 1.9(7) & 1.9(7) & 1.9(7) \\
    H$_2^+$    & 1.2(0) & 1.2(0) & 1.2(0) & 1.2(0) \\
    H$_3^+$    & 2.6(1) & 2.6(1) & 2.5(1) & 2.6(1) \\
    C$^+$      & 1.0(5) & 1.0(5) & 1.0(5) & 1.0(5) \\
    C          & 6.0(5) & 6.0(5) & 8.0(5) & 1.7(6) \\
    CO         & 1.1(6) & 1.1(6) & 6.8(5) & 4.5(4) \\
    CO ice     & 2.1(-3) & 2.1(-3) & 2.6(-3) & 6.0(-8) \\
    O          & 2.7(6) & 2.7(6) & 2.4(6) & 6.6(5) \\
    OH         & 1.9(3) & 1.8(3) & 1.4(3) & 3.7(2) \\
    H$_2$O     & 1.9(4) & 1.5(4) & 9.1(3) & 5.7(3) \\
    H$_2$O ice & 2.6(3) & 8.4(4) & 8.6(5) & 3.2(6) \\
    OH$^+$     & 3.3(1) & 3.3(1) & 3.3(1) & 3.3(1) \\
    H$_2$O$^+$ & 2.3(0) & 2.3(0) & 2.3(0) & 2.2(0) \\
    H$_3$O$^+$ & 2.3(0) & 2.2(0) & 1.7(0) & 8.6(-1) \\
    N$^+$      & 6.6(3) & 6.6(3) & 6.6(3) & 6.6(3) \\
    HCN        & 7.0(2) & 7.3(2) & 1.4(3) & 2.2(3) \\
    HCO$^+$    & 1.5(0) & 1.3(0) & 6.9(-1) & 1.7(-1) \\
    \hline
  \end{tabular}
\end{table}

The masses of the discussed atomic and molecular species for the
time-dependent ($t=10^4$, $10^6$, and $10^8$~yrs) and equilibrium
model are listed in Table \ref{atomic_molecular_masses}. The hydrogen
species (H, H$_2$, H$^+$,H$_2^+$, and H$_3^+$) are not affected by
oxygen depletion, and show constant mass fractions over time. The same
is the case for C$^+$ and N$^+$ that are produced in the
ionized/atomic part of the AGN disk. This is contrary to the total
mass of neutral carbon, that is increasing with time. After 100
million years, it is still a factor of two lower than the mass
resulting from the equilibrium model. For CO the difference is even
larger, but contrary to C, the CO mass is decreasing with time. The CO
mass is more than a factor 10 higher in the time-dependent model after
100 million years. The amount of CO ice is almost negligible, but note
that in the time-dependent model, the CO ice mass is 5 orders of
magnitude higher. Atomic oxygen is also depleted over time, but
similar to gas-phase CO, the mass in atomic oxygen hydrogen is much
higher than the equilibrium model. After $10^8$~yrs, the atomic oxygen
mass is a factor of 4 higher than the equilibrium model. Water is
depleted by a factor 2 after 100 million years with respect to its
initial mass, but a factor 2 higher than the equilibrium model. The
amount of water ice reaches almost a million solar masses after 100
million years, which is a factor 4 below the equilibrium model.

HCN profits from oxygen depletion, and its mass increases by a factor
2 after 100 million years, which is still 30 percent lower than the
equilibrium model. On the other hand, HCO$^+$ is slowly being depleted,
and is reduced by 60 percent over time, but a factor 4 higher than
equilibrium model. The mass ratio of HCN/HCO$^+$ is $\sim 10^4$ for the
entire disk in the equilibrium model. On the other hand, the
time-dependent HCN/HCO$^+$ mass ratio is smaller by more than an order
of magnitude compared to the equilibrium model. This will have
consequences for the observed line emission.

It is not very likely that clouds will survive for times as large as
$10^8$ yrs. Studies of star-forming regions in our galaxy showed that
their ages range between 1 to 15 Myrs
 \citep{Palla2000,Geus1989,Webb1999,Mamajek1999}. Ionic species, such as
C$^+$ and N$^+$, have reached equilibrium by that time. This is not
the case for important molecular species, such as CO and H$_2$O. This
strongly argues that the initial conditions chosen for the molecular
cloud play an important role. The effect of the choice of initial
conditions was investigated by starting with a partially ionized and
partly atomic medium. Although this is probably not a very realistic
initial state, it is illustrative to note that the mass evolution of
several important species is severely altered. For example, the CO
mass in the models with the molecular initial state decreases from
$1.1\times 10^6$ M$_{\odot}$ ($10^4$ yrs) to $6.8\times 10^6$
M$_{\odot}$ ($10^8$ yrs), compared to $1.5\times 10^4$ to $8.7\times
10^5$ M$_{\odot}$ when starting with the ionic/atomic initial
state. In both cases the C$^+$ mass is approximately constant over
time ($10^5$ M$_{\odot}$), which results in a decreasing CO/C$^+$
ratio for the molecular case and an increasing ratio for the
ionic/atomic case. In both cases, the HCN mass increases with time,
but the absolute mass is approximately a factor two higher at all
times, when starting with molecular intitial conditions. HCO$^+$ shows
a similar behaviour as CO, and increases from $3.2\times 10^{-2}$ to
1.1 M$_{\odot}$ in the ionic/atomic case, and decreases from 1.5 to
0.7 M$_{\odot}$ in the molecular case. Although the behaviour is
different, HCN/HCO$^+$ mass ratio is always smaller in the
time-dependent case for ages $\tau_{chem} > 10^4-10^5$~yrs than in the
equilibrium case.

\subsection{Consequences for observables}

We calculated the line emission for a number of fine-structure lines
and rotational transitions of CO, H$_2$O, HCN, and HCO$^+$ for a
face-on disk and using an escape probability method (Table
\ref{atomic_molecular_line_fluxes}).

The [CII], [CI] and [OI] fine-structure lines are hardly affected over
time, even though the [OI] and [CI] mass is varying over time. Both
[OI] and [CI] are optically thick ($\tau > 100$ and 10, respectively)
at those radii in which 90 percent of the emission is being
produced. The line emission of CO in the time-dependent model is much
larger than in the equilibrium model, between a factor 5 to 10, even
after 100 million years. The CO line luminosities are much closer to
the ones observed for Mrk 231 with Herschel \citep{VdWerf2010}. The
water lines are also brighter in the equilibrium model, but they are
not as much enhanced as the CO lines. The H$_2$O lines are close to the
equilibrium model after 100 million years.

The line ratios for the lowest ($J_{\rm up} \le 4$) rotational
transitions of HCN and HCO$^+$ range between 2-10 (between $t=10^4$
and $10^8$~yrs), consistent with current observations of AGN, but
still a factor of 2-4 higher. The time-dependent HCN/HCO$^+$ line
ratios agree better with values observed for galaxies with AGN. The
latter range from 1-2 for $J=1-0$ \citep{Krips2008} and are slightly
larger than 3 for $J=4-3$ \citep{Perez2009}. This suggests that current
HCN and HCO$^+$ observations of AGN disfavor a state that is
characterized by complete chemical equilibrium. This conclusion even
holds for two very different initial states. In a follow-up paper,
line intensities for a larger parameter study will be presented for
the diagnostic molecules identified above.

\begin{table*}
  \centering
  \caption{Line fluxes for commonly observed atomic and molecular species}
  \label{atomic_molecular_line_fluxes}
  \begin{tabular}{lrrrrrrrrr}
    \hline
    Line    &  $\lambda$  & \multicolumn{4}{c}{Luminosity [$L_\odot$]} \\
            &  [$\mu$m]   & $10^4$ & $10^6$ &  $10^8$ & Equil. \\
    \hline
    CO $J=1-0$  & 2600.76 & 2.4(6) & 2.4(6) & 2.3(6) & 2.9(5) \\
    CO $J=2-1$  & 1300.40 & 1.4(7) & 1.4(7) & 1.3(7) & 9.3(5) \\
    CO $J=3-2$  & 866.96  & 3.1(7) & 3.1(7) & 3.0(7) & 1.2(6) \\
    CO $J=4-3$  & 650.25  & 4.7(7) & 4.7(7) & 4.5(7) & 1.8(6) \\
    CO $J=5-4$  & 520.23  & 5.4(7) & 5.4(7) & 5.0(7) & 2.7(6) \\
    CO $J=6-5$  & 433.56  & 5.0(7) & 5.0(7) & 4.4(7) & 4.1(6) \\
    CO $J=7-6$  & 371.65  & 4.0(7) & 4.0(7) & 3.4(7) & 5.9(6) \\
    CO $J=8-7$  & 325.23  & 3.3(7) & 3.3(7) & 2.9(7) & 7.8(6) \\
    CO $J=9-8$  & 289.12  & 3.0(7) & 3.0(7) & 2.7(7) & 9.9(6) \\
    CO $J=10-9$ & 260.24  & 2.9(7) & 2.9(7) & 2.6(7) & 1.2(7) \\
    CO $J=11-10$ & 236.61 & 2.8(7) & 2.8(7) & 2.5(7) & 1.4(7) \\
    CO $J=12-11$ & 216.93 & 2.7(7) & 2.7(7) & 2.4(7) & 1.6(7) \\
    CO $J=13-12$ & 200.27 & 2.6(7) & 2.6(7) & 2.3(7) & 1.7(7) \\
    $[$OI$]$ $^3P_1-^3P_2$ & 63.18  & 4.9(9) & 4.9(9) & 4.8(9) & 4.8(9) \\
    $[$OI$]$ $^3P_0-^3P_1$ & 145.53 & 5.5(8) & 5.5(8) & 5.5(8) & 5.5(8) \\
    $[$CII$]$ $^2P_{3/2}-^2P_{1/2}$ & 157.74 & 2.8(9) & 2.8(9) & 2.8(9) & 2.8(9) \\
    $[$CI$]$ $^3P_1-^3P_0$ & 609.14 & 1.2(8) & 1.2(8) & 1.3(8) & 1.6(8) \\
    $[$CI$]$ $^3P_2-^3P_1$ & 370.42 & 2.5(8) & 2.5(8) & 2.7(8) & 4.2(8) \\
    p-H$_2$O $1_{11}-0_{00}$ & 269.27 & 2.8(7) & 2.4(7) & 1.0(7) & 5.9(6) \\
    p-H$_2$O $2_{02}-1_{11}$ & 303.46 & 1.6(7) & 1.3(7) & 4.2(6) & 3.9(6) \\
    p-H$_2$O $2_{11}-2_{02}$ & 398.64 & 8.9(6) & 7.1(6) & 3.5(6) & 3.5(6) \\
    p-H$_2$O $2_{20}-2_{11}$ & 243.97 & 1.1(7) & 8.5(6) & 6.2(6) & 6.2(6) \\
    o-H$_2$O $3_{12}-3_{03}$ & 273.19 & 7.9(6) & 6.6(6) & 5.4(6) & 5.4(6) \\
    o-H$_2$O $3_{21}-3_{12}$ & 257.79 & 8.6(6) & 7.7(6) & 7.1(6) & 7.1(6) \\
    p-H$_2$O $4_{22}-4_{13}$ & 248.25 & 4.3(6) & 4.2(6) & 4.2(6) & 4.2(6) \\
    o-H$_2$O $5_{23}-5_{14}$ & 212.53 & 4.1(6) & 4.1(6) & 4.1(6) & 4.1(6) \\
    $[$NII$]$ $^3P_1-^3P_0$ & 205.24 & 9.6(8) & 9.6(8) & 9.6(8) & 9.6(8) \\
    HCO$^+$ $J=1-0$ & 3361.33 & 2.8(4) & 2.7(4) & 1.6(4)  & 2.8(3)  \\
    HCO$^+$ $J=4-3$ & 840.38 & 3.4(5) & 3.1(5) & 1.8(5) & 1.4(5) \\
    HCN $J=1-0$ & 3382.46 & 7.7(4) & 8.0(4)  & 1.8(5) & 1.8(5) \\
    HCN $J=4-3$ & 845.66 & 6.6(5) & 7.4(5) & 2.0(6) & 4.6(6) \\
    \hline
  \end{tabular}
\end{table*}    

\section{Summary and conclusions}

We presented a proof-of-concept for the time-dependent chemistry in an
AGN.  We found that the freeze-out of water is strongly suppressed for
durations shorter than 100 Myr. This affects the bulk of the oxygen
and carbon chemistry occurring in AGN. The commonly used AGN tracer
HCN/HCO$^+$ is strongly time-dependent, with ratios of order a few for
dynamical times that are shorter than a million years. This suggests
that high angular resolution ALMA observations, which probe the
narrow-line region on scales smaller than 100 pc, can reveal the
evolutionary state of an AGN as it evolves under strong fluctuations
in the SMBH accretion rate. We should bare in mind though that the
model results do depend on the initial chemical state of the
clouds. Especially the absolute mass of the diagnostic species are
affected by this, and to a lesser extend also the mass ratios of the
species. In order to test whether a result is robust, one should
explore a range of possible initial states. We also note that the AGN
torus is a clumpy medium with likely a complicated velocity
field. This implies that clouds will also see attenuation variations
on time-scales $\tau < 10^6$~yrs at radii $R < 100$ pc from the
center. These effects are beyond the scope of this paper, but should
be investigated carefully as well.

Strong evolutionary trends, occurring on time scales of $10^4-10^8$ years,
are also found in species like, H$_3$O$^+$, CO, and H$_2$O.
These species reflect, respectively, time dependent effects in the
ionization balance, the transient nature of the production of molecular
gas as traced by the X-factor, and the freeze-out/sublimation of water (key
to much of the grain surface chemistry).

In closing, we note that an accurate representation of AGN chemistry requires
time dependence. As a next step to the work presented here, to be pursued in
a follow-up paper, it is necessary to calculate the response of the disk's
thermal structure to radiative and hydrodynamic events and to compute line
intensity maps for the diagnostic species identified above.

\bibliographystyle{apj}

\end{document}